\documentclass[12pt]{article}
\usepackage[authoryear, round]{natbib}
\usepackage{url} 

\usepackage{titling}
\usepackage{authblk} 

\usepackage{times}
\usepackage{bm}

\usepackage{grffile} 

\usepackage{amsfonts, amssymb, amsthm}
\usepackage{amsmath}
\usepackage{enumerate}
\usepackage{nicefrac}       
\usepackage{microtype}      
\usepackage{xcolor}         
\usepackage{booktabs}
\usepackage{caption, subcaption}
\usepackage{multicol, multirow}

\usepackage{float}
\usepackage{adjustbox}
\graphicspath{{./art/}}

\usepackage{algpseudocode}
\usepackage{algorithm}

\usepackage{xr}
\makeatletter
\newcommand*{\addFileDependency}[1]{
  \typeout{(#1)}
  \@addtofilelist{#1}
  \IfFileExists{#1}{}{\typeout{No file #1.}}
}
\makeatother

\usepackage{graphicx}

\newcommand{\blind}{1}

\newtheorem{theorem-new}{Theorem} 

\newtheorem{theorem}{Theorem} 

\newtheorem{remark}{Remark} 

\newcommand{\dx}{\color{black}}

\usepackage{accents, bm, bbm}

\renewcommand{\hat}{\widehat}
\renewcommand{\tilde}{\widetilde}

\renewcommand{\phi}{\varphi}

\newcommand{\M}{{\cal M}}
\newcommand{\I}{{\cal I}}

\newcommand{\ep}{\mathbb{E}}
\newcommand{\pr}{\mathbb{P}}
\newcommand{\var}{\mathrm{Var}}

\newcommand{\ii}{{\mathbbm{i}}}
\newcommand{\td}{{\mathrm{d}}}

\usepackage{xcolor}
\definecolor{yuan-prussianblue}{rgb}{0.0, 0.19, 0.33}
\definecolor{yuancolor-edit}{rgb}{0.29, 0.59, 0.82}

\definecolor{shaocolor}{RGB}{1,132,127}
\newcommand{\shao}[1]{{\color{shaocolor}{#1}}}
\renewcommand{\shao}[1]{{\color{black}{#1}}}

\definecolor{revision-color}{rgb}{0,0.29,0.43}

\definecolor{revision-color-filtered}{RGB}{75, 120, 102}
\newcommand{\revisionfiltered}{\color{black}}

\newcommand{\tOp}{{\tilde{O}_p}}
\renewcommand{\top}{{\tilde{o}_p}}

\newcommand{\smn}{\sigma_{m,n}}

\newcommand{\xiao}{\wedge}
\newcommand{\da}{\vee}

\newcommand{\jma}{{\cal J}_{A;m,\lambda_A}}

\newcommand{\JA}{{\cal J}_A}
\newcommand{\JB}{{\cal J}_B}

\newcommand{\ankA}[1]{a_{m,\lambda_A;#1}}

\newcommand{\tnaijA}{\Theta_{m,\lambda_A;i,j}}

\newcommand{\pool}{{\mathrm{pool}}}

\allowdisplaybreaks

\begin{document}

\def\spacingset#1{\renewcommand{\baselinestretch}%
{#1}\small\normalsize} \spacingset{1}


\if1\blind
{
  \title{\bf Higher-order accurate two-sample network inference and network hashing}
  \author{
  	Meijia Shao\thanks{Ohio State University}, Dong Xia\thanks{Correspondence author: madxia@ust.hk, Hong Kong University of Science and Technology}, Yuan Zhang$^*$, Qiong Wu\thanks{University of Pennsylvania} ~and Shuo Chen\thanks{University of Maryland, Baltimore}
	
  }
  \maketitle
  	\let\thefootnote\relax\footnotetext{Meijia Shao and Yuan Zhang were supported by National Science Foundation grant DMS-2311109.  Dong Xia was partially supported by Hong Kong RGC grant ECS 26302019.}
} \fi

\if0\blind
{
  \bigskip
  \bigskip
  \bigskip
  \begin{center}
    {\LARGE\bf Higher-order accurate two-sample network inference and network hashing}
\end{center}
  \medskip
} \fi

\bigskip
\begin{abstract}
    {\revisionfiltered
    Two-sample hypothesis testing for network comparison presents many significant challenges, including:
    leveraging repeated network observations and known node registration, but without requiring them to operate; 
    relaxing strong structural assumptions;
    achieving finite-sample higher-order accuracy;
    handling different network sizes and sparsity levels;
    fast computation and memory parsimony; controlling false discovery rate (FDR) in multiple testing;
    and theoretical understandings, particularly regarding finite-sample accuracy and minimax optimality. 
    In this paper, we develop a comprehensive toolbox, featuring a novel main method and its variants, all accompanied by strong theoretical guarantees, to address these challenges. 
    Our method outperforms existing tools in speed and accuracy, and it is proved power-optimal. 
    Our algorithms are user-friendly and versatile in handling various data structures (single or repeated network observations; known or unknown node registration). 
    We also develop an innovative framework for offline hashing and fast querying as a very useful tool for large network databases. 
    We showcase the effectiveness of our method through comprehensive simulations and applications to two real-world datasets, which revealed intriguing new structures.
    }
\end{abstract}

\noindent%
\vfill

\newpage
\spacingset{1.5} 

\section{Introduction}

The relational nature of network data presents unique challenges to data analysts.  
It is  therefore very intriguing to ask how to extend statistical concepts, tools and theory in the classical i.i.d. setting to network settings. 
\cite{levin2019bootstrapping} said well: 
``\emph{A core problem in statistical network analysis is to develop network analogues of classical techniques.}''
Indeed, recent years have witnessed significant developments of network analysis toolbox for both point estimation \citep{abbe2017community} and one-sample inference \citep{gao2017testinga, banerjee2017optimal, green2017bootstrapping, jin2018network, levin2019bootstrapping, zhang2020edgeworth}.  
Comprehensive theoretical analysis of these tools has deepened our understanding of their statistical behaviors and performances.
This paper studies a challenging and under-explored topic: two-sample network comparison.  Spoken in plain language, the scientific question is how to compare two network models ${\cal F}_A$ and ${\cal F}_B$ based on their respective observations $A^{(1)},\ldots,A^{(N_A)}$ and $B^{(1)},\ldots,B^{(N_B)}$. 
Though being challenging in general, this problem can be made easier by imposing two popular assumptions. 
The first popular assumption is repeated network observations, that is, $N_A,N_B\geq 2$ and possibly even diverge, such as \cite{ginestet2017hypothesis, ghoshdastidar2018practical, kolaczyk2020averages, ghoshdastidar2020two, chen2020spectral,  maugis2020testing, bravo2021principled, yuan2021practical}.
This assumption is particularly strong in that it implicitly assumes repeated network observations within each group are identically distributed.  However, many real-world data sets clearly exhibit within-group heterogeneity.
Consider the schizophrenia data in Section \ref{subsec::data-example-2-schizophrenia}.
Our analysis discovers subgroups within the patient and healthy people groups as shown in Figure \ref{fig::data-2}.
This is a clear evidence against the i.i.d. assumption on all patients' brain networks. 
The second popular assumption states that all networks share a common node set, with known node correspondence, including \cite{ghoshdastidar2018practical, li2018two,ghoshdastidar2020two,chen2020spectral}.
This assumption is sensible in some applications such as networks between the same set of participants, but it is restrictive in general. 
For instance, this assumption is true for the brain image data we shall analyze in Section \ref{subsec::data-example-2-schizophrenia}, but not for the Google+ data in Section \ref{subsec::data-example-1-google-plus-ego-networks}, where we compare ego-networks of different individuals, which, not only have no node correspondence, but also may vary greatly in network sizes. 
Two-sample network comparison becomes most challenging in the absence of the aforementioned two assumptions, namely, 
$N_A=N_B=1$ and there is no available node correspondence. Prior works typically require additional strong structural assumptions such as low-rankness or strict degree monotonicity \citep{tang2017semiparametric, agterberg2020nonparametric,yang2014nonparametric,sabanayagam2021graphon}. {\revisionfiltered
Moreover, most of these methods depend on strong modeling assumptions such as low-rankness with a known rank and equal network sizes with a known registration between the node sets of any two networks.

In this work, we propose a novel framework based on the network method-of-moments which can effectively tackle various challenges for  network two-sample test. 
The highlights of our contributions are as follows. 
First, we present the first two-sample test procedure with higher-order accurate control of type-I error under the non-degeneracy assumption, a frequently met condition in many network data sets. 
Our method requires far less restrictive structural assumptions and can handle a much wider range of different sizes and sparsity levels than SVD-based methods; and outperforms other network-moment-based methods, such as network bootstraps, in finite-sample risk control accuracy.

Second, our method does not require repeated network observations or known node registration. 
However, the pooled version of our method can effectively leverage such available information to improve statistical efficiency.
Its high versatility in handling various data types, combined with the higher-order accuracy of our inference formula, set our method apart from existing techniques.
Moreover, our one-on-one comparison procedure remains a vital tool for detecting in-group heterogeneity, even with multiple observed networks in each group. 
It plays a crucial role in assessing the validity of pooling networks, addressing a key concern often overlooked in prior studies on comparing groups of networks.

Third, our method enjoys considerable speed advantages in one-on-one test over most existing methods, except for the normal approximation that we also studied in this paper. 
For multiple comparisons, we propose an innovative ``network hashing + fast query'' procedure. 
In the first stage, each network is precomputed into a concise vector of summary statistics; then in the second stage, we run our test using only these statistics. 
Our novel framework is particularly beneficial for large network databases, as it streamlines maintenance, boosts query speeds, and improves data privacy protection. 
While this framework can also enhance several existing methods, its combination with the test we propose in this paper achieves unrivaled memory efficiency after stage 1 and computational speed in stage 2 when facing a large number of queries.

Fourth, our paper also addresses several key issues that previous works have largely overlooked.
The first challenge is computation, which becomes increasingly expensive as motif size grows. 
Another critical problem in practice is handling degeneracy, which remains blank in existing literature on network two-sample test.
We propose a pioneering method in Section \ref{new-section::handling-indeterminate-degeneracy}, leveraging recent advances in reduced U-statistics \citep{shao2023u}, to both speed up computation and adapt automatically to indeterminate degeneracy.
Additionally, we address the unexplored problem of false discovery rate (FDR) control in multiple network comparison tests.
We develop the first provably valid FDR control procedure for the network database query problem, filling a significant gap in the field.

Fifth, our theoretical studies provide rigorous guarantees of all the aforementioned methodological developments in this paper.
Our results are much sharper than counterparts in existing literature.
Furthermore, we present the first lower-bound result (Theorem \ref{thm::inference-optimality}) that delineates the finite-sample fundamental limit in test power and show that our method achieves power-optimality.

In summary, this paper substantially enhances and enriches the existing methodology toolbox and deepens the theoretical understanding of the performance and fundamental limits of network method-of-moments.

}

\section{Graphon model and network moments}

\subsection{Two-sample graphon model}

For narrative simplicity, we state our method and theory for unweighted, undirected networks with no self-loop, but these specifications are not essential.  
We observe two independent networks with $m$ and $n$ nodes, respectively, represented by their adjacency matrices $A\in \{0,1\}^{m\times m}$ and $B\in \{0,1\}^{n\times n}$, where $A=A^T, B=B^T, \mathrm{diag}(A)=0$ and $\mathrm{diag}(B)=0$.
Throughout this paper, we choose graphon model as our base model.  Graphon is a very general wrapper around many famous and widely-used network models \citep{gao2018community, young2007random, hunter2008ergm, bickel2009nonparametric,olhede2014network}. 
Our method, though designed to work for graphons, also works for these models as special cases.
The data generation mechanism is as follows.
First, assign a latent position to each node: $X_1,\ldots,X_m, Y_1,\ldots,Y_n\stackrel{i.i.d.}\sim $~Uniform$[0,1]$.
Second, there exists a latent, symmetric graphon function $f_A(\cdot,\cdot):[0,1]^2\to \mathbb{R}^+$ encoding all network structures, and a sparsity parameter $\rho_A$, such that the edge probability between $(i_1,i_2)$ is $W_{i_1,i_2}^{(A)} = \rho_A \cdot f_A(X_{i_1},X_{i_2})$.  We inherit the regularity condition for model identifiability $\int_{[0,1]^2}f_A(u,v)\td u\td v=1$ from \cite{olhede2014network}. 
Similarly define $\rho_B$ and $f_B(\cdot,\cdot)$.
Finally, our observation $A$ is generated according to $\pr(A_{i_1,i_2}=A_{i_2,i_1}=1) = W_{i_1,i_2}^{(A)}$ for all $1\leq i_1<i_2\leq m$.
Similarly define $\rho_B$, $f_B(\cdot,\cdot)$, $W^{(B)}$ and generate $B$.
General audience may wonder if one can estimate $f_A,f_B$, $X$'es and $Y$'s and use them to compare network structures.  Unfortunately, $f_A,f_B$ are only identifiable up to equivalent classes \citep{olhede2014network,gao2015rate,zhang2017estimating}. The edge probability matrices $W^{(A)}, W^{(B)}$ are fully identifiable, so one naturally considers a graph-matching-based network comparison.
However, graph matching is a difficult problem even with structural assumptions \citep{lyzinski2015graph, arroyo2021maximum}, and it typically requires rather costly computation.

\subsection{Network method of moments}

Our approach is to compare network moments, namely, the frequencies of patterned motifs, e.g. edges, triangles, star-shapes and circles \citep{bickel2011method, jin2018network, levin2019bootstrapping}.
The idea is naturally inspired by its counterpart in classical statistics, where we compare the means, variances and other numerical features of two populations.
Network method of moments enjoys high computational speed and memory efficiency \citep{zhang2020edgeworth}.  Particularly conveniently, they are invariant under node permutations, which exempts our approach from a slow and difficult graph matching step and easily enables it to handle different network sizes.
Interestingly, two potentially very different network models may still share some common numerical features of scientific importance and interest.  This is exactly analogous to the understanding in the {\revisionfiltered classical two-sample test. 
 Moreover, by \cite[Corollary 5.49]{zhao2023graph}, the edit distance between two graphon models can be controlled by finite many moment distances.
Therefore, our devised toolbox provides more flexibility than existing tools that test the exactly equality of two network models under restrictive structural assumptions. }


Now we formally define network moments.
Let $R$ denote a given motif of $r$ nodes and $s$ edges.  
Throughout this paper, we always assume the motif $R$ is connected. 
Following the tradition of \cite{bickel2011method,zhang2020edgeworth,maugis2020testing}, for any $r$-node graph $A_0$, define 
$h(A_0) := 1$ if there exists bijection
$\pi:[1:r]\leftrightarrow[1:r], {\rm s.t.}\; (A_0)_{i,j} \geq R_{\pi(i)\pi(j)}, \forall 1\leq i<j\leq r$; otherwise $h(A_0) := 0$.
The empirical network moment indexed by $R$ for network $A$ is
$$
\hat U_m := \binom{m}r^{-1} \sum_{1\leq i_1<\cdots<i_r\leq m} h(A_{i_1,\ldots,i_r}),
$$
where we omit the dependency of $h$ and $\hat U_m$ on $R$ to simplify notation. Here $A_{\mathcal{I}}$ denotes the sub-matrix with row and column indices from the set $\mathcal{I}$. 
Denote the corresponding population moment by $\mu_m:=\ep[\hat U_m]$, where the expectation is taken w.r.t. the randomness of both network edges and latent positions.  Similarly define $\hat V_n$ and $\nu_n$ for network $B$.  

Notice that, however, the moments $\hat U_m,\mu_m$, $\hat V_n,\nu_n$ are impacted by network sparsity measures $\rho_A$, $\hat\rho_A$, $\rho_B$ and $\hat\rho_B$, respectively. Here the empirical network sparsity is define by 
$$
\hat\rho_A := \binom{m}2^{-1} \sum_{1\leq i_1<i_2\leq m} A_{i_1i_2}
$$ 
and $\hat\rho_B$ similarly.
Networks may exhibit very different sparsity levels \citep{decelle2011inference}. 
To rule out the nuisance of network sparsity, similar to \cite{bickel2011method,bhattacharya2022motif}, we consider the Horvitz–Thompson estimators when designing the population discrepancy measure, which is defined as
$$
    d_{m,n,\rho_A,\rho_B} := \rho_A^{-s}\cdot \mu_m - \rho_B^{-s}\cdot \nu_n.
$$
Our central goal is to perform inference on $d_{m,n,\rho_A,\rho_B}$. For instance, $d_{m,n,\rho_A,\rho_B}=0$ means that the two networks are {\it possibly} generated from the same network model (except sparsity levels), which can be of great scientific values in practice. 
A natural point estimator of $d_{m,n,\rho_A,\rho_B}$ is the plug-in version defined by
\begin{equation}\label{eq:hatD_mn}
\hat D_{m,n} := \hat\rho_A^{-s}\cdot \hat U_m - \hat\rho_B^{-s} \cdot \hat V_n.
\end{equation}
The main task is then to characterize the distribution of $\hat D_{m,n}$, for which its variance estimate is necessary so that it can be studentized. In \citet{zhang2020edgeworth}, the one-sample counterpart of our inference method is simply based on the studentized version of empirical moment $\hat U_m$ where the underlying latent network sparsity only plays a role in theoretical analysis but does not appear in the method formulation. In sharp contrast, the network sparsities are indispensable in our designed estimator (\ref{eq:hatD_mn}). As a result, a higher-order accurate characterization of the distribution of $\hat D_{m,n}$ must carefully handles the estimation error induced by $\hat \rho_A$ and $\hat \rho_B$, 
which will also interact with the estimation error in $\hat U_m$ and $\hat V_n$.
Moreover, the networks can have drastically different sizes and sparsity levels.
All of this should reflect in the Edgeworth expansions.
thus the one-sample result in \citet{zhang2020edgeworth}
needs significantly nontrivial extensions to suit our purpose.


\section{Higher-order accurate method by Edgeworth expansion}\label{sec::single-network}

\subsection{Variance estimation and studentization}
\label{subsec::our-method::studentization}

We quickly outline this section's contents.
We first estimate $\var(\hat D_{m,n})$ and use it to studentize $\hat D_{m,n}$; this is nontrivial as variance estimators that require repeated network observations such as \cite{maugis2020testing, bravo2021principled} cannot function here.
Then we studentize $\hat D_{m,n}$ and formulate a higher-order accurate distributional approximation.  This formulation then leads to our novel network offline hashing and fast querying toolbox.  We also formulate the Cornish-Fisher confidence interval by inverting the CDF approximation.
We conclude this section with our theorems that establish finite-sample accuracy and minimax optimality.

We now brief the rationale in our designed estimator of $\var(\hat D_{m,n})$. Readers not interested in technical details may skip to eq. (\ref{eq:hoeffding-A}). To estimate $\var(\hat D_{m,n})$, we decompose the stochastic variation in $\hat\rho_A^{-s}\cdot \hat U_m$ as follows: 
\begin{equation}\label{eq:decomp_1}
\hat\rho_A^{-s}\cdot \hat U_m = \{\rho_A + (\tilde\rho_A-\rho_A) + (\hat\rho_A-\tilde\rho_A)\}^{-s} \cdot \{\mu_m + (\tilde U_m-\mu_m) + (\hat U_m-\tilde U_m)\},
\end{equation}
where we define $\tilde\rho_A:=\ep[\hat\rho_A|X_1,\ldots,X_m]$ and $\tilde U_m:=\ep[\hat U_m|X_1,\ldots,X_m]$. 
Note that the random variables $\tilde\rho_A$ and $\tilde{U}_m$ are classical noiseless U-statistics of degrees 2 and $r$, respectively.  Their randomness is exclusively driven by the latent positions $X_1,\ldots,X_m$; thus they admit Hoeffding's decompositions \citep{hoeffding1948class}: 
$$
{\dx \tilde U_m} = \mu_m  + \frac{r}{m}\cdot\sum_{i=1}^m g_{A;1}(X_i) + \tOp\Big(\frac{\rho_A^s \log m}{m}\Big)\quad {\rm and}\quad {\dx\tilde \rho_A} =  \rho_A + \frac{2}{m}\cdot\sum_{i=1}^m g_{\rho_A;1}(X_i) +  \tOp\Big(\frac{\rho_A \log m}{m}\Big),
$$
respectively. Here we slightly abuse notation and write
 $h(X_1,\ldots,X_r):=\ep[h(A_{1,\ldots,r})|X_1,\ldots,X_r]$. The Hoeffding's decomposition terms are 
defined by $g_{A;1}(X_1) := \ep[h(X_1,\ldots,X_r)|X_1] - \mu_m$, $g_{\rho_A;1}(X_1) := \rho_A\cdot \ep[f_A(X_1,X_2)|X_1] - \rho_A$. The notation $\tOp$ adopted here is slightly different from the conventional $O_p$ in literature. Basically, we write $a_{m,n} = \tOp(b_{m,n})$ if $\pr(|a_{m,n}/b_{m,n}|\geq C) = O\big( (m\xiao n)^{-1}\big)$ for some absolute constant $C>0$.  

From eq. (\ref{eq:decomp_1}), the variation in $\hat\rho_A^{-s}\cdot \hat U_m$ originates from that in $\hat\rho_A-\tilde\rho_A, \tilde\rho_A-\rho_A$, $\hat U_m-\tilde U_m$ and $\tilde U_m-\mu_m$. 
It turns out (by the proof of Theorem~\ref{thm::main}) that when $\rho_A\gg m^{-1}, \rho_B\gg n^{-1}$, the variation contributed by $\hat\rho_A-\tilde\rho_A$ and $\hat U_m-\tilde U_m$ are dominated by that in $\tilde\rho_A-\rho_A$ and $\tilde U_m-\mu_m$.
In fact, terms $\hat\rho_A-\tilde\rho_A$ and $\hat U_m-\tilde U_m$ are both dominated by a weighted average of $\eta^{(A)}_{ij}:=A_{ij}-W^{(A)}_{ij}$ terms, which encodes the randomness of network edges. Note that $\hat U_m-\tilde U_m$ contains a $\tOp(\rho_A^{-1/2}m^{-1}\log^{1/2}m)$ remainder term while $\hat\rho_A-\tilde\rho_A$ does not.
Putting these understandings together, we have
\begin{align}\label{eq:hoeffding-A}
    \hat\rho_A^{-s}\cdot \hat U_m
    =&~
    \rho_A^{-s}\mu_m + \frac1m\sum_{i=1}^m \alpha_1(X_i)
    + \tOp(\rho_A^{-1/2}\cdot m^{-1}\log^{1/2}m+m^{-1}\log m),
\end{align}
where $\alpha_1(X_i) := r\rho_A^{-s}g_{A;1}(X_i) - 2s\rho_A^{-(s+1)}\mu_mg_{\rho_A;1}(X_i)$. The $B$-index counterparts $g_{B;1}(Y_i)$ and $g_{\rho_B;1}(Y_i)$ are defined in the same fashion so that the variance of $\hat\rho_B^{-s}\cdot \hat V_n$ can be decomposed accordingly. 
The variance decomposition (\ref{eq:hoeffding-A}) is the foundation for us to estimate the variance of $\hat\rho_A^{-s}\cdot \hat U_m$.
We set $\rho_A\gg m^{-1}$ throughout this paper, which is a fairly mild condition of network sparsity. 
Then we write $\var(\hat\rho_A^{-s}\cdot\hat U_m)\approx m^{-1}\var(\alpha_1(X_1))\approx m^{-2}\sum_{i=1}^m\hat \alpha_1^2(X_i)$, where we design the empirical version of $\alpha_1(X_i)$ via
$$
\hat\alpha_1(X_i) := r\hat\rho_A^{-s} \hat g_{A;1}(X_i) - 2s\hat\rho_A^{-(s+1)}\hat U_m\hat g_{\rho_A;1}(X_i).
$$
Here the empirical versions of $g_{A;1}(X_i)$ and $g_{\rho_A;1}(X_i)$ are defined as follows:
$$
\hat g_{A;1}(X_i):=\frac{
\sum_{\{i_1<\cdots<i_m\}\subseteq[1:m]\backslash\{i\}}
h(A_{i,i_1,\ldots,i_{r-1}})}{\binom{m-1}{r-1}}- \hat U_m\quad {\rm and}\quad \hat g_{\rho_A;1}(X_i)  := \frac{\sum_{\substack{1\leq i'\leq m, i'\neq i}} A_{ii'}}{m-1} - \hat\rho_A.
$$
The $B$-indexed counterparts $\beta_1(Y_i)$ and $\hat \beta_1(Y_j)$ are defined similarly.  We are now in position to estimate $\var(\hat D_{m,n})\approx \smn^2:=m^{-1}\ep[\alpha_1^2(X_1)]+n^{-1}\ep[\beta_1^2(Y_1)]$ by 
\begin{align}
    \hat S_{m,n}^2 := m^{-2}\sum_{i=1}^m \hat\alpha_1^2(X_i) + n^{-2}\sum_{j=1}^n \hat\beta_1^2(Y_j).
    \label{def::S_m,n}
\end{align}
Finally, we studentize $\hat D_{m,n}$ as follows
\begin{align}
    \hat T_{m,n}
    &:=
    \hat S_{m,n}^{-1}\cdot \{\hat D_{m,n} - (\rho_A^{-s}\cdot \mu_m - \rho_B^{-s}\cdot \nu_n)\}.
    \label{def::T_m,n}
\end{align}
One can equivalently use jackknife \citep{maesono1997edgeworth} to design $\hat S_{m,n}$, but our design is more convenient for analysis and faster to compute; 
see Theorem 3.3 in \cite{zhang2020edgeworth} and the paragraph beneath it.

\subsection{Characterizing the distribution via Edgeworth expansion}
\label{subsec::characterizing the distribution::alpha}

In this section, we present Edgeworth expansions as a higher-order accurate means to approximate $F_{\hat T_{m,n}}$, i.e., the c.d.f. of $\hat T_{m,n}$.
To start, we set up some auxiliary quantities as building bricks of the coefficients in our expansion formula.  Define the second-order terms in Hoeffding decomposition  
$$
g_{A;2}(X_1,X_2):=\ep[h(X_1,\ldots,X_r)|X_1,X_2]-g_{A;1}(X_1)-g_{A;1}(X_2)-\mu_m,
$$
and similarly define $g_{\rho_A;2}(X_1,X_2):=\rho_Af_A(X_1,X_2)-g_{\rho_A;1}(X_1)-g_{\rho_A;1}(X_2)-\rho_A$. Moreover, we set their variances and covariance by
$$
\xi_{A;1}^2:=\var(g_{A;1}(X_1)),\quad \xi_{\rho_A;1}^2:=\var(g_{\rho_A;1}(X_1))\quad {\rm and} \quad \xi_{A,\rho_A;1} := \ep[g_{A;1}(X_1)g_{\rho_A;1}(X_1)].
$$
Recall we defined $\alpha_1(X_i)$ in Section \ref{subsec::our-method::studentization}. The higher-order accurate characterization of $F_{\hat T_{m,n}}$ requires auxiliary quantities $\alpha_0$, $\alpha_1(X_i)$, $\alpha_2(X_{i_1},X_{i_2})$, $\alpha_3(X_i)$ and $\alpha_4(X_{i_1},X_{i_2})$, $1\leq \{i;i_1<i_2\}\leq m$.  Due to page limit, we sink their formal definitions to Section \ref{calculation alpha} in Supplementary Material.
Define $\xi_{B;1}^2$, $\xi_{\rho_B;1}^2$, $\xi_{B,\rho_B;1}$, $\beta_0$, $\beta_2$ through $\beta_4$ similarly for graphon $B$.  
The formal definitions of $\alpha_0$ to $\alpha_4$ and $\beta_0$ to $\beta_4$ are involved, but they are all $\asymp_p 1$ and can be fast calculated empirically. Similarly, the quantities $\xi_{A;1}, \xi_{\rho_A;1}, \xi_{B;1}$ and $\xi_{\rho_B;1}$ are all $\asymp 1$. {\revisionfiltered Note that we focus on the non-degenerate case where $\xi_{A;1}, \xi_{B;1}\geq {\rm const}>0$. 
We address degeneracy in Section~\ref{new-section::handling-indeterminate-degeneracy}.}

Denote $\varphi(u)$ and $\Phi(u)$ the density and distribution function of a standard normal random variable. The higher-order accurate characterization of $F_{\hat T_{m,n}}$ is decided by the following Edgeworth expansion (population version)
\begin{equation}\label{eq:Gmn}
G_{m,n}(u):=\Phi(u)-\varphi(u)\cdot \big\{Q_{m,n,\rho_A,\rho_B;1}+Q_{m,n,\rho_A,\rho_B;2}(u^2+1)+\I_0 \big\},    
\end{equation}
where the expansion terms are defined by 
\begin{align}
    \I_0
    := &~
    \smn^{-1}(m^{-1}\alpha_0-n^{-1}\beta_0),
    \notag\\
    Q_{m,n,\rho_A,\rho_B;1}
    := &~
    \frac12\smn^{-3}\Big\{
    -m^{-2}\ep[\alpha_4(X_1,X_2)\alpha_1(X_2)]
    -m^{-2}\ep[\alpha_1(X_1)\alpha_3(X_1)]
    \notag\\
    &
    +n^\shao{-2}\ep[\beta_1(Y_1)\beta_3(Y_1)]
    +n^{-2}\ep[\beta_4(Y_1;Y_2)\beta_1(Y_2)]
    \Big\},
    \notag\\
    Q_{m,n,\rho_A,\rho_B;2}
    :=&~
    \smn^{-3}\Big\{
        m^{-2}\big( \ep[\alpha_1^3(X_1){\dx /6}
        +\alpha_1(X_1)\alpha_1(X_2)\alpha_2(X_1,X_2)] \big)
        \notag\\
        &\shao{-}n^{-2}\big( \ep[\beta_1^3(Y_1){\dx /6}\shao{+}\beta_1(Y_1)\beta_1(Y_2)\beta_2(Y_1,Y_2)] \big)
    \Big\}
    \notag\\
    +\frac12\smn^{-5}
    \Big\{
        \big( &-m^{-3}\xi_{\alpha;1}^2-m^{-2}n^{-1}\xi_{\beta;1}^2 \big)
        \cdot \ep[\alpha_1(X_1)\alpha_3(X_1)+\alpha_4(X_1;X_2)\alpha_1(X_2)]\notag\\
         +
        \big( m^{-1}n^{-2}\xi_{\alpha;1}^2&+n^{-3}\xi_{\beta;1}^\shao{2} \big)
        \cdot \ep[\beta_1(Y_1)\beta_3(Y_1)+\beta_4(Y_1;Y_2)\beta_1(Y_2)]
    \Big\}.
    \notag
\end{align}
Here we write $\xi_{\alpha;1}^2:={\rm Var}(\alpha_1(X_1))$ and $\xi_{\beta;1}^2:={\rm Var}(\beta_1(Y_1))$, both of which are $\asymp 1$. 
Therefore, $\sigma_{m,n}^2\asymp m^{-1}+n^{-1}$, and all of ${\cal I}_0$, $Q_{m,n,\rho_A,\rho_B;1}$ and $Q_{m,n,\rho_A,\rho_B;2}$ are $\asymp (m\wedge n)^{-1/2}$. They characterize the first order terms in Edgeworth expansion. Basically, the goal is to show that $G_{m,n}(u)$ approximates the distribution of $\hat T_{m,n}$ (up to certain continuity correction to be clarified soon) with an accuracy sharper than $(m\wedge n)^{-1/2}$, i.e., more precise than a Berry-Esseen bound. The following Theorem~\ref{thm::main} confirms that such a higher-order accuracy can be achieved by $G_{m,n}(u)$ under mild conditions.

The population version $G_{m,n}(u)$ is not immediately applicable since it involves unknown quantities like $\sigma_{m,n}^2, \alpha_0$, $\beta_0$  $\ep[\alpha_4(X_1,X_2)\alpha_1(X_2)]$ and so on.  Fortunately, all these quantities can be accurately estimated and fast computed. Due to space limit, the definitions of corresponding estimators are relegated to Section~\ref{calculation alpha} in the Supplementary Material. These estimators lead to empirical Edgeworth expansion terms $\hat \I_0$, $\hat Q_{m,n,\rho_A,\rho_B;1}$ and $\hat Q_{m,n,\rho_A,\rho_B;2}$. The empirical Edgeworth expansion (EEE) $\hat G_{m,n}(u)$ is defined by replacing $\I_0$, $Q_{m,n,\rho_A,\rho_B;1}$ and $Q_{m,n,\rho_A,\rho_B;2}$ by their estimators in $G_{m,n}(u)$. 
It is worth noting that all the involved quantities are computed separately for network $A$ or network $B$, respectively, and no cross-network quantity is needed. 

There are cases in which $\alpha_1(X_i)$ and $\beta_1(Y_j)$ admit discrete-type distributions (see \citep{zhang2020edgeworth} fore more discussions). If that happens, the dominating terms in $\hat T_{m,n}$ such as $\tilde \rho_A^{-s}\cdot\tilde U_m$ are discrete-type random variables, and it becomes generally impossible to achieve the higher-order approximation to a discrete-type distribution by a continuous function like $G_{m,n}(u)$. 
A common assumption in the analysis of Edgeworth expansion of noiseless and noisy U-statistics is the Cram\'er's condition.  \citet{zhang2020edgeworth} discovered that the observational errors in networks can contribute a surprising self-smoothing effect that waives the Cram\'er's condition in most cases, but it does not cover very dense networks.  In the context of this paper, this corresponds to assuming that either $\rho_A=O(\log m)^{-1}$ or $\limsup_{t\to\infty}|\ep[e^{\ii t g_{A;1}(X_1)/\xi_{A;1}}]|<1$ for network $A$; and a similar assumption for network $B$.
Here, we propose a simple remedy that completely waives Cram\'er's condition.  Let $\delta_T\sim N\big( 0, C_\delta\cdot  (\log m/m)+ (\log n/n) \big)$  be an artificial Gaussian noise term independent of the observed data with a sufficiently large constant $C_\delta>0$. Simply put, the artificial Gaussian noise is strong enough to smooth the potentially discrete distribution of the dominating terms in $\hat T_{m,n}$. 
We shall use $\hat T_{m,n}+\delta_T$ instead of $\hat T_{m,n}$ for statistical inference.

We call $R$ acyclic if it is a tree; otherwise call it cyclic. The approximation accuracy of Edgeworth expansion $G_{m,n}(u)$ depends on both the network size and sparsity level. More precisely, it is characterized by
\begin{equation}
	{\cal M}_A={\cal M}_A(\rho_A,m; R):=
	\begin{cases}
		 \left(\rho_A\cdot m \right)^{-1}{\cdot \log^{1/2} m+m^{-1}\cdot \log^{3/2}m}, & \textrm{ For acyclic }R,\\
		 \rho_A^{-r/2}\cdot m^{-1}{\cdot \log^{1/2} m+m^{-1}\cdot \log^{3/2}m}, & \textrm{For cyclic }R
	\end{cases}
	\label{MA}
\end{equation}
We can define ${\cal M}_B$ similarly for network B.  We now present our main theorem. Note that the Kolmogorov distance is defined by $\|F(u)-G(u)\|_{\infty}:=\sup_{u\in \mathbb{R}}|F(u)-G(u)|$. 

\begin{theorem}[Population and empirical Edgeworth expansions]\label{thm::main}
    Assume:
    \begin{enumerate}
        \item Network sizes satisfy: $\log(m\da n)/(m\xiao n)\to 0$;
        \item Network sparsities satisfy: $\rho_A=\omega(m^{-1})$ and $\rho_B=\omega(n^{-1})$ if $R$ is acyclic; and $\rho_A=\omega(m^{-1/r})$ and $\rho_B=\omega(n^{-1/r})$ if $R$ is cyclic.
        \item Non-degeneracy: $\big\{\var(\alpha_1(X_1))\xiao \var(\beta_1(Y_1)) \big\} \geq \textrm{Constant}>0$.
    \end{enumerate}
    Define the population Edgeworth expansion $G_{m,n}(u)$ for $\hat T_{m,n}+\delta_T$ as in (\ref{eq:Gmn}). Let $\hat G_{m,n}$ be its empirical version  defined above. Then we have
    \begin{align}
        \big\|
            F_{\hat T_{m,n}+\delta_T}(u) - G_{m,n}(u)
        \big\|_\infty
        &=
        O\Big((m\xiao n)(m^{-1}\M_A+n^{-1}\M_B)\Big),
        \label{main-theorem::oracle}\\
        \pr\Big\{
        \big\|
            F_{\hat T_{m,n}+\delta_T}(u) - \hat G_{m,n}(u)
        \big\|_\infty
        &\geq (m\xiao n)(m^{-1}\M_A+n^{-1}\M_B)
        \Big\}
        =
        O(m^{-1}+n^{-1}).
        \label{main-theorem::empirical}
    \end{align}
\end{theorem}

Theorem \ref{thm::main} addresses a much wider range of $(m,n,\rho_A,\rho_B)$ than all existing similar works.  For example, \cite{ghoshdastidar2017two} assumes $\rho_A,\rho_B\asymp 1$;
and \cite{agterberg2020nonparametric} makes the restrictive assumption that $m/n\to$~constant for dense networks or $\rho_Am/(\rho_Bn)\to$~constant for sparse networks. 
Moreover, under the same parameter configuration assumptions, our method achieves much stronger results.
For instance, under the settings of \cite{ghoshdastidar2017two}, our eq. 
(\ref{main-theorem::empirical}) 
gives a higher-order accurate $O\big((m\wedge n)^{-1}\big)$ distribution approximation error bound; to our best knowledge, this is the first result of its kind. 
Note that eq. (\ref{main-theorem::oracle}) also implies the canonical Berry-Essen bound $\|F_{\hat T_{m,n}}(u)-\Phi(u)\|_{\infty}=O\big((m\wedge n)^{-1/2}\big)$ under appropriate network sparsity conditions. {\revisionfiltered
The network sparsity requirement in Theorem~\ref{thm::main} matches those in the classical literature on network method-of-moments \citep{bickel2011method, bhattacharyya2015subsampling}. }
Also, compared to other approaches to improve risk control accuracy such as iterative bootstrap \citep{hall1988bootstrap, beran1987prepivoting, beran1988prepivoting}, which though has not yet been formulated for the network setting, our empirical Edgeworth expansion method computes much faster and has a much better provable error bound.

\subsection{Two-sample test and Cornish-Fisher confidence interval}

We now present our approach to two-sided test. Parallel results for one-sided alternatives can be easily derived. Our goal is to test 
\begin{equation}
    H_0: d_{m,n,\rho_A,\rho_B}=0,\quad \textrm{ vs }\quad H_a: d_{m,n,\rho_A,\rho_B}\neq 0.
    \label{hypothesis-test}
\end{equation}
The empirical p-value produced by our method is 
$$
\hat p_{\textsf{\tiny val}}:=2\cdot \min\big\{ \hat G_{m,n}(\hat T_{m,n}^{\rm (obs)}) , 1-\hat G_{m,n}(\hat T_{m,n}^{\rm (obs)}) \big\}, 
$$
where we define the observed statistic by $\hat T_{m,n}^{\rm (obs)}:=\hat D_{m,n}/\hat S_{m,n}+\delta_T$. Given a significance level $\alpha$, we reject the null hypothesis $H_0$ if $\hat p_{\textsf{\tiny val}}<\alpha$ and do not reject $H_0$ otherwise.

We also formulate the Cornish-Fisher confidence interval by inverting the Edgeworth expansion.  Define $q_{\hat T_{m,n};\alpha}:=\arg\min_{q\in\mathbb{R}} \big\{F_{\hat T_{m,n}}(q)\geq \alpha\big\}$ to be the true lower-$\alpha$ quantile of the distribution of $\hat T_{m,n}$.  We can approximate $q_{\hat T{m,n};\alpha}$ by
$\hat q_{\hat T_{m,n};\alpha}:=  z_\alpha + \hat \I_0 + \hat Q_{m,n,\rho_A,\rho_B;1} + \hat Q_{m,n,\rho_A,\rho_B;2}(z_\alpha^2-1)$ 
where $z_\alpha:=\Phi^{-1}(\alpha)$.  The two-sided Cornish-Fisher CI for estimating $d_{m,n,\rho_A,\rho_B}$ is 
\begin{equation}
    \Big(
        \hat D_{m,n} - \big(\hat q_{\hat T_{m,n};1-\alpha/2}-\delta_T\big)\cdot \hat S_{m,n},\ \ 
        \hat D_{m,n} - \big(\hat q_{\hat T_{m,n};\alpha/2}-\delta_T\big)\cdot \hat S_{m,n}
    \Big).
    \label{confidence-interval}
\end{equation}
The higher-order accuracy of our distribution approximation leads to accurate controls of type-I error in the hypothesis test and the confidence level associated with the confidence interval. 
\begin{theorem}
    \label{thm::inference-error-bounds}
    Under the conditions of Theorem \ref{thm::main}, we have
    \begin{enumerate}
        \item The type I error of the test \eqref{hypothesis-test} with nominal level $\alpha$ has an actual type I error probability of $\alpha + O\big( (m\xiao n)(m^{-1}\M_A+n^{-1}\M_B) \big)$.
        The type II error of the test \eqref{hypothesis-test} is $o(1)$ when the true separation satisfies $d_{m,n,\rho_A,\rho_B} = \omega(m^{-1/2}+n^{-1/2})$.
        \item The Cornish-Fisher confidence interval \eqref{confidence-interval} with nominal confidence level $1-\alpha$ has an actual coverage probability of $1-\alpha + O\big( (m\xiao n)(m^{-1}\M_A+n^{-1}\M_B) \big)$.
        The length of the confidence interval is $O(m^{-1}+n^{-1})$.
    \end{enumerate}
\end{theorem}
The next theorem shows that our method, while enjoying higher-order accurate risk controls, simultaneously achieves minimax optimality.

\begin{theorem}
    \label{thm::inference-optimality}
    \begin{enumerate}
        \item (Rate-optimality of the separation condition of our test) For any fixed $\alpha\in(0,1)$, there exists population models:
        \begin{itemize}
            \item[*] Under $H_0:$ network 1 $\sim\rho_A\cdot f_0(u,v)$,  network 2 $\sim\rho_B\cdot f_0(u,v)$;
            \item[*] Under $H_a:$ network 1 $\sim\rho_A\cdot f_a(u,v)$,  network 2 $\sim\rho_B\cdot f_b(u,v)$
        \end{itemize}
        for some $\rho_A$, $\rho_B$, $f_0$, $f_a$, $f_b$, such that
        $
            d_{m,n,\rho_A,\rho_B} = O(m^{-1/2}+n^{-1/2})
        $ under $H_a$,
        while any procedure ${\cal T}$ for the hypothesis test \eqref{hypothesis-test} suffers 
        \begin{equation}
            \pr_{\cal T}(\textrm{Reject }H_0|H_0) + \pr_{\cal T}(\textrm{Keep }H_0|H_a)
            \geq C>0
        \end{equation}
        for some constant $C>0$ as $m,n\to \infty$.

        \item (Rate-optimality of the expected length of our confidence interval)  For any fixed $\alpha\in(0,1)$, 
        \begin{equation}            \min_{\substack{\I:\pr(d_{m,n,\rho_A,\rho_B}\in \I)\to 1-\alpha}}
            \;
           \max_{\rho_A,\rho_B,f_a,f_b}\ep[|\I|] \geq C(m^{-1/2}+n^{-1/2})
        \end{equation}
        for some constant $C>0$.
    \end{enumerate}
\end{theorem}

\begin{remark}
    By Theorem \ref{thm::inference-error-bounds} and Theorem \ref{thm::inference-optimality}, respectively, our method enjoys both higher-order accurate risk control and power-optimality.
    Importantly, we emphasize that achieving either of these two desirable properties is not difficult.  For instance, a simple normal approximation also enjoys the optimality properties in Theorem \ref{thm::inference-optimality}, but not that in Theorem \ref{thm::inference-error-bounds}.
    However, achieving both higher-order accurate risk control and power-optimality, is challenging.  To our best knowledge, our method is the first to provably achieve both.
\end{remark}

\section{Network hashing and fast querying}
\label{section::network-hashing}

Suppose we query \footnote{The goal is to search for a network in a large database of networks that, potentially, is generated similarly as $A$ ignoring the possibly different sparsity levels and network sizes.} a graph $A\in\{0,1\}^{m\times m}$ in a graph database that contains $K$ network entries $\{B_k\}_{k=1,\ldots,K}$ of sizes $n_1,\ldots,n_K$ and sparsity parameters $\rho_{B_1},\ldots,\rho_{B_k}$, respectively. Here any or all of $m,K,n_1,\ldots,n_K$ could potentially be very large.
Our method effectively addresses this challenge by hashing database entries (each entry corresponds to a network) with offline computation into just a few summary statistics.
When querying a keyword (network) $A$, we only need to compare the hash of $A$ to entries' hashes.
This provides a very fast screening algorithm to quickly narrow down the search range. 
Our hashing-query procedure also protects data privacy, since all parties only need to disclose network summary statistics, not the entire networks.  
The fast speed and enhanced privacy may cater to the urgent needs of multi-institutional clinical research collaboration on Alzheimer \citep{chen2022four} and other diseases, where privacy protection and high communication cost are two paramount concerns when sharing sensitive patient-level data \citep{chen2022privacy}.  Using our method, researchers can conveniently build, maintain and query a large similarity graph between patient brain images across multiple hospitals for disease subtyping with improved privacy protection and high communication efficiency.
We do clarify that mathematical quantification of privacy protection would be an interesting future work but not the purpose of this paper.

\begin{algorithm}[h!]
    \caption{Network hashing}\label{alg::hashing}
    {\bf Input:} Network database: $\{B_1,\ldots,B_K\}$ \\
    {\bf Output:} Each $B_k$ hashes into a short vector of its own summary statistics \\
    {\bf Steps:}
        For $k=1,\ldots,K$ and 
        for each motif, compute and output 
        $\hat\rho_B$ and $\hat V_n$ defined earlier; and 
        $\hat\beta_0$, 
        $\hat\xi_{B;1}^2$, 
        $\hat \ep[\beta_1^3(Y_1)]$,
        $\hat \ep[\beta_1(Y_1)\beta_3(Y_1)]$, 
        $\hat \ep[\beta_4(Y_1,Y_2)\beta_1(Y_2)]$, 
        $\hat \ep[\beta_1(Y_1)\beta_1(Y_2)\beta_2(Y_1,Y_2)]$ 
        using 
        Section~\ref{calculation alpha} of Supplementary Material.
        Also output the motif name and $n_k$.
\end{algorithm}

\begin{algorithm}[h!]
    \caption{Fast querying}\label{alg::querying-screening}
    {\bf Input:} Queried ``keyword'' network $A$; Hashed database (output of Algorithm \ref{alg::hashing}); level $\alpha$\\
    {\bf Output:} A list of $B_k$'s with similar network moments to $A$; or ``not found'' (empty query result)
    {\bf Steps:}
    \begin{enumerate}
        \item\label{algstep::query-screening-1} 
        Compute 
        $\hat\rho_A$ and $\hat U_n$ defined earlier; and 
        $\hat\alpha_0$, 
        $\hat\xi_{A;1}^2$, 
        $\hat \ep[\alpha_1^3(Y_1)]$,
        $\hat \ep[\alpha_1(Y_1)\alpha_3(Y_1)]$, 
        $\hat \ep[\alpha_4(Y_1,Y_2)\alpha_1(Y_2)]$, 
        $\hat \ep[\alpha_1(Y_1)\alpha_1(Y_2)\alpha_2(Y_1,Y_2)]$ 
        using 
        Section~\ref{calculation alpha} of Supplementary Material using $A$.
        \item For $k=1,\ldots,K$,
        for network pair $(A,B_k)$, compute 
        \begin{align}
            \hat p_k
            &:=
            2\min\Big\{
                \hat G_{m,n_k}\big( \hat T_{m,n_k}^{\rm (obs)} + \delta_T\big),
                1-\hat G_{m,n_k}\big( \hat T_{m,n_k}^{\rm (obs)} + \delta_T \big)
            \Big\}
        \end{align}
        using statistics computed by Algorithm \ref{alg::hashing} and Step \ref{algstep::query-screening-1} of Algorithm \ref{alg::querying-screening}
        \item Extract entries satisfying $\{B_k: \hat p_k\geq \alpha\}$ as candidates passing our screening algorithm, whose match with the queried entry is to be further inspected.
    \end{enumerate}
\end{algorithm}

Algorithms \ref{alg::hashing} and \ref{alg::querying-screening} show the offline nature of our hashing procedure, which can prepare database entries for fast query without knowing the queried keyword.  This is very different from most existing methods such as \citet{ghoshdastidar2017two, tang2017semiparametric, agterberg2020nonparametric}, which do not have a hashing stage and their query methods unavoidably require at least $\Omega(mn_k)$ cross-computation, where $m$ is the size of the queried keyword network. 
The hashing cost on each individual entry is $O(\rho_n^{r-1}\cdot n^r)$, see Section III.A of \citet{ahmed2015efficient}.
To see the distinction in computational cost, suppose all entries have same sizes $n_1=\cdots=n_K=n$ and density levels $\rho_{B_1}=\cdots=\rho_{B_K}=\rho_B$ and there are $K_q$ incoming queries.
Then our method costs $O(K \rho_B^{r-1}n^r + K K_q)$. 
In sharp constrast, existing methods that requires cross-computation cost $O(K K_q mn)$, which will soon  burn out the query system as $K_q$ becomes very large.


Next, we give a finite-sample error bound for type I and type II query errors.
\begin{theorem}
    \label{thm::query-accuracy}
    Under the conditions of Theorem \ref{thm::main}, there exist positive constants $C_1,C_2,C_3,C_4>0$,
    such that for some 
    $\delta:=\delta_{m,n_k}:= C_1(m^{-1/2}\log m+n_k^{-1/2}\log n_k)$,
    setting ${\cal K}_{d,\delta}:=\{k\in\{1,\ldots,K\}, |d_{m,n_k,\rho_A,\rho_{B_k}}^{(k)}|\geq \delta_{m,n_k}\}$,
    the query result produced by Algorithms \ref{alg::hashing} and \ref{alg::querying-screening} satisfy
    \begin{align}
        \textrm{(Type I error)}
        \sum_{k\in [1:K]\backslash{\cal K}_{d,\delta}}
        \pr(\hat p_k< \alpha)
        =&~
        \sum_{k\in [1:K]\backslash{\cal K}_{d,\delta}}
        \Big(\alpha+O\big( m^{-1/2}\log m + n_k^{-1/2}\log n_k \big)\Big),
        \notag
        \\
        \textrm{(Type II error)}
        \hspace{1.7em}
        \sum_{k\in {\cal K}_{d,\delta}}
        \pr(\hat p_k\geq \alpha)
        =&~
        \sum_{k\in {\cal K}_{d,\delta}}
        O\big(m^{-C_3} + n_k^{-C_3}\big).
        \notag
    \end{align}
\end{theorem}

{\revisionfiltered
Our proposed network hashing framework can also benefit several existing methods.
However, these methods cannot match our approach in terms of memory efficiency and query speed\footnote{
    In this scenario, hashing is computed only once, while the speed of handling continually incoming queries is the paramount concern. 
    For the computation cost in the hashing stage: SVD methods require $O(m^2d)$; our method costs $O(\rho_A^{r-1}m^r)$ (full computation) or $O(d_R m^{\lambda_A})$  (reduced computation, see Section \ref{new-section::computation-acceleration-U-statistic-reduction}) for each motif; bootstraps need at least $B$ times the total cost of our method.}. 
Let us first consider memory cost. 
For each network, after stage 1, SVD-based methods such as \cite{agterberg2020nonparametric} require $O(md)$ memory, where $d$ is the network model's known rank;
bootstrap methods need $O(B)$ memory, with $B$ being the number of bootstrap iterations; 
the $O(B)$ cost is shared by ad-hoc inference procedures for other benchmarks \citep{wills2020metrics, tsitsulin2018netlsd}, as they also utilize bootstrap. 
As \cite{zhang2020edgeworth} highlights, to maintain accurate inference, bootstrap methods must increase $B$ at least as fast as $m$ increases. 
In stark contrast, our method requires only $O(d_R)$ memory, where $d_R$ is the number of motifs considered, which can either stay fixed or grow rather slowly.
Then we compare their query speeds.
SVD-based methods are very slow due to their required cross-term computations -- they cost $O(mnd)$;
bootstraps cost $O(B)$;
whereas our method only costs $O(d_R)$.
This shows our method's clear superiority in both memory and computational costs.

}

{\color{revision-color}

{\revisionfiltered

\section{Pooling over multiple networks in the same group}
\label{new-section::pooling-multiple-networks}

As introduced in Section~\ref{sec::single-network}, our study initially focuses on the challenging scenario of $N_A=N_B=1$, where each group has only one observed network. 
This prompts the question: can our method be adapted for less challenging scenarios where $N_A$ and/or $N_B$ exceed 1? 
Although our method can be applied in a one-on-one manner across $N_A N_B$ tests, an extension that allows group-wise pooling would more effectively exploit available information in practice, when networks within a group are known to originate from the same model.
To simplify presentation, we assume that all networks in group $A$ not only share the same graphon function but also have identical density adjustments $\rho_A$ and network sizes $m$; the same goes for group $B$. 
While relaxing this simplifying assumption is not difficult, it would significantly complicate the formulas.
We differentiate between two sub-cases: (i) networks in the same group sharing a common node set, and (ii) networks in each group independently generating their own node set.

\subsection{Common node set}

When networks $A^{(1)},\ldots,A^{(N_A)}$ share a common set of $m$ nodes, it is natural to assume that these nodes have identical latent positions $X_1,\ldots,X_m$ across all $N_A$ networks. 
This implies a shared common probability matrix $W_A$ for all adjacency matrices. 
A natural treatment is to average the adjacency matrix as follows:
\begin{align}
    A^\pool
    :=&~
    N_A^{-1}\big\{
        A^{(1)} + \cdots + A^{(N_A)}
    \big\}
\end{align}
and use $A^\pool$ to estimate $\mu_m$ and $\rho_A$. 
Denote the estimators as $\hat U_m^\pool$ and $\hat \rho_A^\pool$, defined by:
\begin{align}
    \hat U_m^{\pool} := \binom{m}r^{-1} \sum_{1\leq i_1<\cdots<i_r\leq m} h\big(A_{i_1,\ldots,i_r}^{\pool}\big)\quad {\rm and}\quad 
    \hat\rho_A^{\pool} := \frac{1}{N_A}\binom{m}2^{-1} \sum_{1\leq i_1<i_2\leq m} A_{i_1i_2}^{\pool}.
    \label{pooling::matched-nodes::estimators}
\end{align}
In comparison to $\hat U_m$ and $\hat \rho_A$ from the $N_A=1$ case, following \citet{zhang2020edgeworth} and this paper's methodology, we decompose the variations in $(\hat\rho_A^\pool)^{-s} \hat U_m^\pool$ into two components: one due to randomness in $X_{[1:m]}$ and the other attributable to edge-wise observational errors. 
Remarkably, the formula of our method is calculated based on the first component, which remains unaffected by pooling. 
Therefore, no modifications to our algorithm are required, other than using $A^\pool$ as the input instead of $A$ in the $N_A=1$ case; the same applies to group $B$. 
Theoretically, as anticipated, pooling modifies the error term.

\begin{theorem}
    \label{new-theorem::pooled::matched-nodes}
    Under the conditions of Theorem~\ref{thm::main}, the Edgeworth expansion formulas $G_{m,n}$ and $\hat G_{m,n}$, as well as the bounds (\ref{main-theorem::oracle}), (\ref{main-theorem::empirical}), remain the same, except that ${\cal M}_A$ is now replaced by
    \begin{align}
        \tilde {\cal M}_A
        :=&~
        {\cal M}_A(\rho_A,m, N_A; R)
        \notag\\
        :=&~
    	\begin{cases}
    		\left(\rho_A\cdot m \right)^{-1}\cdot N_A^{-1/2}{\cdot \log^{1/2} m+m^{-1}\cdot \log^{3/2}m}, & \textrm{ For acyclic }R;\\
    		\rho_A^{-r/2}\cdot m^{-1}\cdot N_A^{-1/2}{\cdot \log^{1/2} m+m^{-1}\cdot \log^{3/2}m}, & \textrm{For cyclic }R.
    	\end{cases}
        \label{eqn::tilde-MA}
    \end{align}
    and similarly define $\tilde {\cal M}_B$ and replace ${\cal M}_B$ in the original bound with it.
\end{theorem}
With a fixed $m$, the term $m^{-1}\log^{3/2}m$ in \eqref{eqn::tilde-MA} does not vanish as $N_A\to\infty$, because it stems from the noiseless U-statistic part in the decomposition of the test statistic.
This variation cannot be suppressed by repeated network observations, since these networks all share the same $X_{[1:m]}$.

\subsection{Independently selected node sets}

Slightly abusing notation, we redefine the estimators $\hat\rho_A^\pool$ and $\hat U_m^\pool$ for this scenario as
\begin{align}
    \hat\rho_A^\pool
    :=&~
    N_A^{-1}\sum_{\ell=1}^{N_A}
    \hat\rho_A^{(\ell)}
    \quad
    \textrm{and}
    \quad
    \hat U_m^\pool
    :=
    N_A^{-1}\sum_{\ell=1}^{N_A}
    \hat U_m^{(\ell)}.
    \label{pooling::unmatched-nodes::estimators}
\end{align}
In \eqref{pooling::unmatched-nodes::estimators}, the formula for $\hat\rho_A^\pool$ aligns with \eqref{pooling::matched-nodes::estimators} due to its linearity, yet the expression for $\hat U_m^\pool$ has altered. 
Similarly define estimators for group $B$. 
The studentization now becomes
\begin{align}
    \hat T_{m,n;N_A,N_B}^\pool
    :=&~
    \dfrac
        {
            (\hat\rho_A^\pool)^{-s}
            \cdot
            \hat U_m^\pool
            -
            (\hat\rho_B^\pool)^{-s}
            \cdot
            \hat V_n^\pool
        }
        {
            \hat S_{m,n,N_A,N_B}^\pool
        },
\end{align}
where
$
    \big(
        \hat S_{m,n;N_A,N_B}^\pool
    \big)^2
    :=
    (N_A m)^{-2}
    \sum_{\ell=1}^{N_A}
    \sum_{i=1}^m
    \hat\alpha_1^2(X_i^{(\ell)})
    +
    (N_B n)^{-2}
    \sum_{\ell'=1}^{N_B}
    \sum_{j=1}^n
    \hat\beta_1^2(Y_j^{(\ell')}).
$

\begin{theorem}
    \label{new-theorem::pooled::unmatched-nodes}
    Under the conditions of Theorem~\ref{thm::main}, the Edgeworth expansion formula (\ref{main-theorem::oracle}) and (\ref{main-theorem::empirical}) and the error bounds all maintain their form, with the exception that in the definitions of $G_{m,n}$, $\hat G_{m,n}$, $\delta_T$, and the right-hand side of \eqref{main-theorem::oracle} and \eqref{main-theorem::empirical}, we substitute $m$ and $n$ with $mN_A$ and $nN_B$, and replace ${\cal M}_A$ and ${\cal M}_B$ by ${\cal M}_A\cdot N_A^{-1/2}$ and ${\cal M}_B\cdot N_B^{-1/2}$, respectively.
\end{theorem}

}

{\revisionfiltered

\section{Computation acceleration and adapting to degeneracy}

\subsection{Computation acceleration by U-statistic reduction}
\label{new-section::computation-acceleration-U-statistic-reduction}

Computing a network moment for $r$ nodes and $s$ edges on a single network has a computational cost of $O(\rho_A^{r-1} m^r)$. 
This is notably higher compared to low-rank decomposition methods, which cost $O(m^2 \cdot \mathrm{rank}(\ep[A]))$ with a known rank \citep{tang2017semiparametric}, particularly when $r\geq 3$ and $\rho_A$ slowly diminishes. 
To speed-up our method, we use the \emph{U-statistic reduction} technique \citep{chen2019randomized, shao2023u}. 
For a tuning parameter $\lambda_A\in(1,3)$, let 
$
    \jma
    :=
    \big\{I_{A;r}^{(\ell)}: \ell\in |\jma|\big\}\subset {\cal C}_m^r,
$  
be randomly selected with replacement from all $r$-tuples ${\cal C}_m^r$, with $\big|\jma\big|=O(m^{\lambda_A})$. 
The \emph{incomplete network U-statistic} is
\begin{align}
    \hat U_{\JA}
    :=&~
    |\jma|^{-1}
    \sum_{I_{A;r}\in \jma} h(A_{I_{A;r}})
    =
    |\jma|^{-1}
    \sum_{I_{A;r}\in {\cal C}_m^r}
    \ankA{r}(I_{A;r}) h(A_{I_{A;r}}),
\end{align}
where for any $k\in[1:r]$ and $k$-tuple $I_k$, define $\ankA{k}(I_k):=\sum_{\ell=1}^{|\jma|} \mathbbm{1}_{[I_k \subset I_{A;r}^{(\ell)}]}$.
Similarly, we estimate the sparsity by
$
    \hat\rho_{A,\JA}
    :=
    \big\{
        \binom{r}2 |\jma|
    \big\}^{-1}
    \cdot
    \sum_{\ell=1}^{|\jma|}
    \sum_{\{i_1,i_2\}\subseteq I_{A;r}^{(\ell)}}
    A_{i_1,i_2}.
$
The reduction in computation necessitates modifications to the inference procedure, including updates to the variance estimator and the Edgeworth expansion formula. 
While \cite{shao2023u} provides a framework for higher-order accurate one-sample inference using a non-degenerate $\hat U_{\JA}$, adapting this to the two-sample test statistic is a complex task. 
Consequently, our study will concentrate on asymptotic methods. 
To save space, we will devise an accelerated method that also tackles indeterminate degeneracy in the subsequent subsection, deferring the exploration of the more complex higher-order accurate approximation to future work.

\subsection{Handling indeterminate degeneracy}
\label{new-section::handling-indeterminate-degeneracy}

Much of the literature on network moments relies on the non-degeneracy conditions, i.e., $\xi_{A;1}, \xi_{B;1}\geq {\rm const}>0$. 
However, in real-world applications, network U-statistics may not always fulfill this criterion. 
For instance, consider a stochastic block model with two equal-sized communities and connection probabilities $[a,b;b,a]$; all network moments here are inherently degenerate due to symmetry.
The existing literature on degenerate network moment statistics is rather limited \citep{gao2017testinga,gao2017testingb,hladky2021limit}. 
Furthermore, many previous works, such as \citep{maugis2020central, bickel2011method}, that do not explicitly mention the non-degeneracy condition actually require it\footnote{
    For example, both \cite{maugis2020central} and \cite{bickel2011method} assumed non-degeneracy but missed the statement in their theorems.
    In \cite{maugis2020central}, the proof of Theorem 2 used Theorem 15.1 in \cite{dasgupta2008asymptotic};
    and in \cite{bickel2011method}, the proof of Lemma 1 (needed by their main theorem) used the asymptotic normality of noiseless U-statistics \cite{serfling2009approximation}.
    Both premises require non-degeneracy.
    }.

We propose a novel, unified formula that can consistently estimate the variance in both non-degenerate and degenerate scenarios. 
The automatic adaptivity sets our approach apart from all existing methods.
For clarity, we will focus on the one-sample case, later extending the result to the two-sample test problem.
The first nontrivial step is to establish that under mild conditions,
\begin{align}
    \hat\rho_{A,\JA}^{-s} \cdot U_{\JA}
    -
    \rho_A^{-s} \cdot \mu_m
    =
    \big(
        \Gamma_{1,\JA}
        +
        \Gamma_{2,\JA}
        +
        \Gamma_{3,\JA}
    \big)
    \big\{
        1 + \top(1)
    \big\},
    \label{method::degeneracy::step-1}
\end{align}
where define
$
    \Gamma_{1,\JA}
    :=
    \frac1m
    \sum_{i=1}^m
    \alpha_1(X_i),
$
and
$
    \Gamma_{2,\JA}
    :=
    \mathbbm{1}_{[\lambda_A>2]}
    \cdot
    |\jma|^{-1}\sum_{1\leq i_1<i_2\leq m}
    \Theta_{m,\lambda;i_1,i_2}
    \eta_{i_1,i_2}^{(A)},
$
and
$
    \Gamma_{3,\JA}
    :=
    |\jma|^{-1}
    \sum_{\ell=1}^{|\jma|} 
    \big\{
        \rho_A^{-s}\cdot 
        \prod_{\{i',j'\}\subseteq I_r^{(\ell)}}\eta_{i',j'}^{(A)}
    \big\},
$
with 
$
    \tnaijA 
    := 
    \sum_{\ell=1}^{|\jma|} 
    \big( 
        \mathbbm{1}_{[\{i,j\}\subseteq I_r^{(\ell)}]} \cdot \rho_A^{-s} \prod_{\{i',j'\}\subseteq I_r^{(\ell)}: \{i',j'\}\neq \{i,j\}} W_{i',j'}^{(A)} 
    \big) 
    - 
    s\rho_A^{-(s+1)}\mu_m \binom{r}2
$.
If $\lambda_A \leq 2$, a modified version of '``$\Gamma_2$'' replaces $\Gamma_{2,\JA}$ in equation \eqref{method::degeneracy::step-1}, but fortunately, the alternative term is invariably dominated by $\Gamma_{1,\JA} + \Gamma_{3,\JA}$. 
This simplifies our variance estimation.
Our individual variance estimators are
\begin{align}
    \hat \sigma_{\Gamma_{1,\JA}}^2
    :=
    \frac1m
    \sum_{i=1}^m
    \hat\alpha_{1;\JA}(X_i)^2; & \quad
    \hat \sigma_{\Gamma_{2,\JA}}^2
    :=
    |\jma|^{-2}
    \sum_{1\leq i_1<i_2\leq m}
    \big(
    \hat \Theta_{m,\lambda;i_1,i_2}^2
    \cdot
    A_{i,j}
    \big),
    \notag
    \\
    \textrm{and}\quad
    \hat \sigma_{\Gamma_{3,\JA}}^2
    :=&~
    |\jma|^{-1}\hat U_{\JA},
    \label{def::var_est::Gamma_3}
\end{align}
where
$
    \hat\alpha_{1;\JA}(X_i)
    :=
    r \hat\rho_{A,\JA}^{-s}
    \hat g_{A,\JA;1}(X_i)
    -
    2s \hat\rho_A^{-(s+1)}
    \hat U_{\JA}
    \hat g_{\rho_A,\JA;1}(X_i),
$
in which,
$
    \hat g_{A,\JA;1}(X_i)
    :=
    \big\{
        \sum_{I_r\in \jma: i\in I_r}
        h(A_{I_r})
    \big\}
    /
    \big|
        \{I_r\in \jma: i\in I_r\}
    \big|,
$
and $\hat g_{\rho_A,\JA;1}(X_i)$ is defined similarly to $\hat g_{A,\JA;1}(X_i)$, except that ``$h(A_{I_r})$'' in the formula of $\hat g_{A,\JA;1}(X_i)$ is replaced by ``$\sum_{i'\in I_r\backslash \{i\}} A_{i,i'}$''.
In practice, identifying the dominant term among $\sigma_{\Gamma_{1,\JA}}$, $\sigma_{\Gamma_{2,\JA}}$, and $\sigma_{\Gamma_{3,\JA}}$ is challenging due to its dependence on factors including $m, \rho_A$, and $\lambda_A$ in complex ways. 
Fortunately, we can achieve a consistent overall variance estimator by simply summing the estimated variances of these terms.

\begin{theorem}[Asymptotic normality with automatic adaptation to indeterminate degeneracy]
    \label{new-theorem::degeneracy::one-sample-universal-asymptotic-normality}
    Under the conditions of Theorem~\ref{thm::main}, we further assume 
    (i) $\rho_A\log m\to 0$,
    (ii) $\rho_A^{s} m^{\lambda_A}\to \infty$; (iii) select $\lambda_A\in (1,3)$.
    To circumvent boundary cases, we impose two technical conditions:
    (iv) $\rho_A^{s-1}m^{\lambda_A}\to\infty$ or $\to 0$;
    and
    (v) $\rho_A^{s-1}m^{\lambda_A-2}\to 0$ or $\gg \log^{2s-2}m$.
    Assume similar conditions for $B$-indexed terms. 
    We also assume $\var(\alpha_1(X_1))\geq \textrm{Constant}>0$ or it equals 0; the same goes for $\var(\beta_1(Y_1))$.
    Then we have
    \begin{align}
        \frac
            {
                \big(
                    \hat\rho_{A,\JA}^{-s}\hat U_{\JA}
                    - 
                    \hat\rho_{B,\JB}^{-s}\hat V_{\JB}
                \big)
                -
                (\rho_A^{-s}\mu_m - \rho_B^{-s}\nu_n)
            }
            {\sqrt{
                \hat \sigma_{\Gamma_{1,\JA}}^2
                +
                \hat \sigma_{\Gamma_{2,\JA}}^2
                \cdot
                \mathbbm{1}_{[\lambda_A>2]}
                +
                \hat \sigma_{\Gamma_{3,\JA}}^2
                +
                \hat \sigma_{\Gamma_{1,\JB}}^2
                +
                \hat \sigma_{\Gamma_{2,\JB}}^2
                \cdot
                \mathbbm{1}_{[\lambda_B>2]}
                +
                \hat \sigma_{\Gamma_{3,\JB}}^2
            }}
        \stackrel d\to N(0,1),
    \end{align}
    as $m, n\to\infty$,
    regardless of the respective degeneracy statuses of $\hat U_{\JA}$ and $\hat U_{\JB}$.
\end{theorem}

The selection of $\lambda_A,\lambda_B\in (1,3)$ in Theorem \ref{new-theorem::degeneracy::one-sample-universal-asymptotic-normality} follows the rationale partially outlined in \cite{shao2023u}. 
Specifically, choosing $\lambda_A<1$ is using only a diminishing fraction of input data; while in the non-degenerate case, setting $\lambda_A>3$ complicates variance estimation but will not improve inference accuracy beyond what is achieved at $\lambda_A=3$. 
The technical conditions (iv) and (v) are manageable through careful selection of $\lambda_A$ and are empirically verifiable.
The proof of Theorem \ref{new-theorem::degeneracy::one-sample-universal-asymptotic-normality}, though seemingly straightforward, is quite intricate. 
It involves examining regimes where each term dominates and demonstrating the consistency and dominance of the corresponding variance estimator in all regimes. 
Ultimately, Theorem \ref{new-theorem::degeneracy::one-sample-universal-asymptotic-normality} puts a neat wrapper around these complexities and presents a user-friendly result.

}

{\revisionfiltered

\section{False Discovery Rate (FDR) control}\label{sec::FDR}

In this section, we provide an FDR control algorithm for the network query problem (Section \ref{section::network-hashing}).
Given a query network $A$, our goal is to identify networks from a database that are dissimilar to $A$. 
Following the convention of FDR control literature, we work under the assumption that most networks in the database resemble $A$. 
To this end, we assess the covariance between test statistics and implement factor-adjusted False Discovery Rate (FDR) control, as described in \cite{fan2012estimating} and \cite{friguet2009factor}. 
For any chosen threshold $t>0$, the false discovery proportion (FDP), i.e., the ratio of incorrectly screened-out networks that actually share the same moment as network $A$, is
\begin{align}\label{eq:FDP_rel}
    {\rm FDP}(t):=\frac{\hat V(t)}{\hat R(t)}:=\frac{\sum_{k\in \{\textrm{true nulls}\}} \mathbbm{1}\big(\hat p_k\leq t\big)}{\sum_{k\in[K]} \mathbbm{1}(\hat p_k\leq t)},
\end{align}
where the denominator is known in practice but the numerator is unknown, and \{true nulls\} denotes the set of network $B_k$ satisfying $d_{m,n_k,\rho_A,\rho_{B_k}}=0$. 

\begin{theorem}\label{thm:FDP}
    Suppose the conditions in Theorem~1 hold for all query  the query network and database networks and 
    $$
    \lim_{K,m,n_{\min}\to\infty}\sqrt{\frac{\log (mK)}{m}}+\sqrt{\frac{\log(n_{\min}K)}{n_{\min}}}+\frac{\log^{1/2}(mK)}{\rho_A m}+\max_{j\in[K]}\frac{\log^{1/2}(n_jK)}{\rho_{B_j} n_j}=0
    $$ 
    Define $\gamma=\big(\sigma_{m,n_1}^{-1},\cdots,\sigma_{m,n_K}^{-1}\big)'\cdot m^{-1/2} {\rm Var}^{1/2}\big(\alpha_1(X_1)\big)$ and let $a_k:=\big(1-\gamma_k^2\big)^{-1/2}$. Denote $\iota_k :=d_{m,n_k,\rho_A,\rho_{B_k}}/\sigma_{m,n_k}$. 
    Then, 
    \begin{align*}
     \lim_{K,m, n_{\min}\to\infty}\bigg[{\rm FDP}(t)-\frac{\sum_{k\in \{\textrm{true nulls}\}}\big[\Phi\big(a_k(z_{t/2}+\eta_k)\big)+\Phi\big(a_k(z_{t/2}-\eta_k)\big)\big]}{\sum_{K\in[K]}\big[\Phi\big(a_k(z_{t/2}+\eta_k+\iota_k)\big)+\Phi\big(a_k(z_{t/2}-\eta_k-\iota_k)\big)\big]}\bigg]=0,\ {\rm a.s.},
    \end{align*}
    where $z_t:=\Phi^{-1}(t)$,$\eta_k:=\gamma_k W$ for $\forall k\in[K]$ with $W\sim N(0,1)$ and $n_{\min}:=\min_{k\in [K]} n_k$.
\end{theorem}
In Theorem \ref{thm:FDP}, the vector $\gamma\in\mathbb{R}^K$ characterizes the correlation between the test statistics. 
The theorem approximates the actual FDP in \eqref{eq:FDP_rel} by
$
\big[\sum_{k\in{\{\textrm{true null}\}}}
    \big\{\Phi\big(\hat a_k(z_{t/2}+\hat \eta_k)\big)+\Phi\big(\hat a_k(z_{t/2}-\hat \eta_k)\big)\big\}
\big]
\big/
\big\{
    \sum_{k\in[K]}\mathbbm{1}(\hat p_k\leq t)
\big\}, 
$
utilizing the estimated $\gamma$ and $W$. The procedure for network screening is outlined in Algorithm~\ref{alg::FDP}, where the tuning parameter $\zeta\in(0,1)$ should be less than the anticipated proportion of true null hypotheses.

}

\begin{algorithm}[ht!]
    \caption{Network screening}\label{alg::FDP}
    {\bf Input:} Network database: $\{B_1,\ldots,B_K\}$, query keyword $A$, target FDR $\alpha\in(0,1)$, threshold $\zeta$\\
    {\bf Output:}  $B_k$'s which have different moments compared with $A$  \\
    {\bf Procedure:}
        For $k=1,\ldots,K$, compute the test statistics and $p$-values
        $$
            \hat T_{m,n_{k}}=\frac{\hat \rho_A^{-s}\cdot \hat U_m-\hat \rho_{B_k}^{-s}\cdot \hat V_{n_k}}{\hat S_{m,n_k}}\quad {\rm with}\quad \hat S_{m,n_k}^2=\frac{1}{m^2}\sum_{i=1}^m \hat\alpha_1^2(X_i)+\frac{1}{n_k^2}\sum_{j=1}^{n_k}\hat\beta_1^2(Y_j^{(k)})
        $$
        $$
        \hat p_k=2\min\big\{\Phi(\hat T_{m,n_k}),\ 1-\Phi(\hat T_{m,n_k})\big\}.
        $$
        Compute the principal component:
        $$
        \hat \gamma_k=\frac{\sqrt{\frac{1}{m^2}\sum_{i=1}^m \hat a_1^2(X_i)}}{\hat S_{m,n_k}}\quad {\rm and}\quad \hat a_k=\frac{1}{\sqrt{1-\hat \gamma_k^2}},\quad k\in [K].
        $$
        Denote $\mathcal{A}\subset [K]$ such that $|\mathcal{A}|=\zeta \cdot K$ and $|\hat T_{m,n_j}|\leq |\hat T_{m,n_k}|$ for all $j\in\mathcal{A}$ and $k\notin \mathcal{A}$. Estimate the factor by
        $$
        \hat\omega\leftarrow \underset{\omega\in\mathbb{R}}{\arg\min} \sum_{k\in\mathcal{A}}\big(\hat T_{m, n_k}-\hat\gamma_k \omega\big)^2,\quad \textrm{then set}\quad \hat \eta_k:=\hat\gamma_k \hat\omega,\quad \forall k\in[K].
        $$
        Solve 
        $$
        \hat t\leftarrow \max\Bigg\{t\in(0,1): \frac{\sum_{k=1}^K \big[\Phi\big(\hat a_k(z_{t/2}+\hat \eta_k)\big)+\Phi\big(\hat a_k(z_{t/2}-\hat \eta_k)\big)\big]}{\sum_{k\in [K]}\mathbbm{I}\big(\hat p_k\leq t\big)}\leq \alpha\Bigg\}.
        $$

        Screen out (reject null hypotheses for) those networks $B_k$ with $\hat p_k\leq t$. 
\end{algorithm}

}

\section{Simulations}

\begin{figure}[ht!]
\centering
\begin{adjustbox}{width=1.05\linewidth,center}
    \centering
    \includegraphics[width=0.19\linewidth]{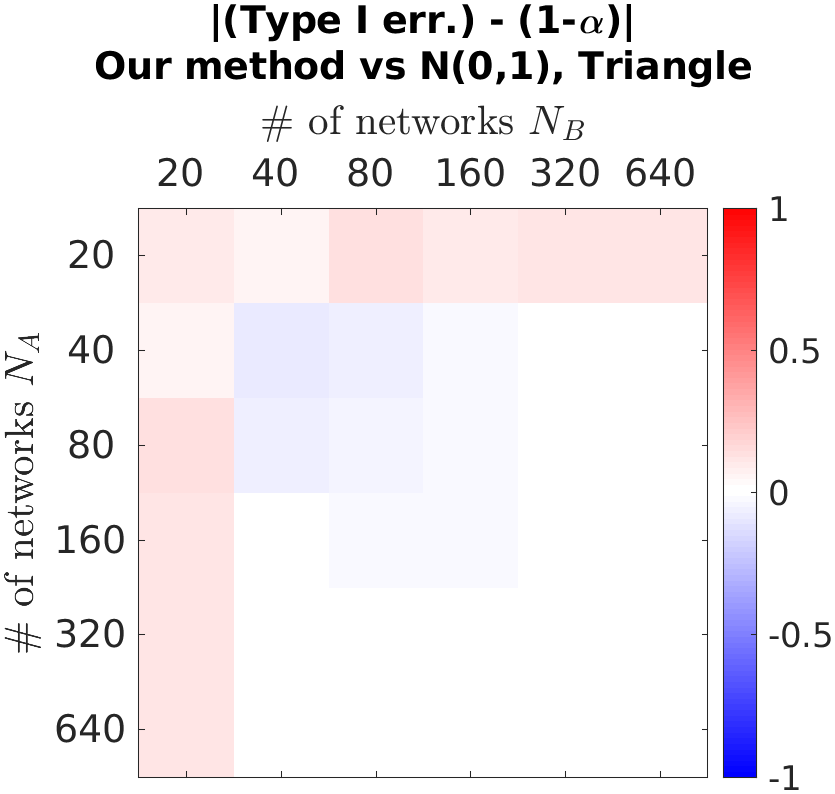}
    \includegraphics[width=0.19\linewidth]{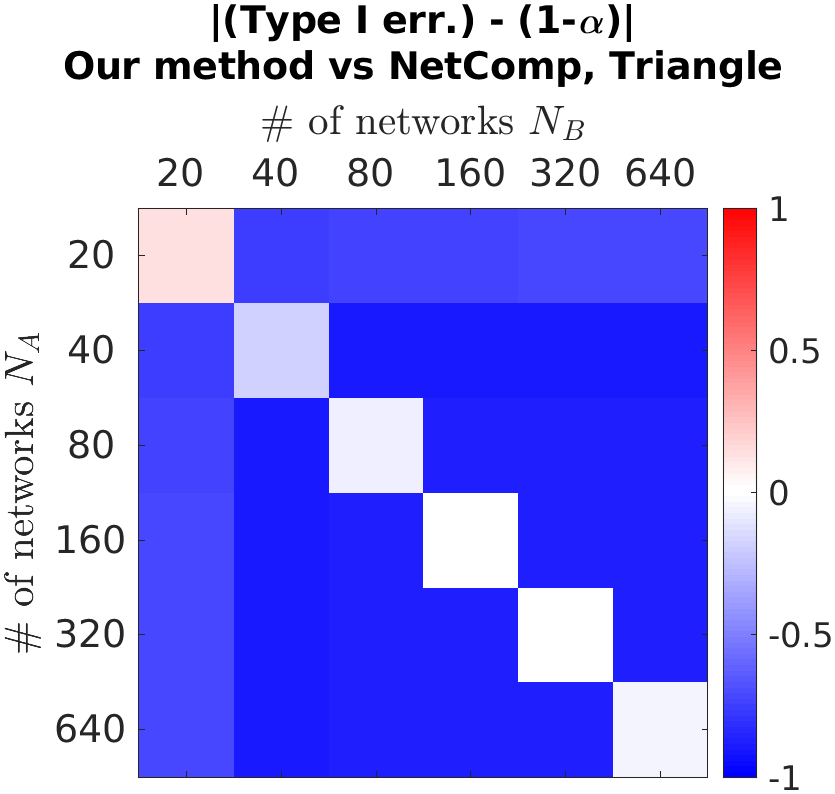}
    \includegraphics[width=0.19\linewidth]{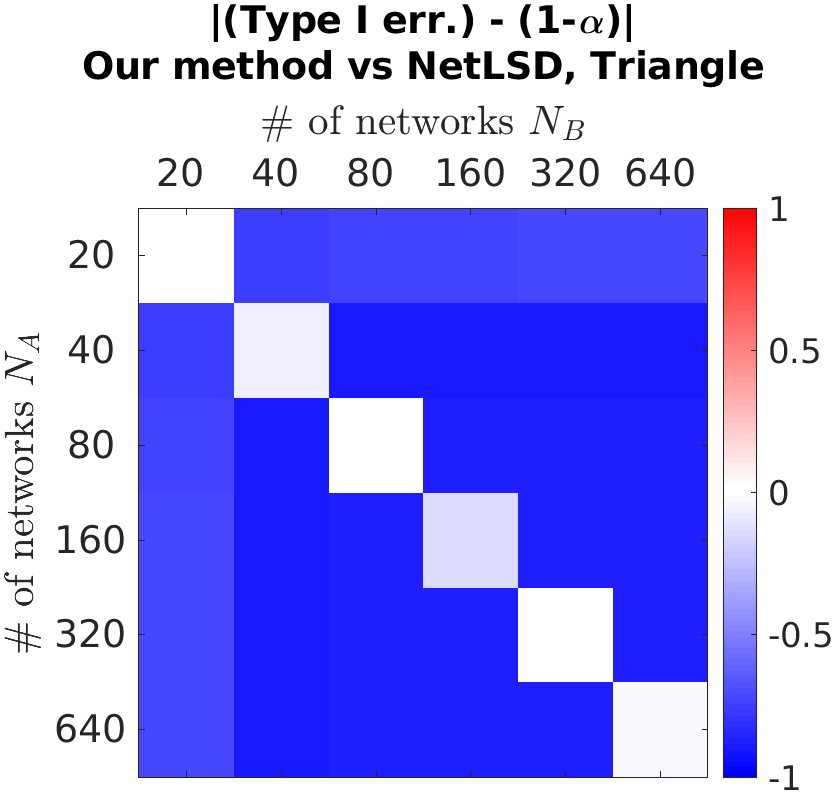}
    \includegraphics[width=0.19\linewidth]{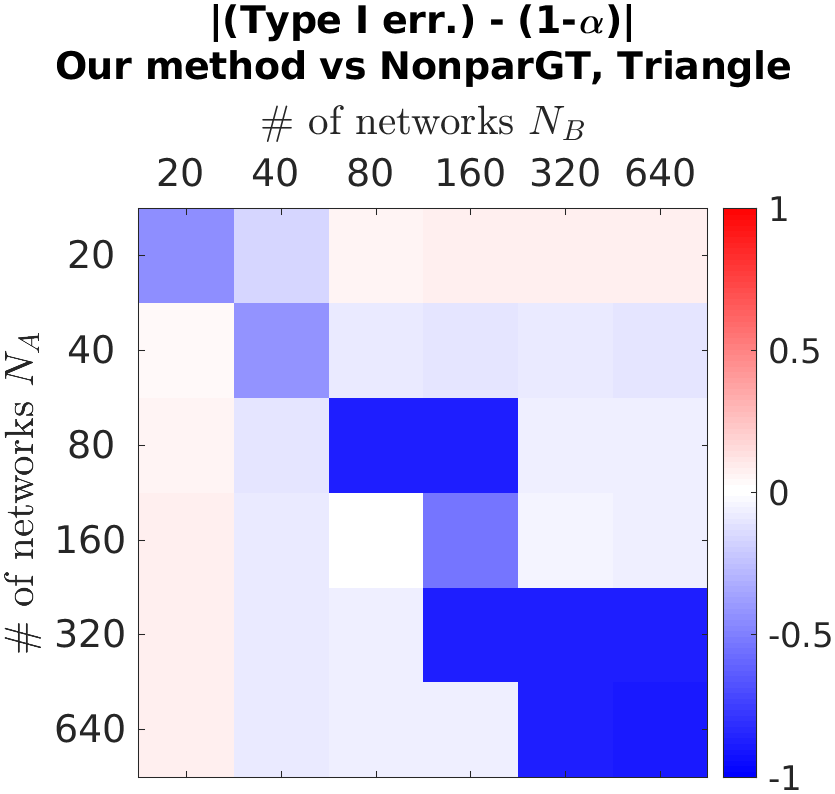}
    \includegraphics[width=0.19\linewidth]{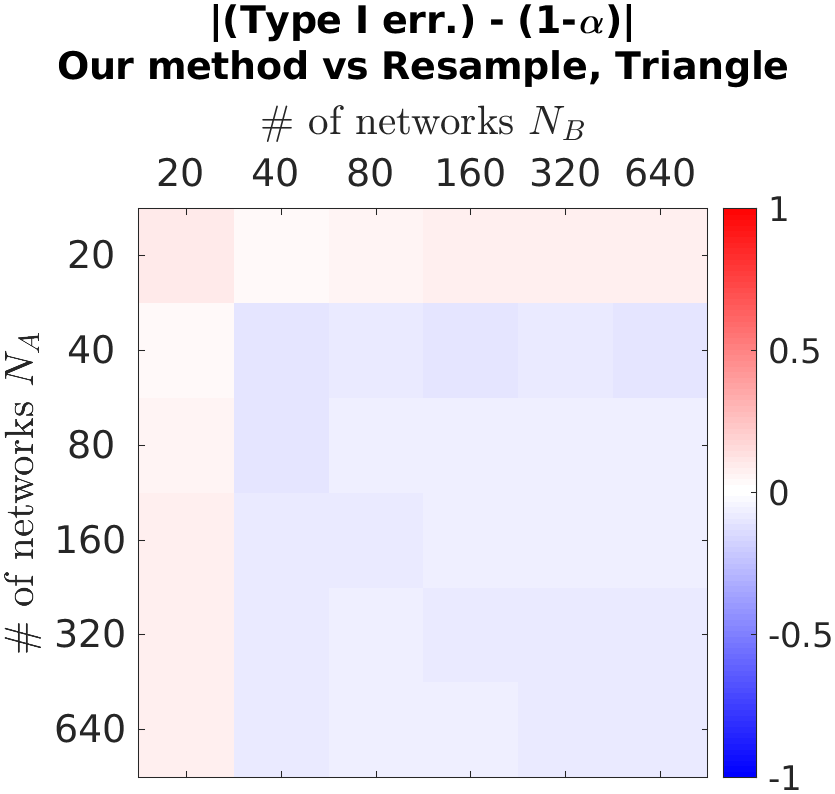}
    \includegraphics[width=0.19\linewidth]{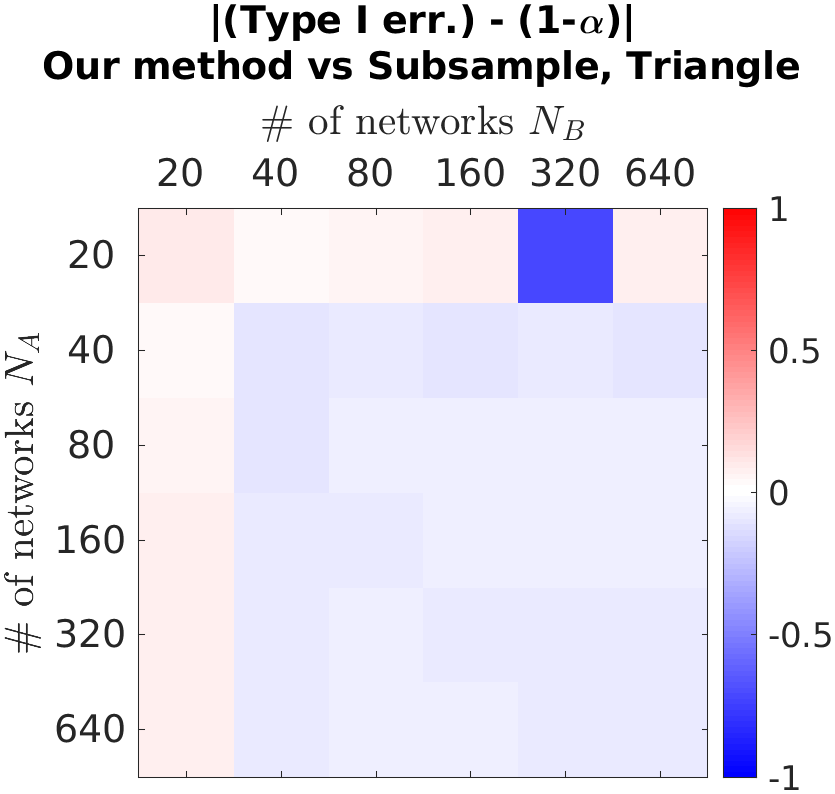}
\end{adjustbox}
\begin{adjustbox}{width=1.05\linewidth,center}
    \includegraphics[width=0.19\linewidth]{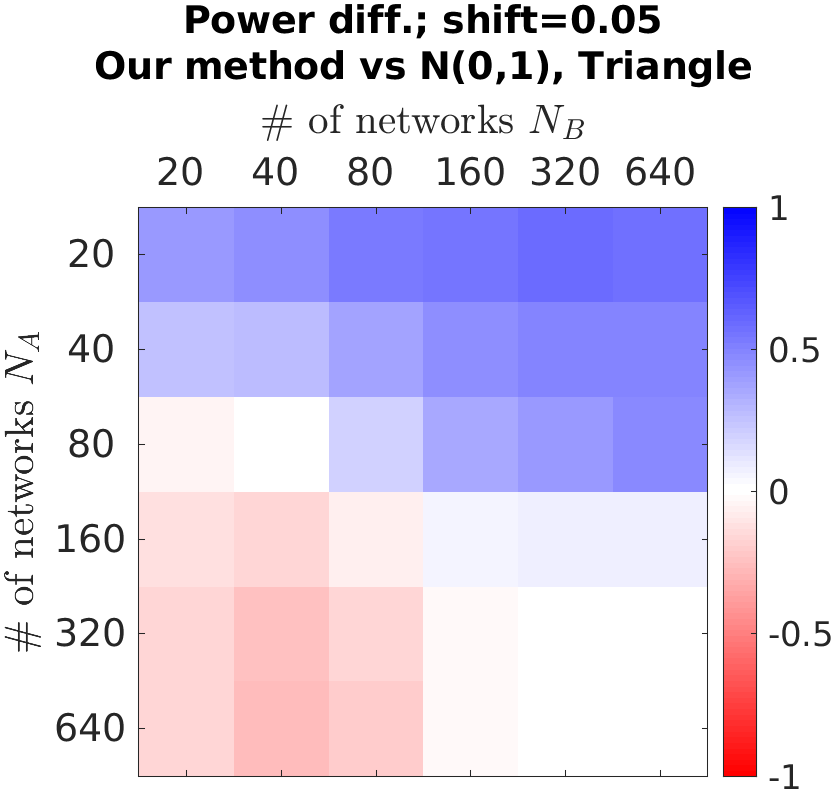}
    \includegraphics[width=0.19\linewidth]{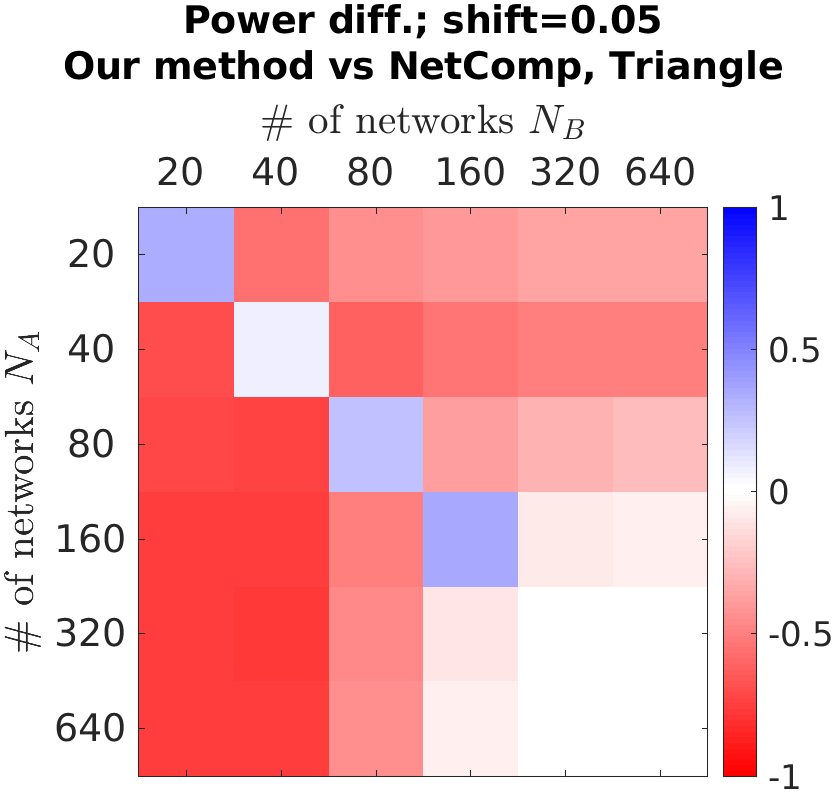}
    \includegraphics[width=0.19\linewidth]{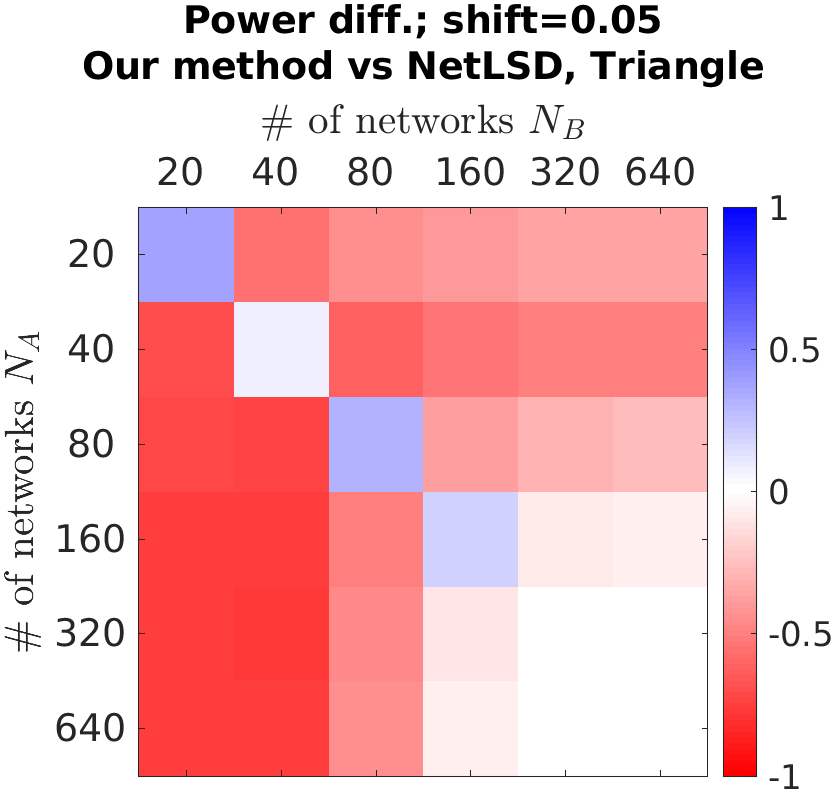}
    \includegraphics[width=0.19\linewidth]{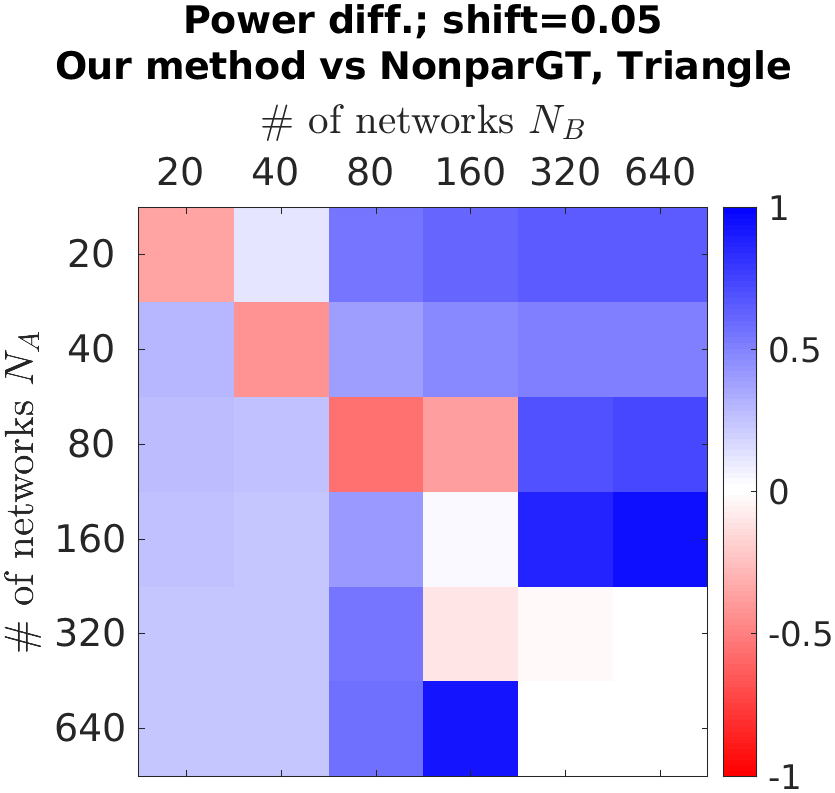}
    \includegraphics[width=0.19\linewidth]{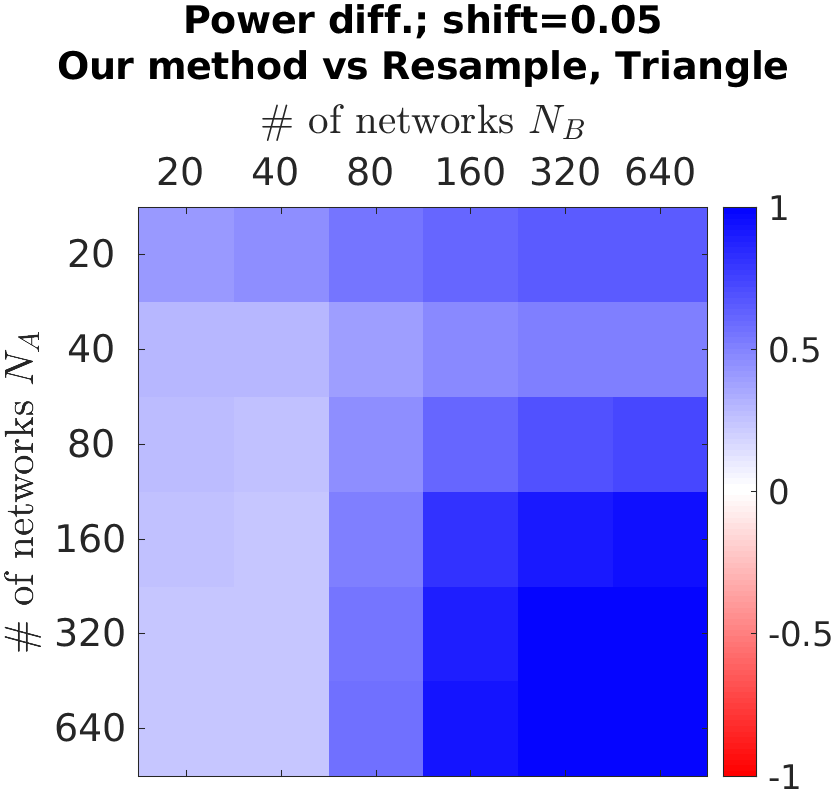}
    \includegraphics[width=0.19\linewidth]{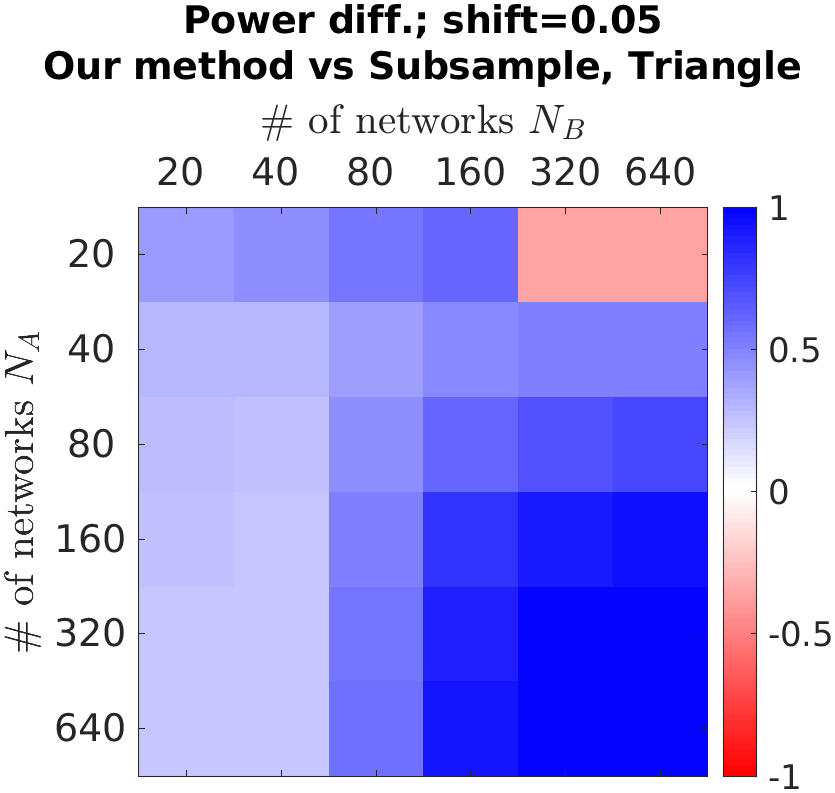}
\end{adjustbox}
\begin{adjustbox}{width=1.05\linewidth,center}
    \includegraphics[width=0.19\linewidth]{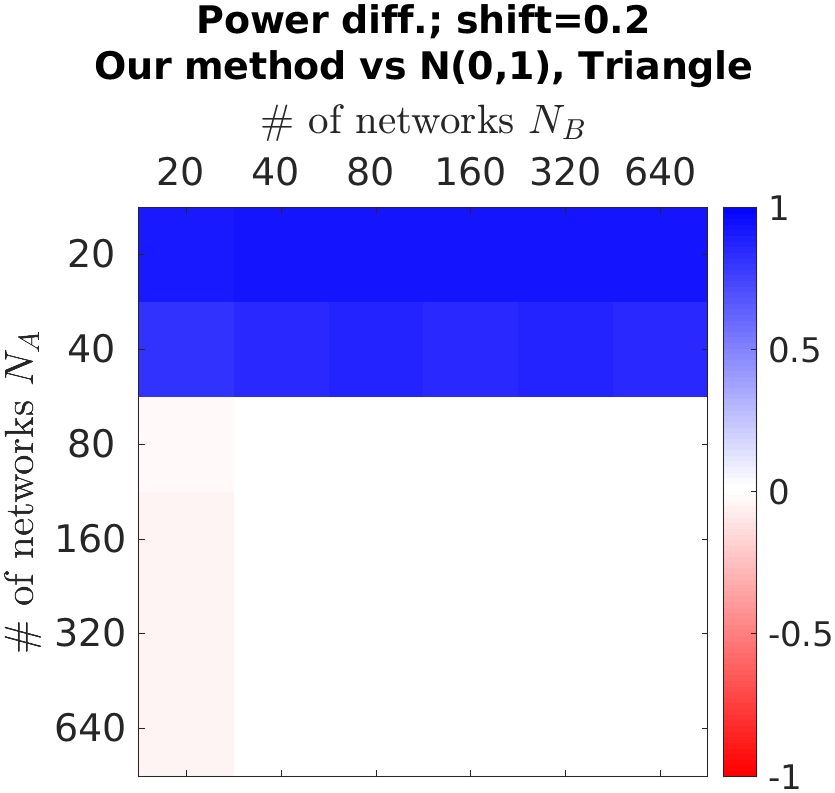}
    \includegraphics[width=0.19\linewidth]{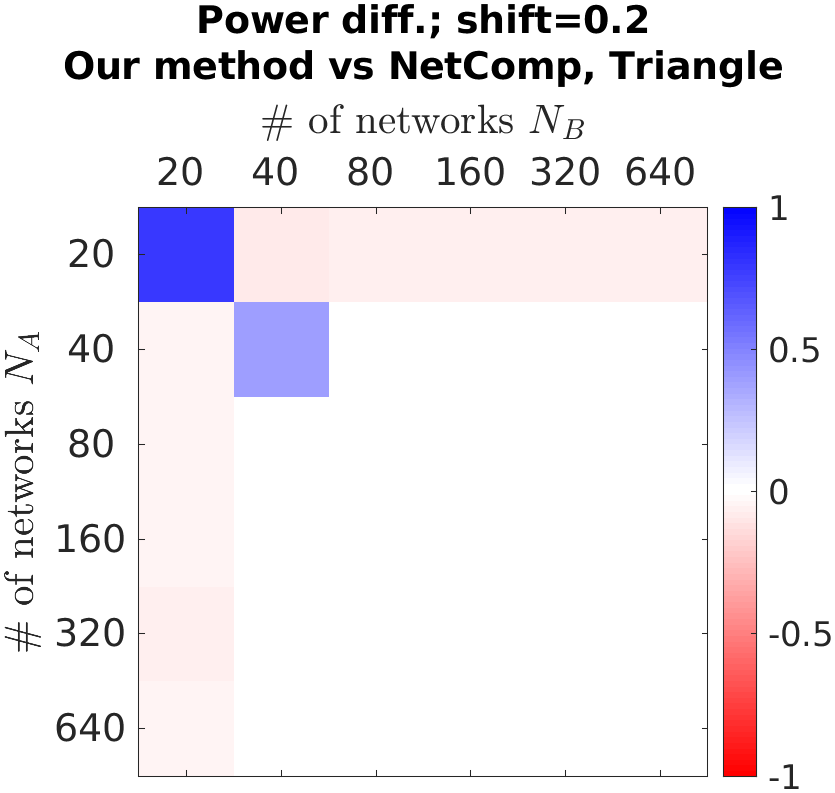}
    \includegraphics[width=0.19\linewidth]{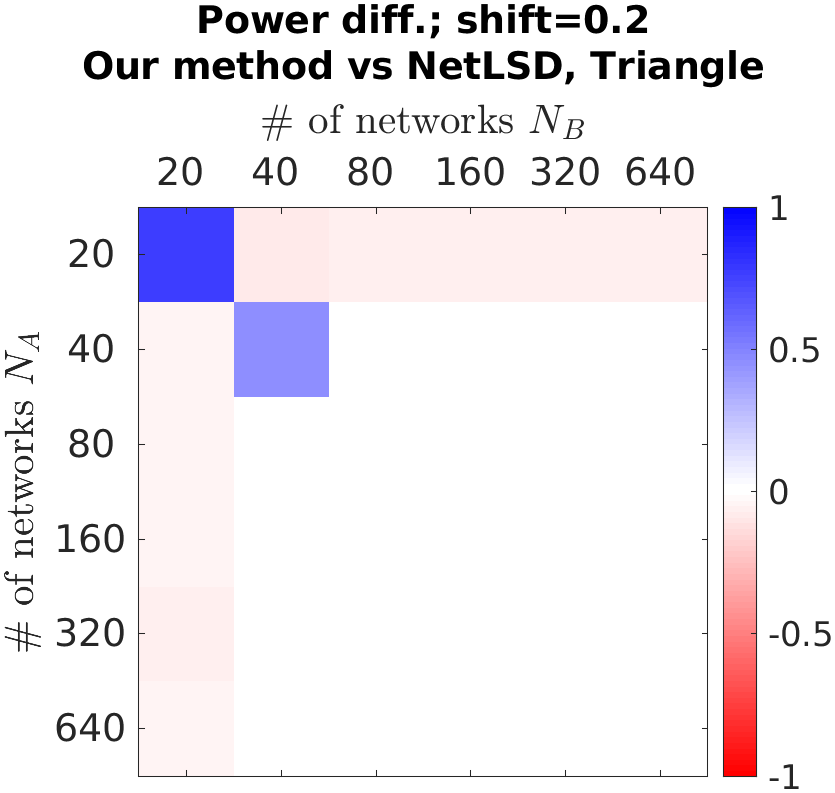}
    \includegraphics[width=0.19\linewidth]{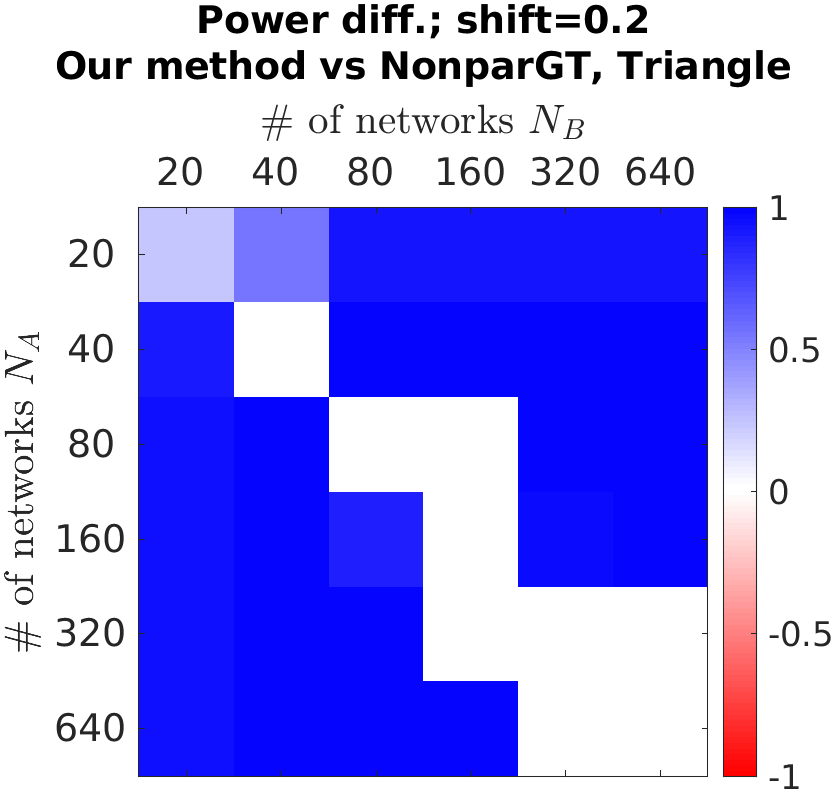}
    \includegraphics[width=0.19\linewidth]{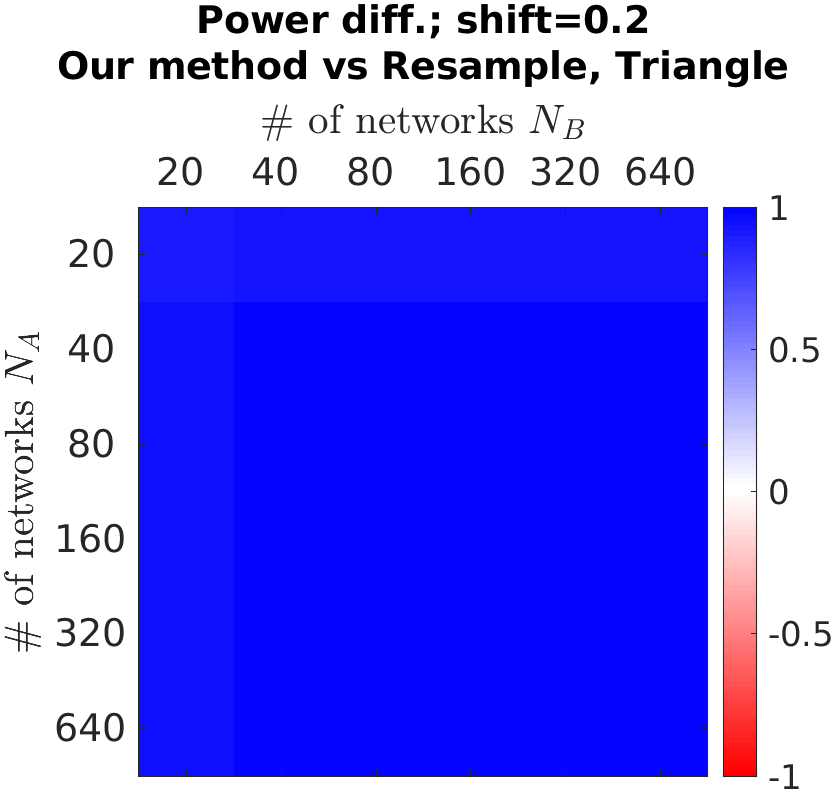}
    \includegraphics[width=0.19\linewidth]{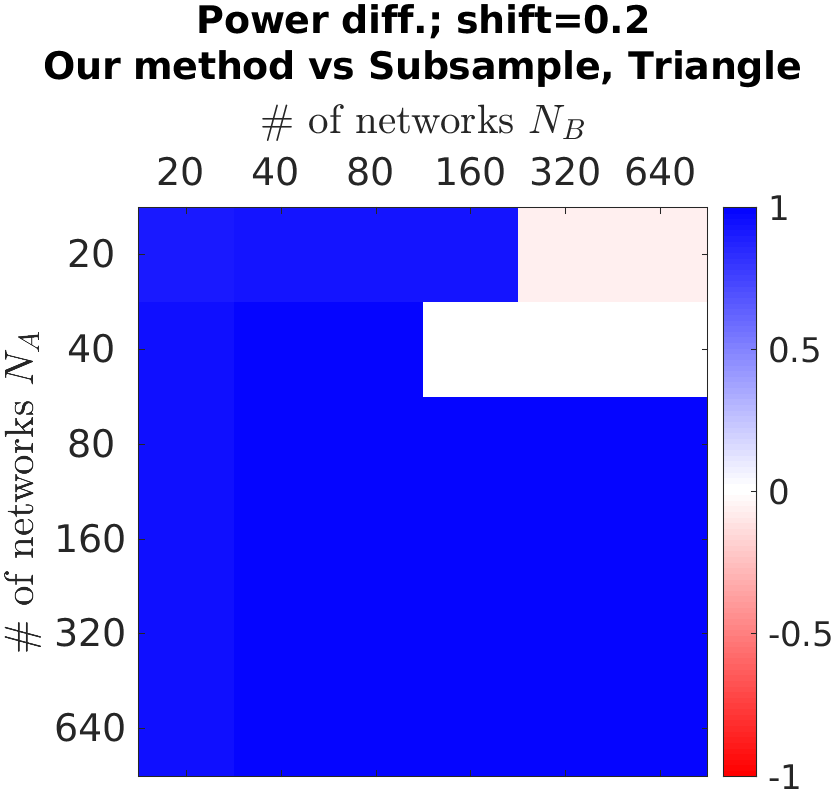}
\end{adjustbox}
\caption{Comparison of type I error control (Row 1) and power difference (Row 2: $\varpi=0.05$ and Row 3: $\varpi=0.20$).  Blue in Row 1 and green in Rows 2 and 3 indicate performance advantage of our method; red and brown indicate disadvantageous comparisons.}
\label{new-fig::new-simu-1-type-I-error-and-power}
\end{figure}

{\revisionfiltered
\subsection{Simulation 1: Type I error and power comparison}
\label{subsec::simu-1::type-I-error-and-power}

In our first experiment, we evaluate our method's ability to accurately control type-I error at the nominal level and its power. 
We generated data from two graphons: $f_A(u,v) = 1.38\cdot \exp(-(u+v)/3)$, and $f_B(u,v) = f_A(u,v) + \varpi$, with $\varpi\in{0,0.05,0.2}$ indicating the location shift. 
The observed rejection rate of any method is denoted by $\varrho$. 
Under $H_0$ (when $\varpi=0$), we assess performance by $\big| (1-\varrho)-(1-\alpha) \big|$, the smaller the better. 
Under $H_a$ (when $\varpi>0$), $\varrho$ represents the method's power, where larger values are preferable.
We compare our method against several benchmarks: normal approximation; NetComp \citep{wills2020metrics}; NetLSD \citep{tsitsulin2018netlsd}; NonparGT \citep{agterberg2020nonparametric}; Resampling bootstrap; and Subsampling bootstrap. 
Benchmarks NetComp and NetLSD, originally only providing heuristic dissimilarity measures, were adapted using the approach from Section 2.6.1 of \citet{wills2020metrics} to generate ad-hoc p-values. 
Due to their varying computational costs, we set different Monte Carlo repetitions for each method: $10^4$ for our method, $10^3$ for subsampling, 100 for resampling, 30 for NetLSD and NonparGT, and 25 for NetComp.

Figure \ref{new-fig::new-simu-1-type-I-error-and-power} presents our results.
Our method notably outperforms bootstrap approaches in both type-I error control and power across most settings. 
Compared to normal approximation, our method demonstrates superior higher-order accuracy in type-I error control for moderate network sizes and consistently higher power in a majority of settings. 
It is important to note that, as shown in Row 1, plots 2 and 3, the ad-hoc testing procedures using NetComp and NetLSD do not effectively control type-I error in networks of varying sizes. Therefore, their apparent power advantages in these contexts are not meaningful when compared to our method.

}

\subsection{Simulation 2: Network hashing and querying}
\label{subsec::simulation3::10differentgraphon}
In this experiment, we simulate a large network database consisting of 10 different graphon models, each corresponds to 100 adjacency matrices as database entries.  
We use our Algorithm \ref{alg::hashing} to hash database entries, and then evaluate the accuracy of our Algorithm \ref{alg::querying-screening} on querying two representative keyword networks.
We compare the time costs of hashing and querying with 30 repeated experiments.  To save memory, in each experiment we only re-generate the queried keyword network but reuse the database generated by the first repetition.
We conduct two sub-simulations.  Sub-simulation 1 aims to compare different methods' AUC curves and time costs, therefore, we generate one keyword network from one of the constituent graphons in the database.
Sub-simulation 2 aims to visualize a sketch of our method's query results, therefore, we compare the query results of two different keyword networks from graphons inside and outside the database, respectively.
Due to page limit, we {\revisionfiltered relegate} the formulation of database graphons and the two keyword networks to Section \ref{supple::table for coverage probability} in Supplementary Material.
When querying, we perform steps 1 \& 2 of Algorithm \ref{alg::querying-screening} to evaluate the ROC curves and AUC scores that measure the accuracy of our method in ranking database entries by similarity to the queried keyword network.
Due to page limit, we only compare AUC scores for all methods.
For simplicity, we equate all network sizes.
We repeat each experiment 30 times and cap the total running time for each method for each network size at 12 hours.

\begin{figure}[h!]
    \centering
    \includegraphics[width=0.3\textwidth]{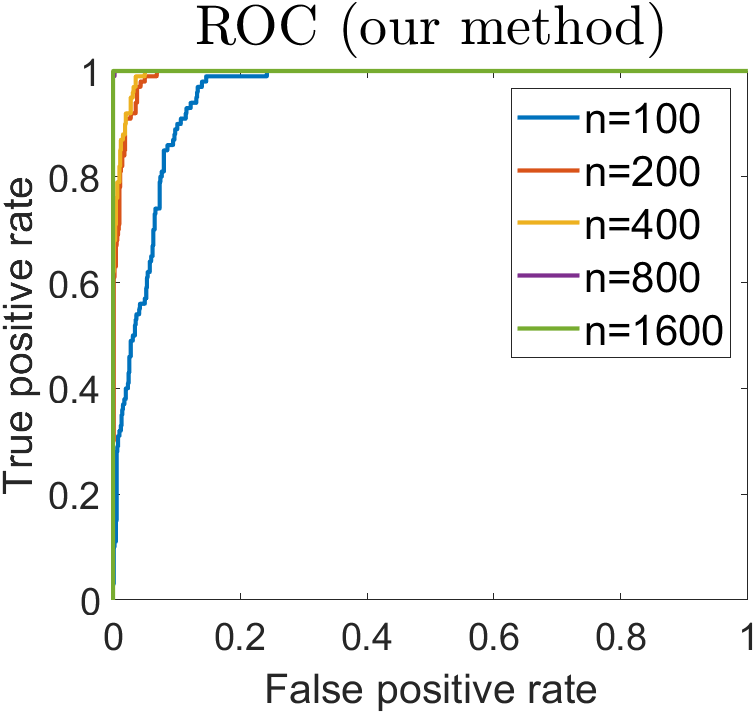}
    \includegraphics[width=0.31\textwidth]{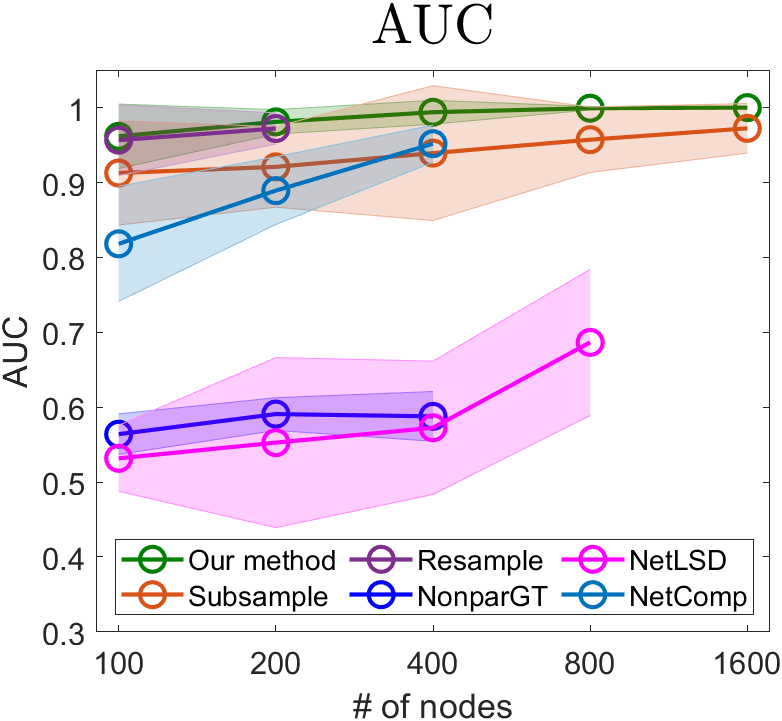}
    \includegraphics[width=0.3\textwidth]{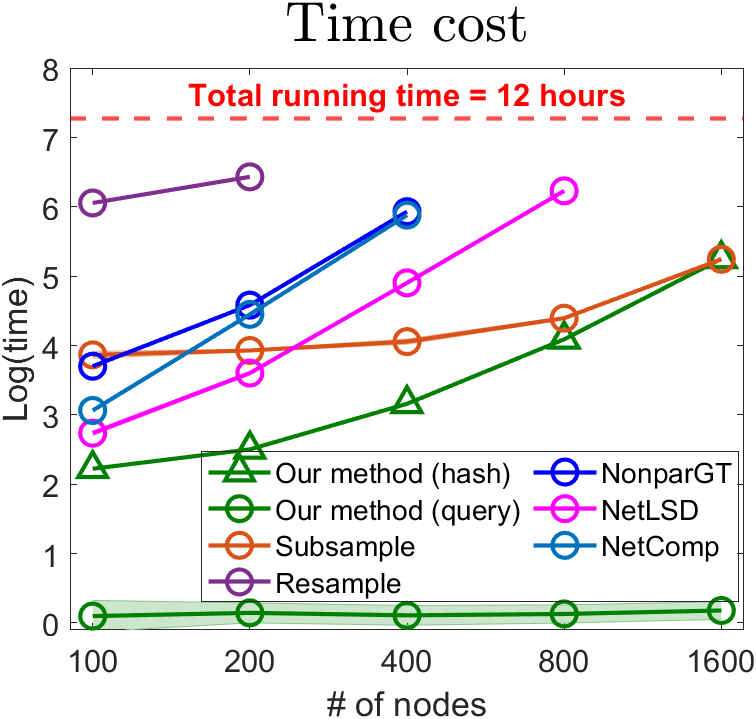}
    \\
    \includegraphics[width=0.3\textwidth]{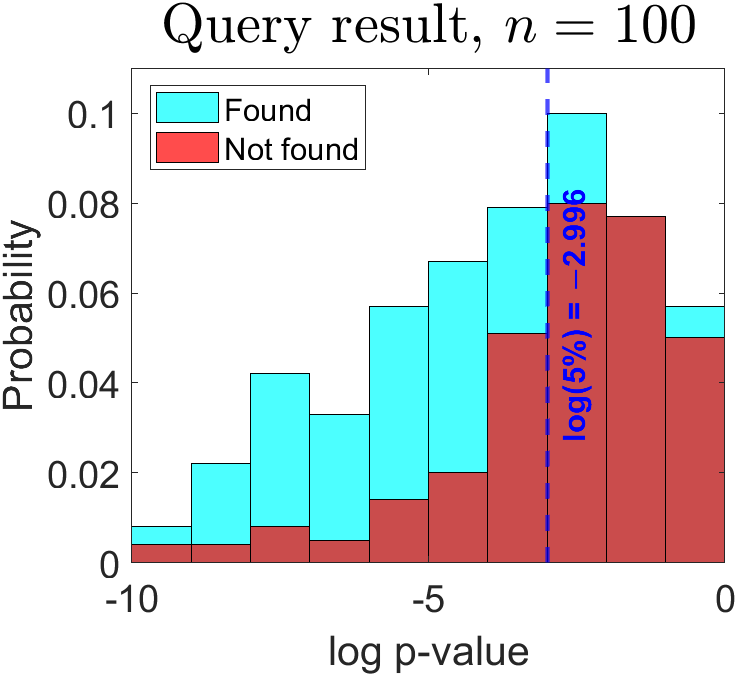}
    \includegraphics[width=0.3\textwidth]{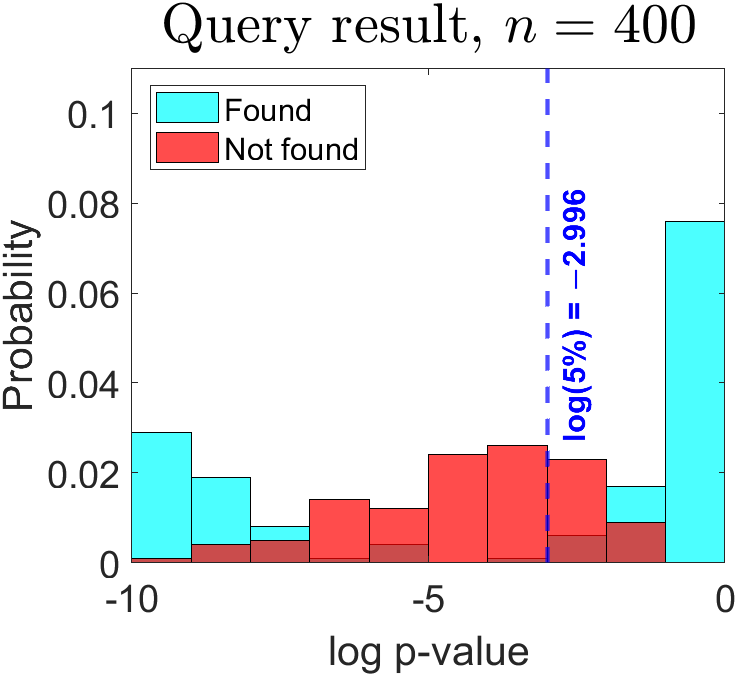}
    \includegraphics[width=0.3\textwidth]{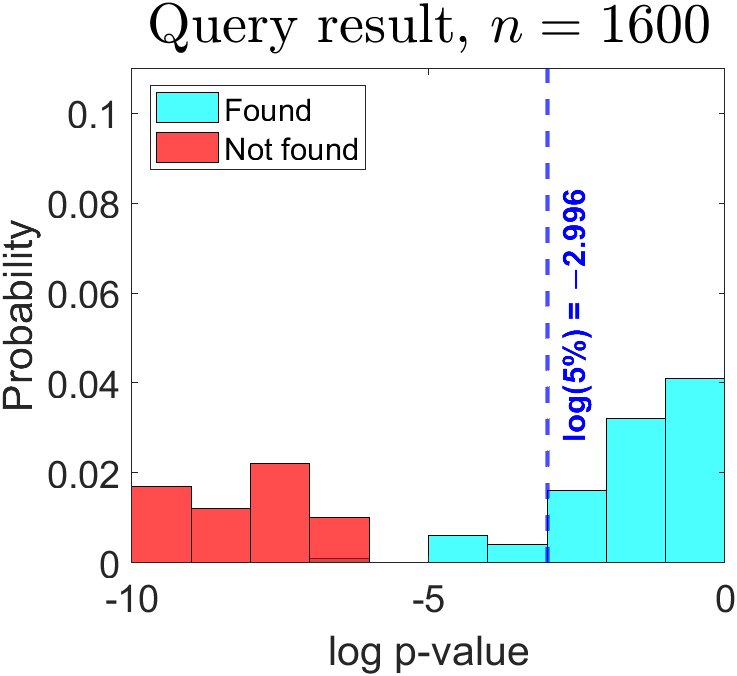}
    \caption{Database offline hashing and querying.
    Row 1: comparison of methods on query accuracy and time cost.
    In row 2, we kept the $X$-axis range consistent, but this cuts out some cyan bars on the far left.
    For plots with complete $X$-axes, see Section \ref{supple::different graphon} in Supplementary Material.
    }
    \label{fig::simulation-3}
\end{figure}

Row 1 of Figure \ref{fig::simulation-3} shows the result for sub-simulation 1.  {\revisionfiltered In this example, the motifs considered are triangles and V-shapes.} 
The ROC and AUC plots confirm our Algorithm \ref{alg::querying-screening}'s high accuracy in screening database entries similar to the queried keyword.  
The time cost plot clearly shows the speed advantage of our method.
Importantly, to query a new keyword,
our method only costs the time described by the pink curve tagged query time, and we would not need to repeat the hashing step;
in stark contrast, all the other methods would need to rerun, which incurs the same time costs shown on their curves.  This experiment therefore demonstrate our method's significant advantage in scalability.
Row 2 of Figure \ref{fig::simulation-3} shows the result for sub-simulation 2 by
comparing the p-value distributions associated with both keywords.  
We marked the typically used $5\%$ significance line in dashed blue.  Red bars to the right of it are making type I error.  As $n$ increases, we see the expected result that all red bars move to the left of the dashed blue threshold line.
But the cyan bars, which corresponds to power, should be interpreted with much more carefulness.
Notice that by construction, the keyword network corresponding to cyan bars only matches $10\%$ database entries but does not match the remaining $90\%$, so the majority chunk of cyan bars should still remain distant from 0 as $n$ increases, see also the full-X-scale plots in Section \ref{supple::different graphon} in Supplementary Material.

{\revisionfiltered

\subsection{FDR control in multiple testing}
\label{subsec::simu-4::pooling}

This simulation assesses the effectiveness of our Algorithm \ref{alg::FDP} in controlling the FDR for keyword querying (Section \ref{section::network-hashing}), where $N_A=1$ and $N_B=300$. 
The performance of FDR control typically hinges on several factors:
(a) proportions of true null and alternative hypotheses;
(b) 'gap' between these hypotheses;
and
(c) any tuning parameters in the algorithm.
In each test, we use two graphons $f_A$ and $f_B+\varpi$, designating $f_A$ as the queried keyword network. 
The database entries are generated with a $\mathfrak{q}$ proportion following $f_A$ (true $H_0$'s) and $1-\mathfrak{q}$ following $f_B$ (true $H_a$'s). 
We vary $\mathfrak{q}$ from 0.1 to 0.9 and $\varpi$ from 0 to 0.05. 
Our goal is to maintain an FDR under $\alpha=0.1$, setting $\zeta=0.1$.
We consider four graphon models:
(1) stochastic block model (SBM) with 3 equal-sized communities, connection probabilities of $[0.1, 0.1, 0.2; 0.1, 0.4, 0.1; 0.2, 0.1, 0.9]$;
(2) SBM variant with probabilities $[0.1, 0.7, 0.2; 0.7, 0.5, 0.6; 0.2, 0.6, 0.3]$;
(3) smooth graphon $f(u,v)=(x+y)^2/4$.
(4) smooth graphon $f(u,v)=e^{3(x-1)}+e^{3(y-1)}$.
Power is quantified by the ratio of correctly rejected null hypotheses. 
As our method uniquely offers FDR control among network two-sample test methods, no benchmark comparisons are included in this simulation.

\begin{figure}[h]
    \centering
    \begin{adjustbox}{width=1.05\linewidth,center}
        \centering
        \includegraphics[width=0.27\linewidth]{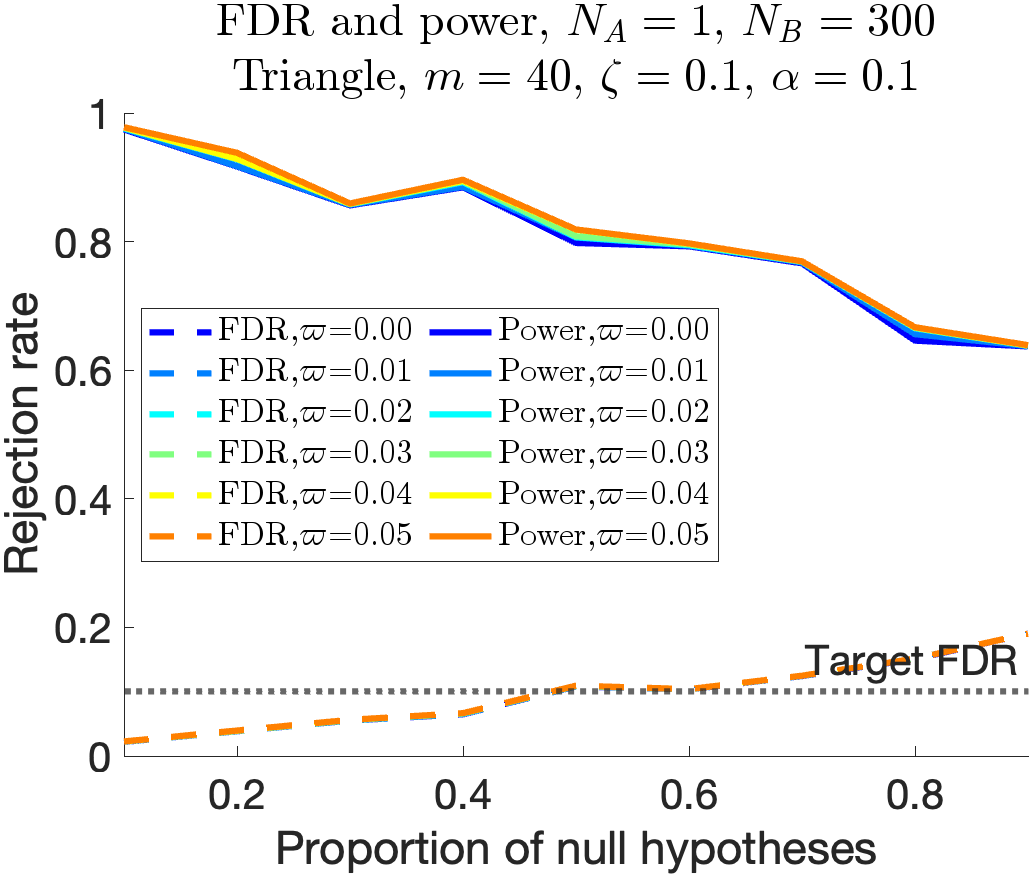}
        \includegraphics[width=0.27\linewidth]{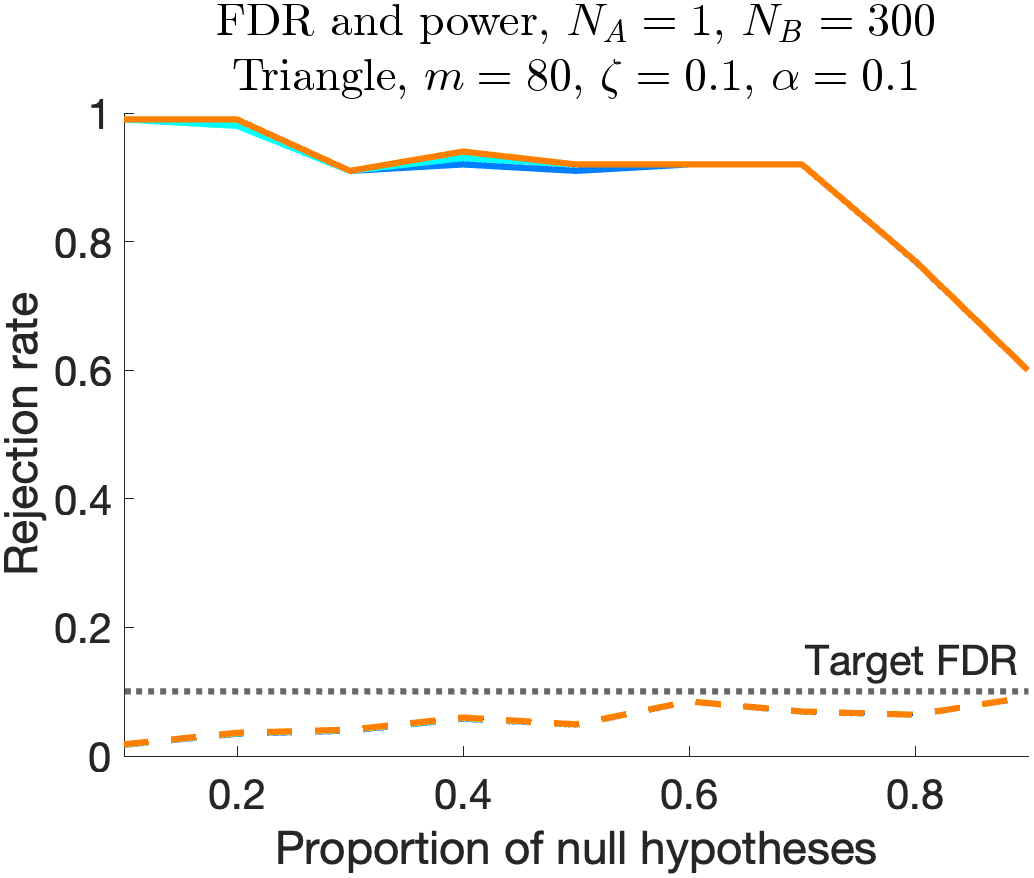}
        \includegraphics[width=0.27\linewidth]{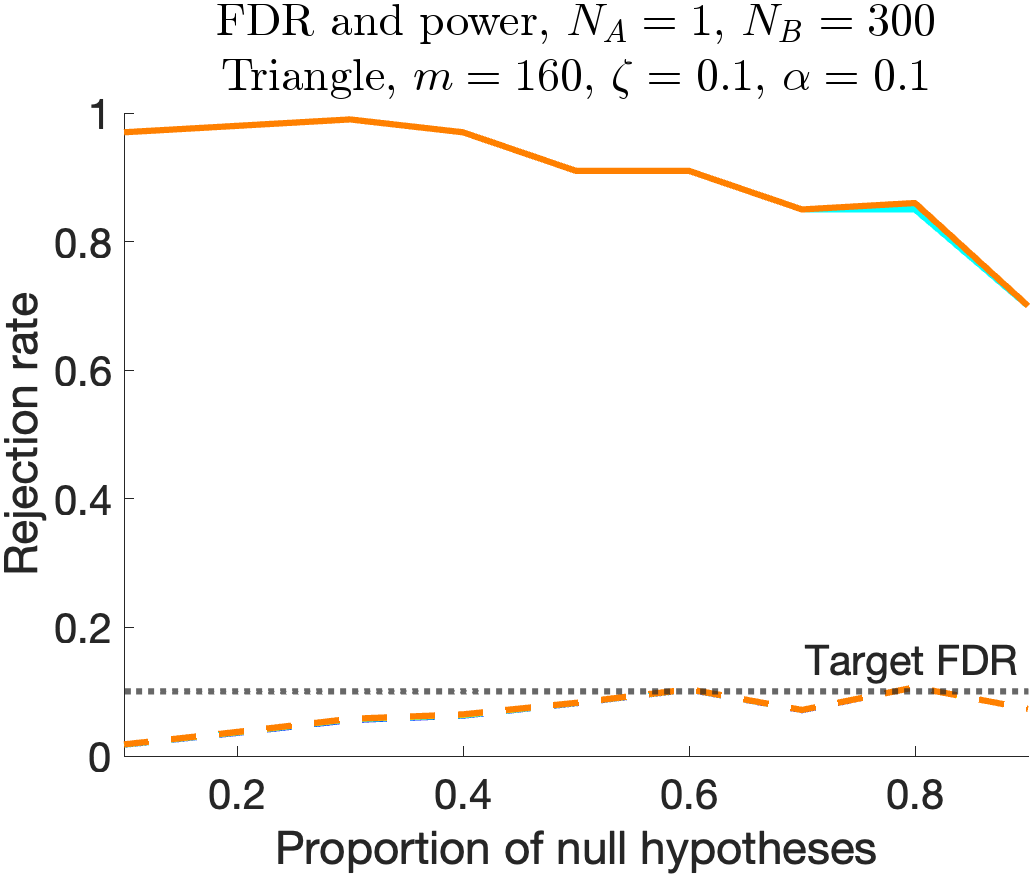}
        \includegraphics[width=0.27\linewidth]{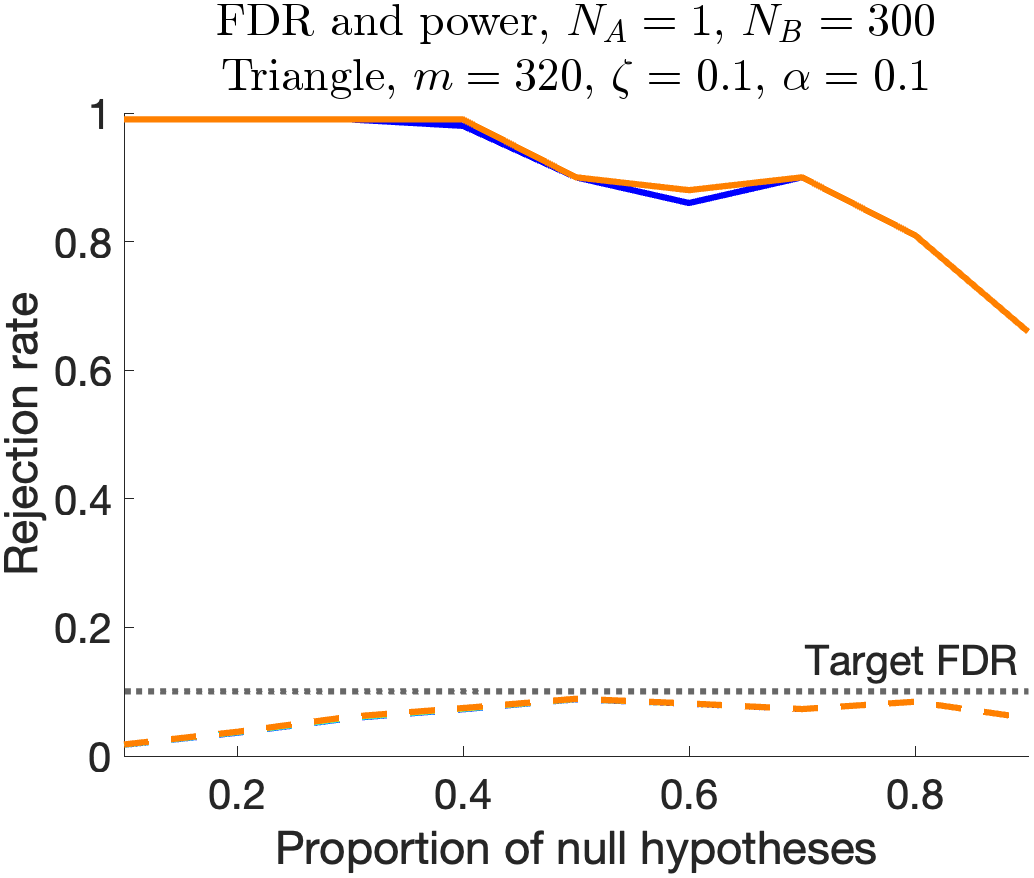}
    \end{adjustbox}
    \begin{adjustbox}{width=1.05\linewidth,center}
        \centering
        \includegraphics[width=0.27\linewidth]{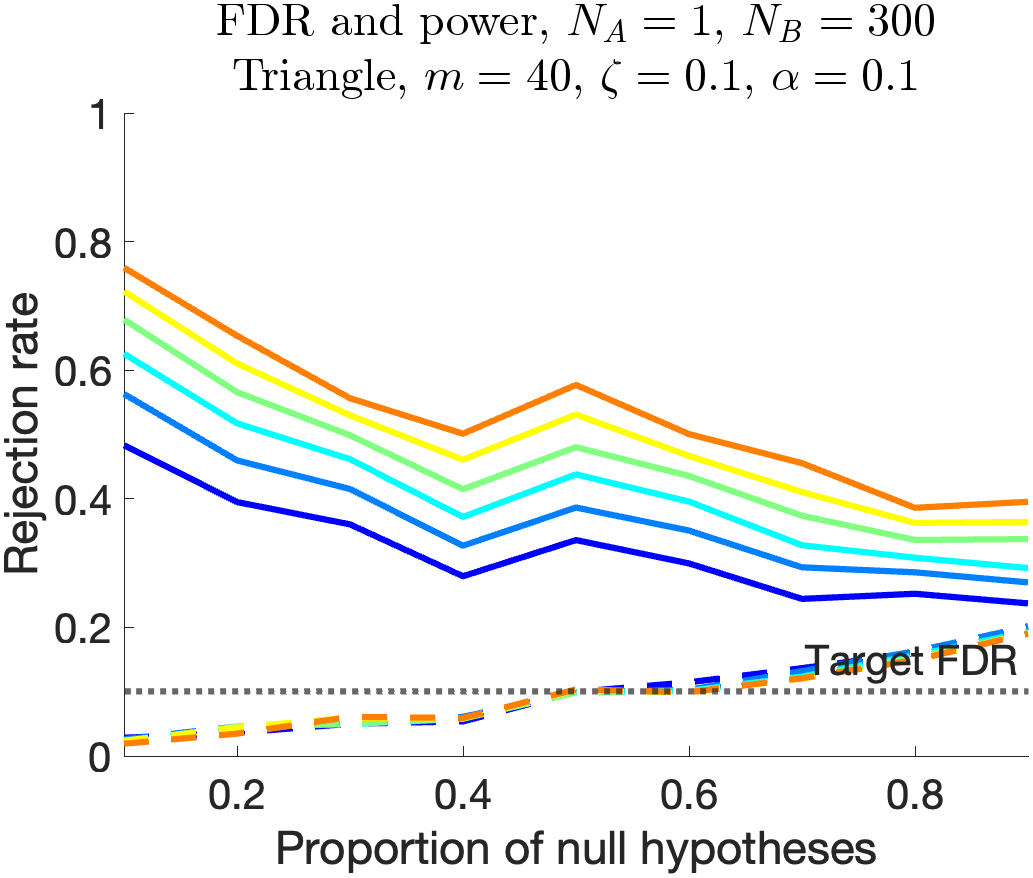}
        \includegraphics[width=0.27\linewidth]{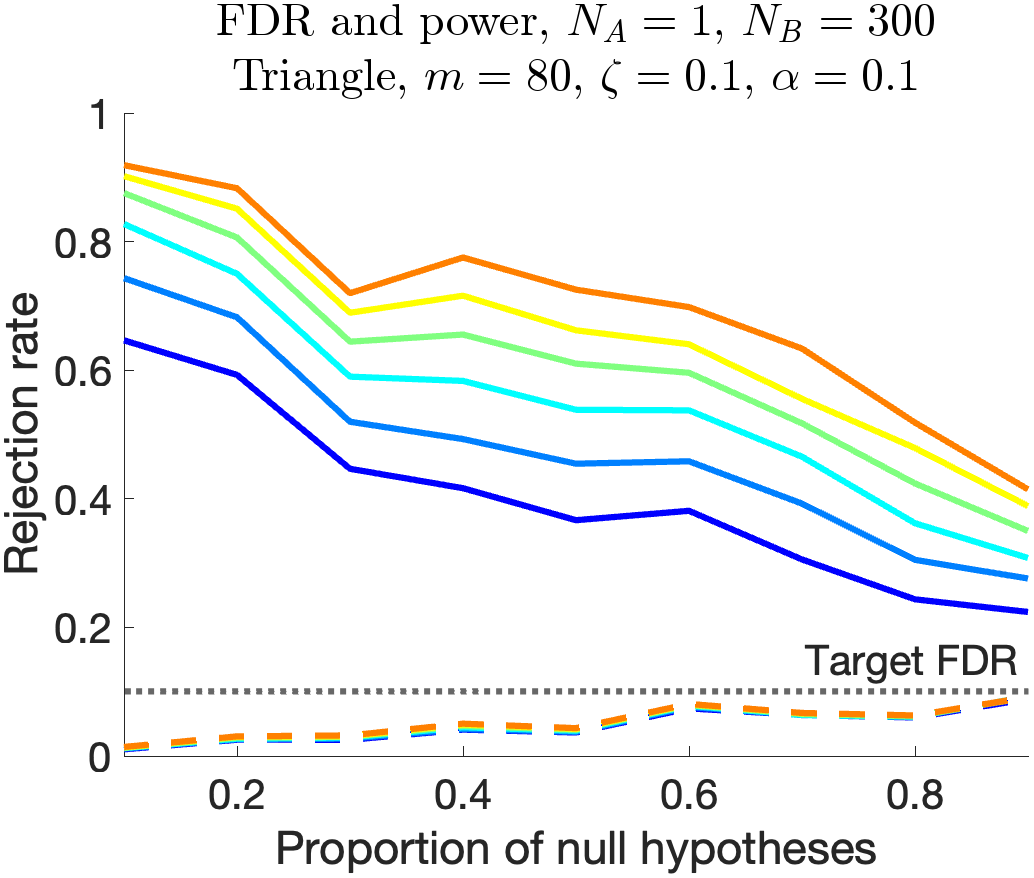}
        \includegraphics[width=0.27\linewidth]{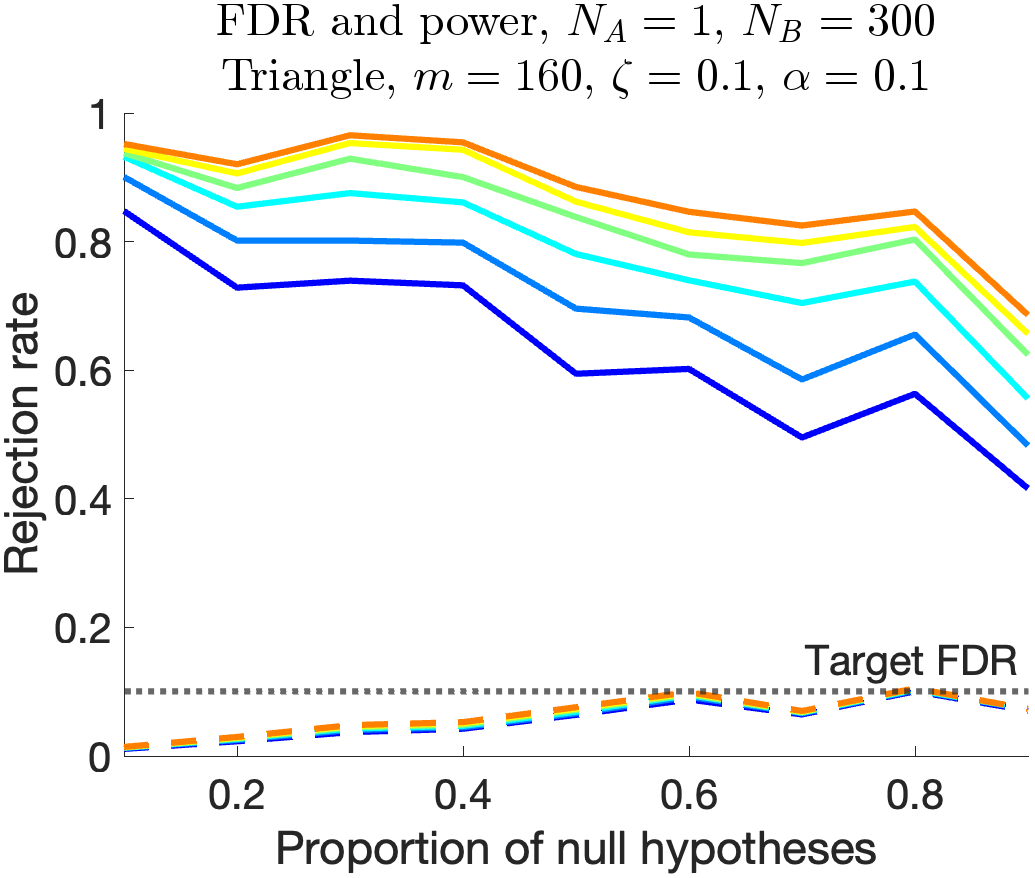}
        \includegraphics[width=0.27\linewidth]{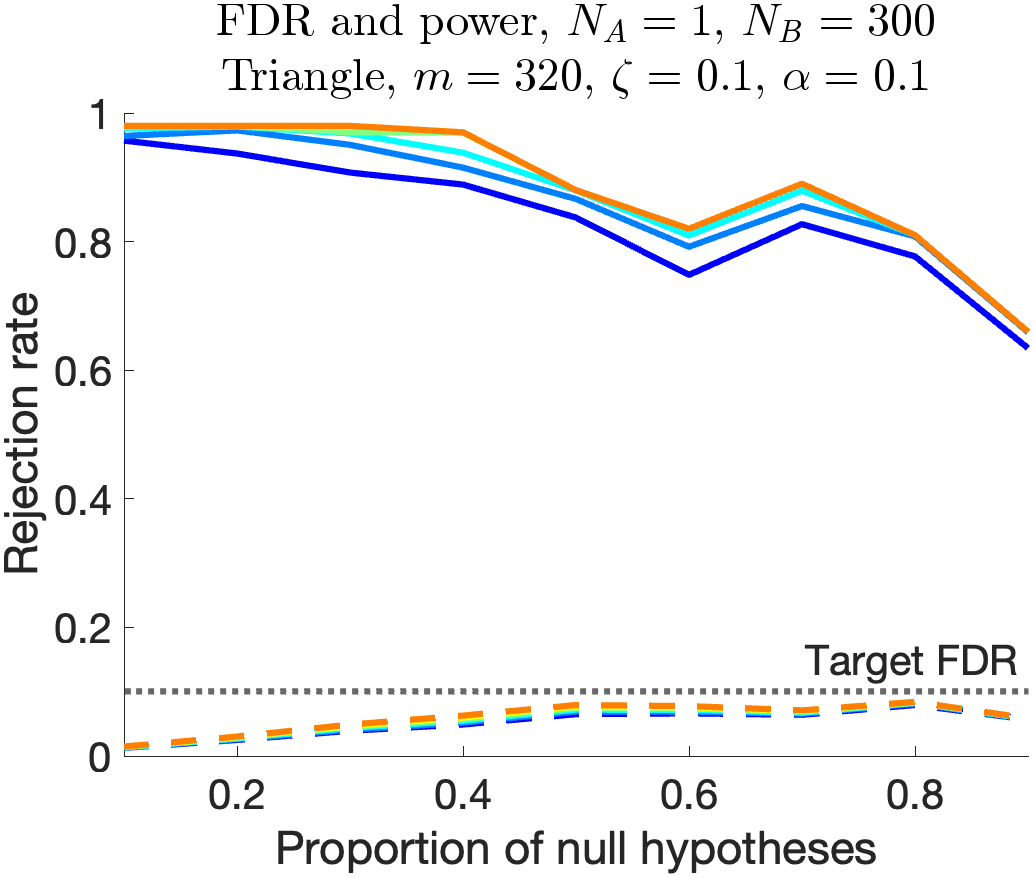}
    \end{adjustbox}
    \begin{adjustbox}{width=1.05\linewidth,center}
        \centering
        \includegraphics[width=0.27\linewidth]{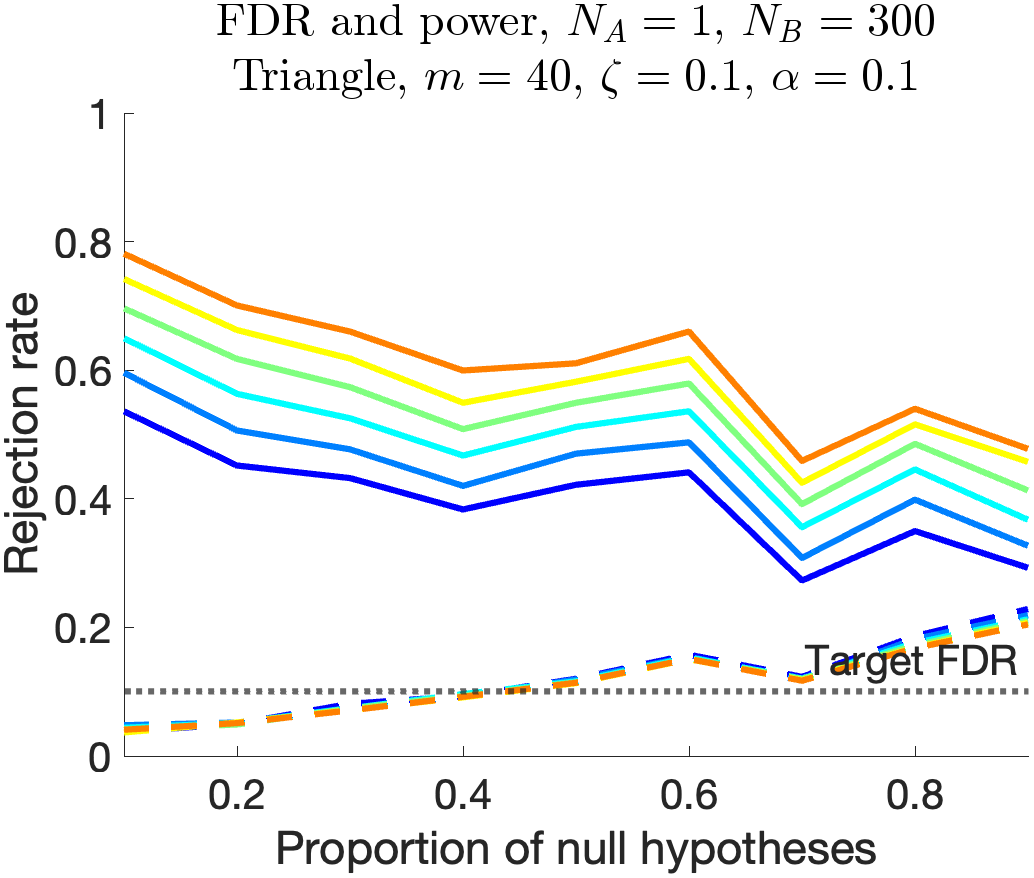}
        \includegraphics[width=0.27\linewidth]{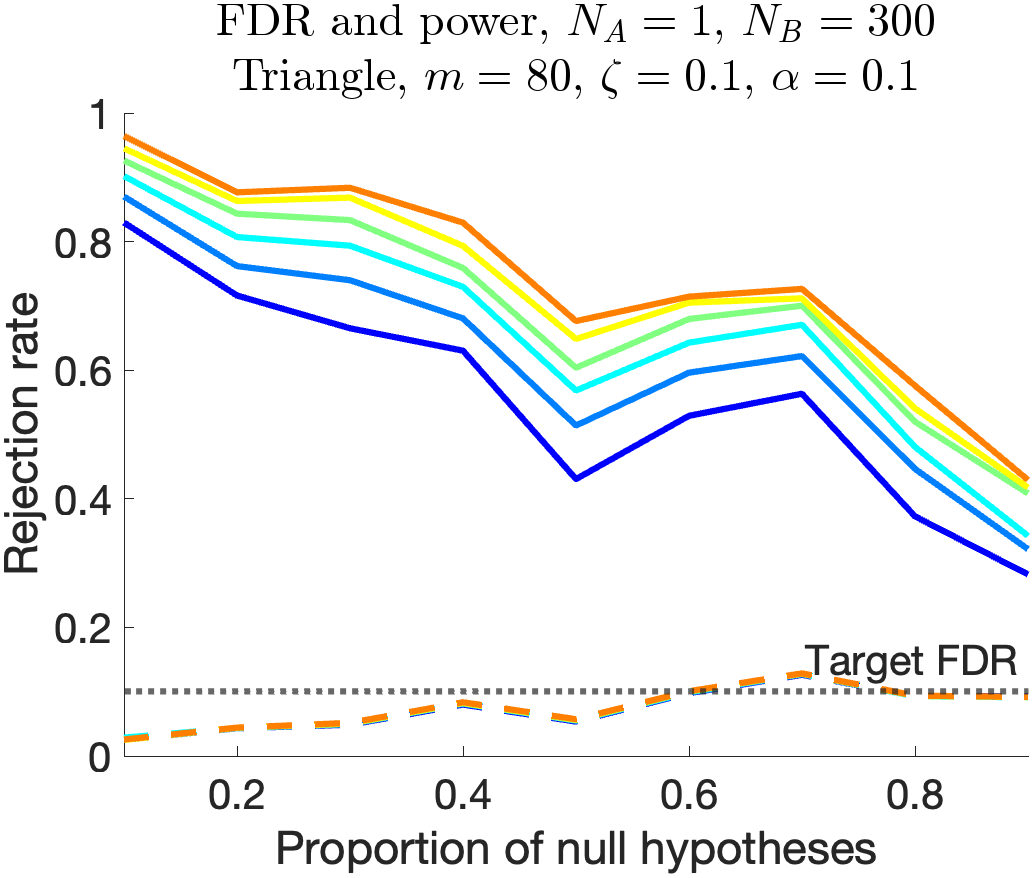}
        \includegraphics[width=0.27\linewidth]{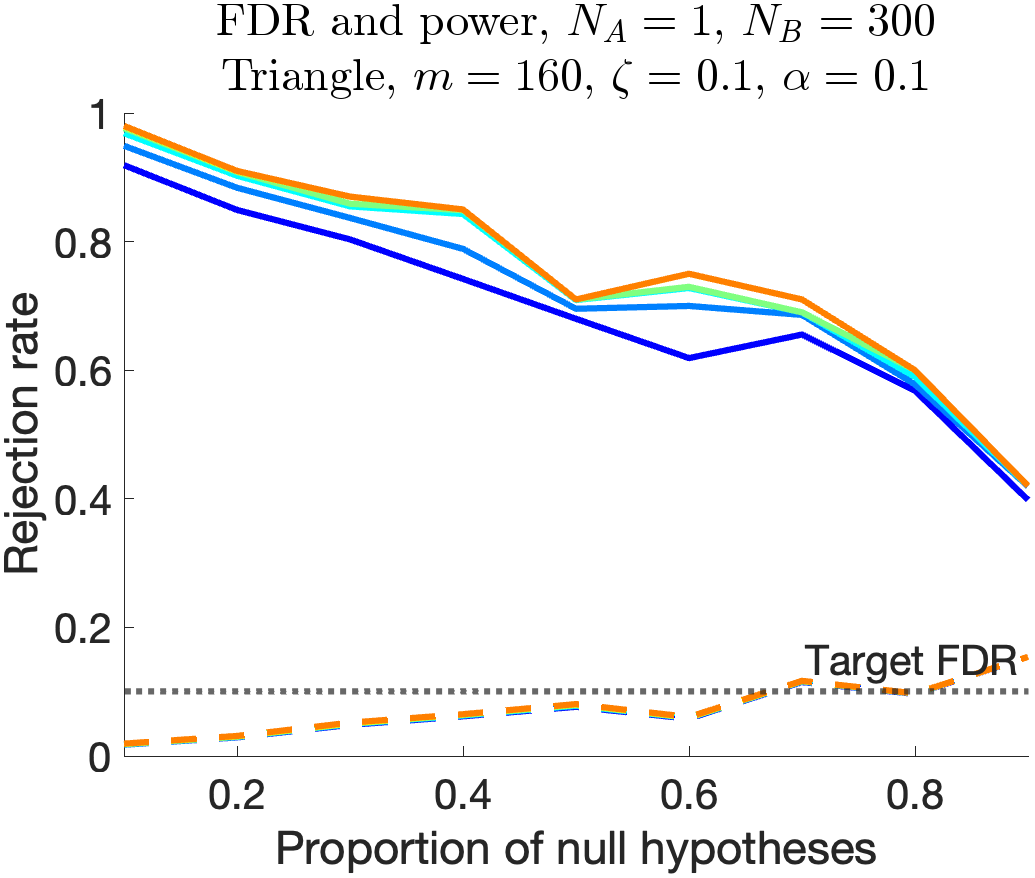}
        \includegraphics[width=0.27\linewidth]{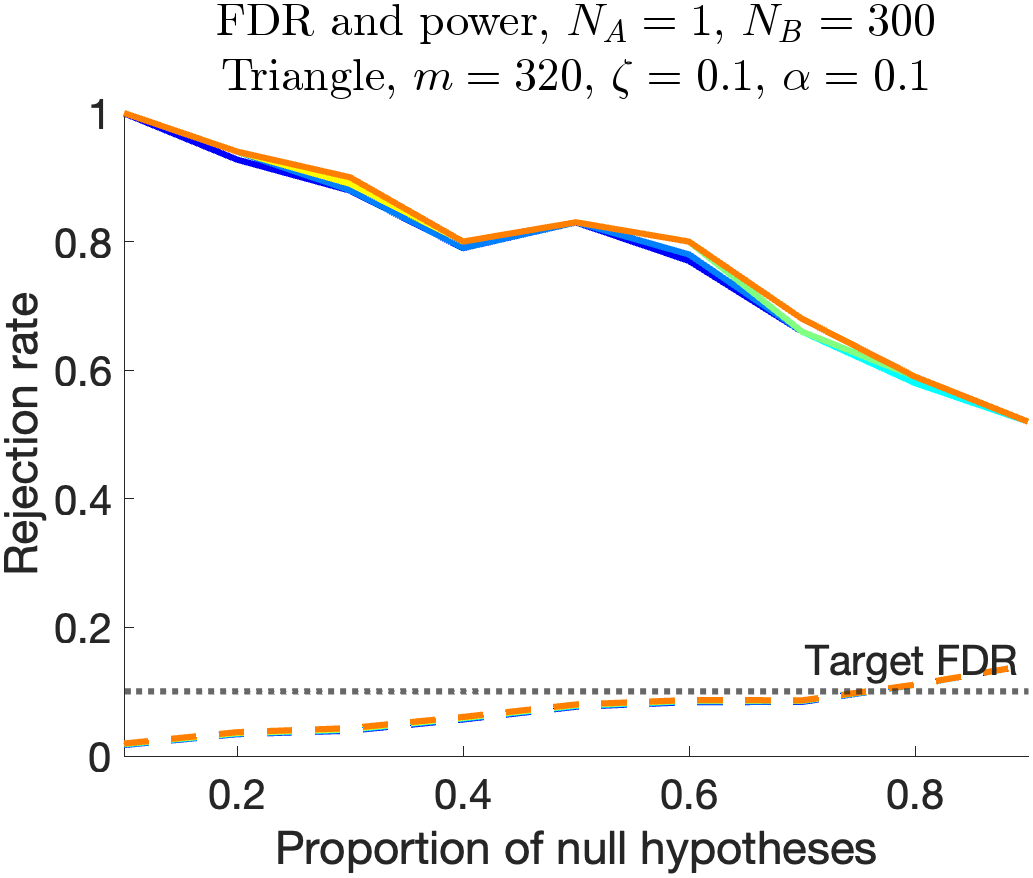}
    \end{adjustbox}
    \caption{Control of FDR (dashed curves) and test power (solid curves) under different $\mathfrak{q}$ ($H_0$ proportion) and gaps between hypotheses ($\varpi$, marked as ``shift'' in the plots).
    Row 1: model 1 (keyword, $m$ nodes) vs. model 2 ($n$ nodes); row 2: model 1 vs model 3; row 3: model 3 vs model 4.
    Columns 1--4 are increasing network sizes $m=n\in\{40,80,160,320\}$.
    }
    \label{new-fig::new-simu-3-FDR}
\end{figure}

Figure \ref{new-fig::new-simu-3-FDR} displays our results, indicating that Algorithm \ref{alg::FDP} effectively controls the FDR in almost all scenarios while maintaining strong performance in power. 
We observe that as $\mathfrak{q}$ nears 1, the empirical FDP converges towards the target FDR. 
This is because, in the formula for $\hat t$, the numerator conservatively assumes all test pairs could potentially be $H_0$.
The approach becomes less conservative with higher $\mathfrak{q}$. 
Conversely, a smaller $\mathfrak{q}$ inflates the numerator, leading to a smaller $\hat t$ and more conservative FDR control. 
Additionally, a well-documented trend in FDR literature \citep{genovese2002operating} is validated: power decreases as $\mathfrak{q}$ increases. As expected, power improves with larger $\varpi$ values and increased network sizes. 
These trends align well with our theoretical predictions.

\subsection{Pooling over repeated network observations}

Building on Section \ref{new-section::pooling-multiple-networks}, we simulate two sub-scenarios: 
(i) common node set, and (ii) independent node sets. 
For simplicity and a more principled comparison, we inherit the graphon setting from Section \ref{subsec::simu-1::type-I-error-and-power}, varying $N_A, N_B \in{1,2,5,10,20,40}$.
In scenario (i), we generate $W^{(A)}$ and $W^{(B)}$ only once, while in scenario (ii), they are independently generated for each network. 
We use the same performance metrics as in Section \ref{subsec::simu-1::type-I-error-and-power}.
Most benchmarks from other simulations unfortunately lack pooled versions, but we can compare our results with a normal approximation approach, which closely resembles our method but omits higher-order correction terms.

\begin{figure}[ht!]
    \centering
    \begin{adjustbox}{width=1.05\linewidth,center}
        \centering
        \includegraphics[width=0.29\linewidth]{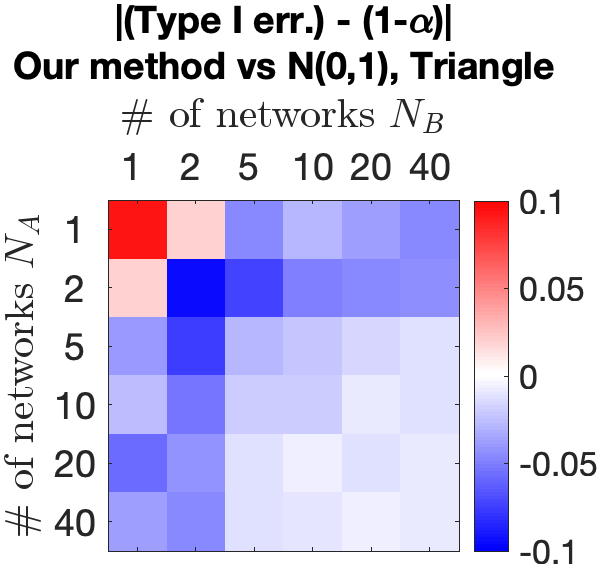}
        \includegraphics[width=0.29\linewidth]{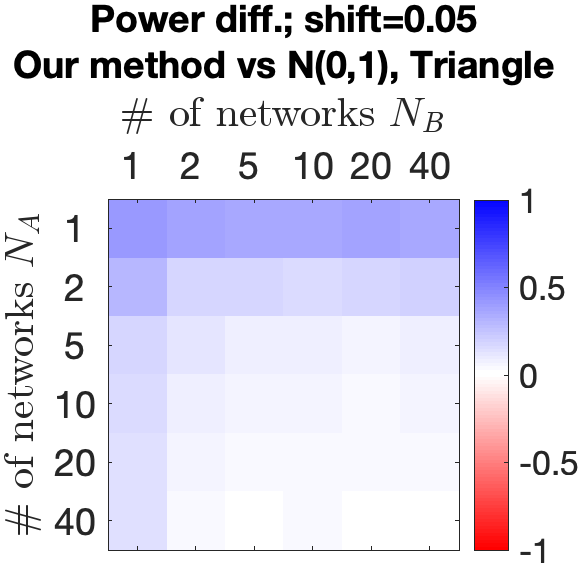}
        \includegraphics[width=0.29\linewidth]{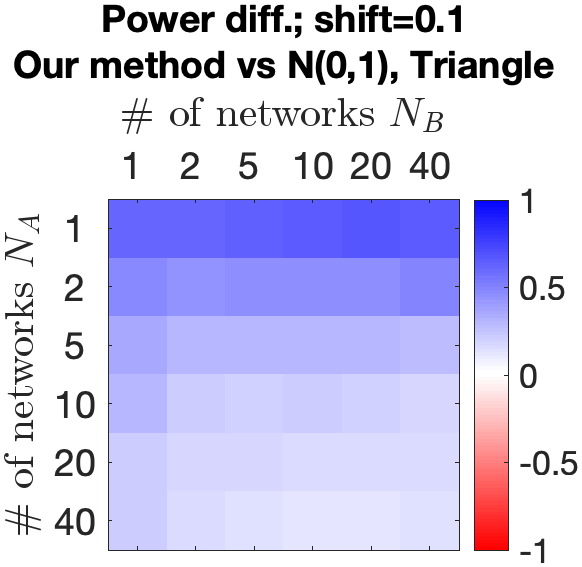}
        \includegraphics[width=0.29\linewidth]{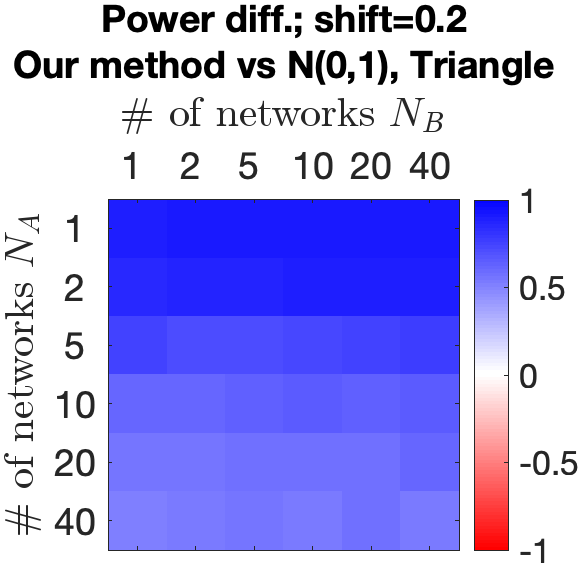}
    \end{adjustbox}
    \begin{adjustbox}{width=1.05\linewidth,center}
        \includegraphics[width=0.29\linewidth]{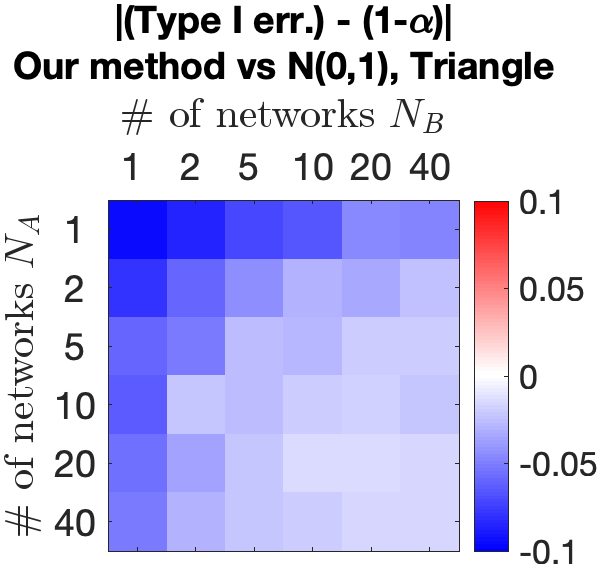}
        \includegraphics[width=0.29\linewidth]{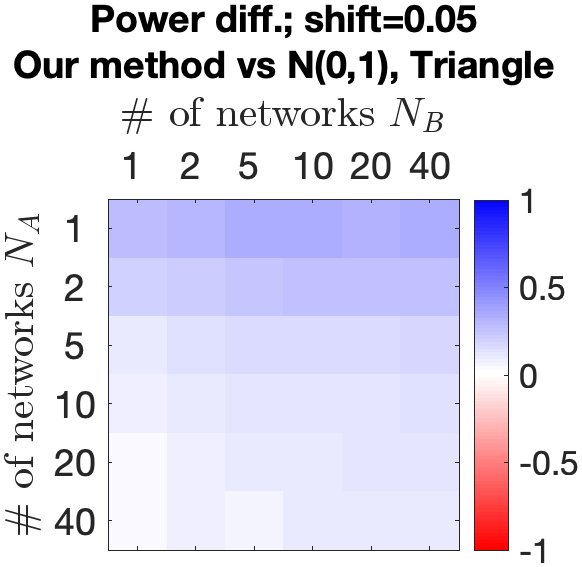}
        \includegraphics[width=0.29\linewidth]{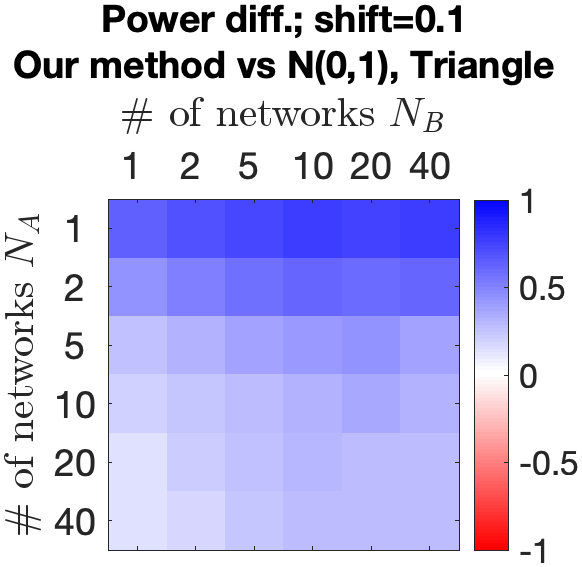}
        \includegraphics[width=0.29\linewidth]{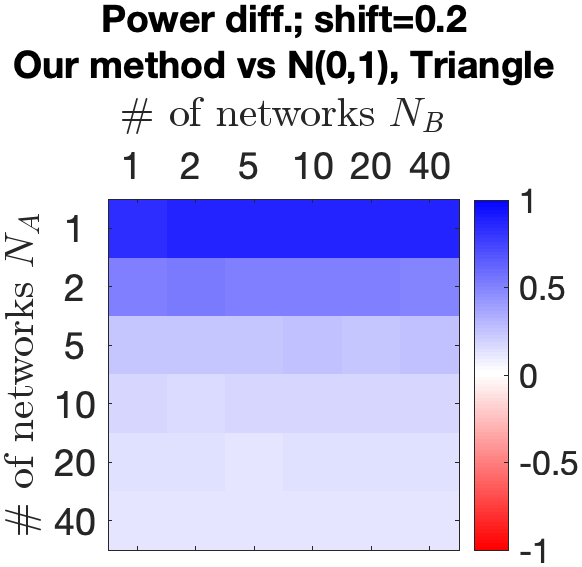}
    \end{adjustbox}
    \caption{Scenario 1: common node set.  Row 1: $m=n=20$; row 2: $m=n=40$.}
    \label{new-fig::simu-4::common-node-set}
\end{figure}
\begin{figure}[ht!]
    \centering
    \begin{adjustbox}{width=1.05\linewidth,center}
        \centering
        \includegraphics[width=0.29\linewidth]{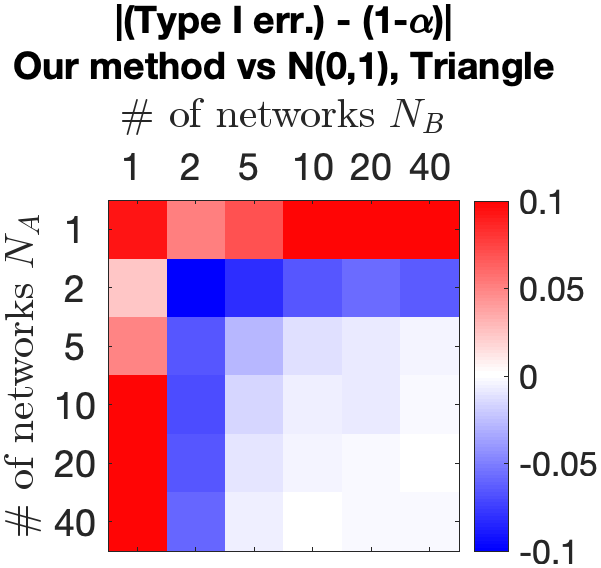}
        \includegraphics[width=0.29\linewidth]{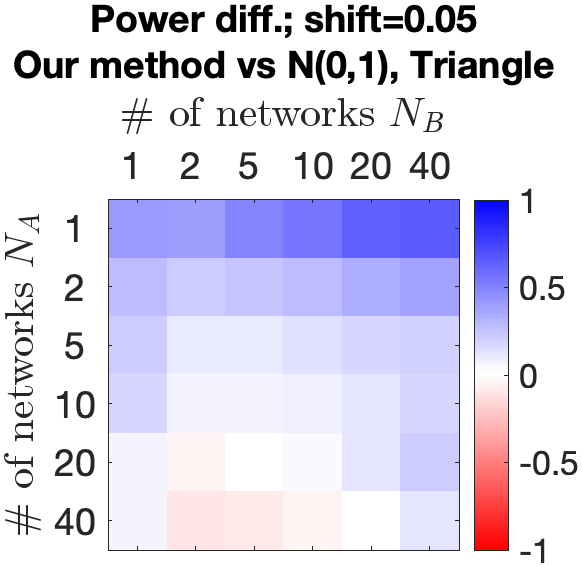}
        \includegraphics[width=0.29\linewidth]{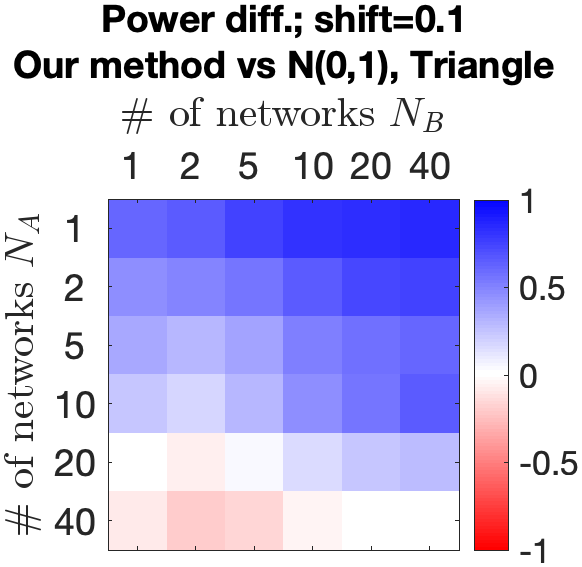}
        \includegraphics[width=0.29\linewidth]{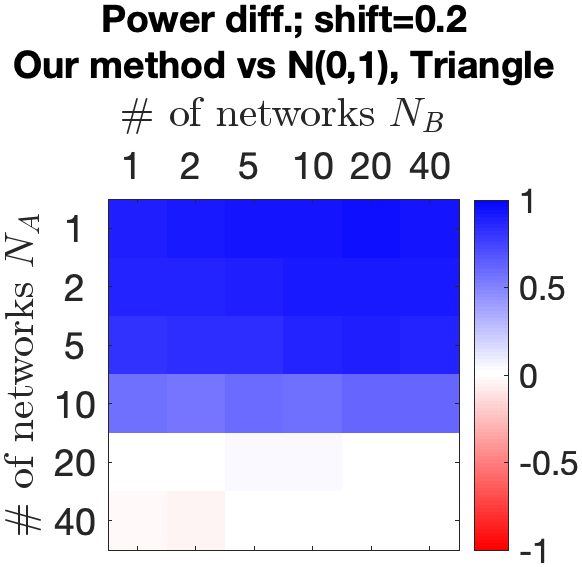}
    \end{adjustbox}
    \begin{adjustbox}{width=1.05\linewidth,center}
        \includegraphics[width=0.29\linewidth]{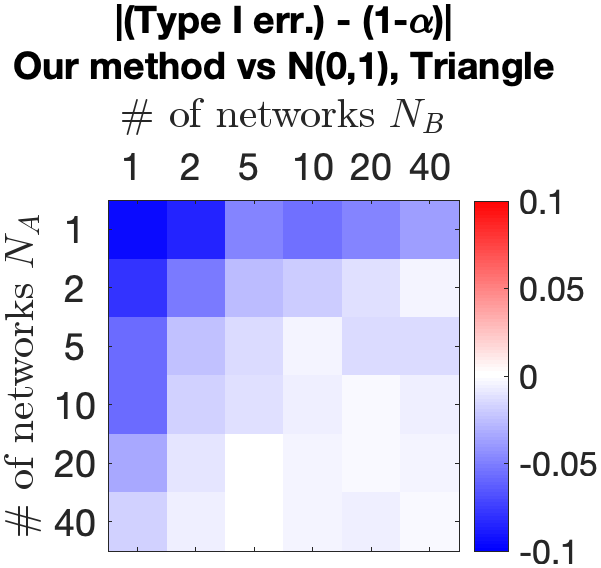}
        \includegraphics[width=0.29\linewidth]{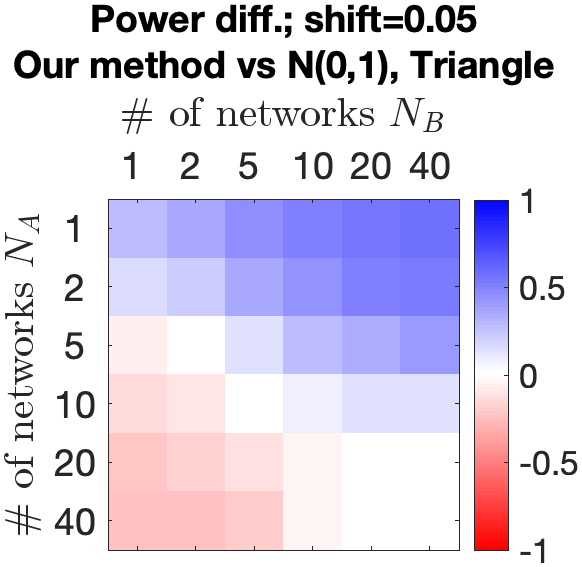}
        \includegraphics[width=0.29\linewidth]{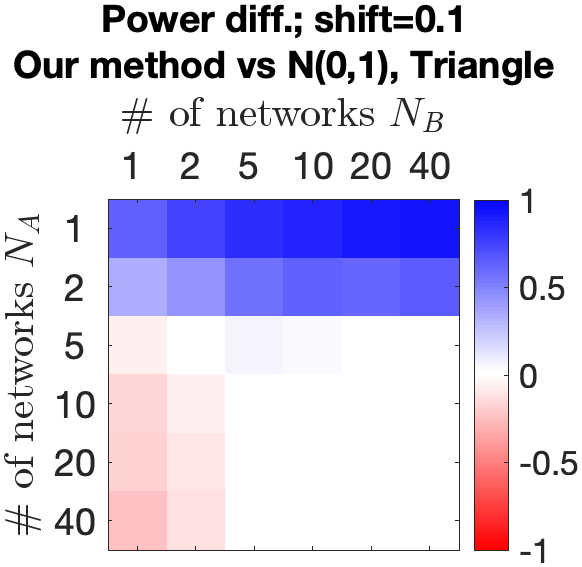}
        \includegraphics[width=0.29\linewidth]{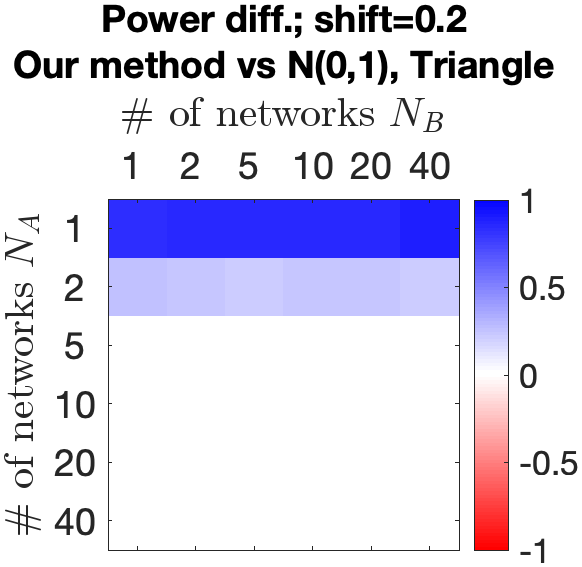}
    \end{adjustbox}
    \caption{Scenario 2: independent node sets.  Row 1: $m=n=20$; row 2: $m=n=40$.}
    \label{new-fig::simu-4::independent-node-sets}
\end{figure}

Figures \ref{new-fig::simu-4::common-node-set} and \ref{new-fig::simu-4::independent-node-sets} present the result. 
Across most scenarios, our approach shows a marked improvement in both accuracy and power. 
In Row 1 of both figures, where the network size is small ($m=n=20$), normal approximation outperforms our method in type-I error control when either group $A$ or $B$ has only one network. 
However, our method quickly gains a significant advantage when $N_A$ and $N_B$ increase to at least 2. 
This improvement can be attributed to the enhanced accuracy in estimating empirical Edgeworth expansion coefficients. 
This aligns with our theoretical understanding that accuracy is bottle-necked by the group with less information.

\subsection{Computational acceleration and handling indeterminate degeneracy}
\label{subsec::simulation-5::degeneracy}

Since our method is the first to automatically adapt to indeterminate degeneracy in network method-of-moments. 
this simulation demonstrates our method's validity and validate the predictions of our Theorem \ref{new-theorem::degeneracy::one-sample-universal-asymptotic-normality}. 
Due to page limit, we present selected results under: 
model 1, an SBM with two equal-sized communities and connection probabilities $[0.5, 0.2; 0.2, 0.5]$, leading to degeneracy; 
and model 2, as described in Section \ref{subsec::simu-4::pooling}, producing non-degenerate network moments.

\begin{figure}[ht!]
    \centering
    \begin{adjustbox}{width=1.2\linewidth,center}
        \includegraphics[width=0.29\linewidth]{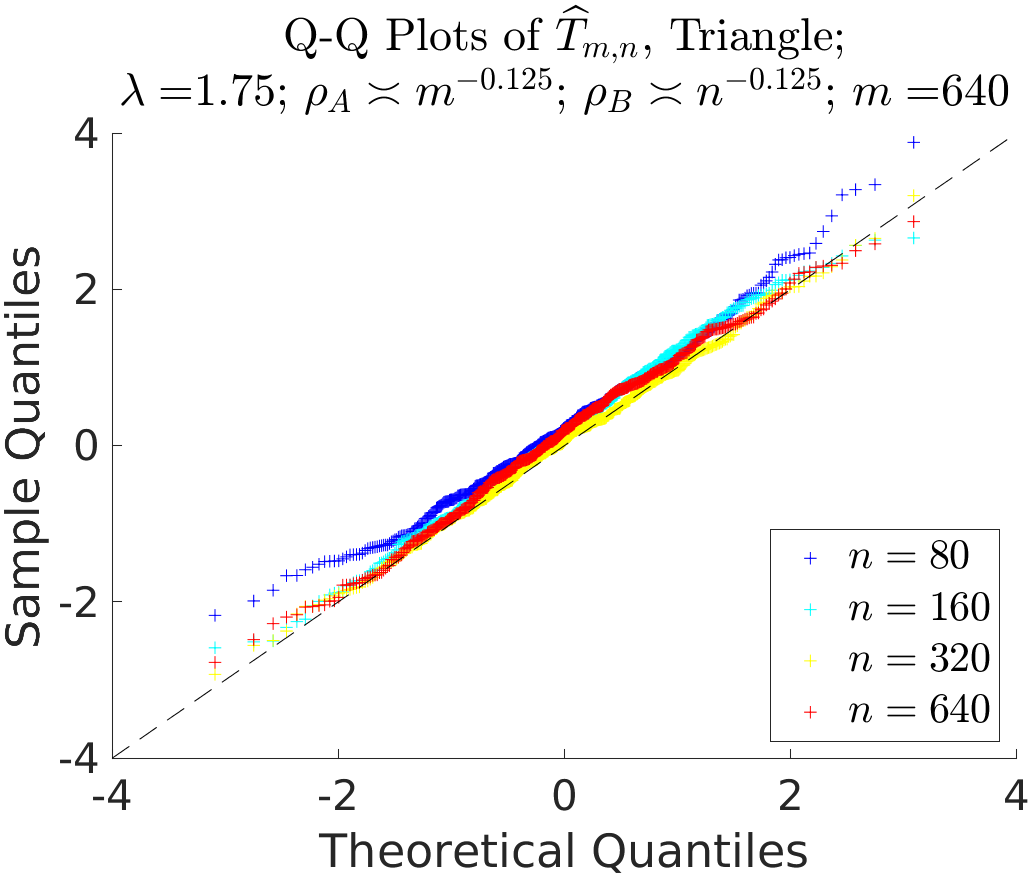}
        \includegraphics[width=0.29\linewidth]{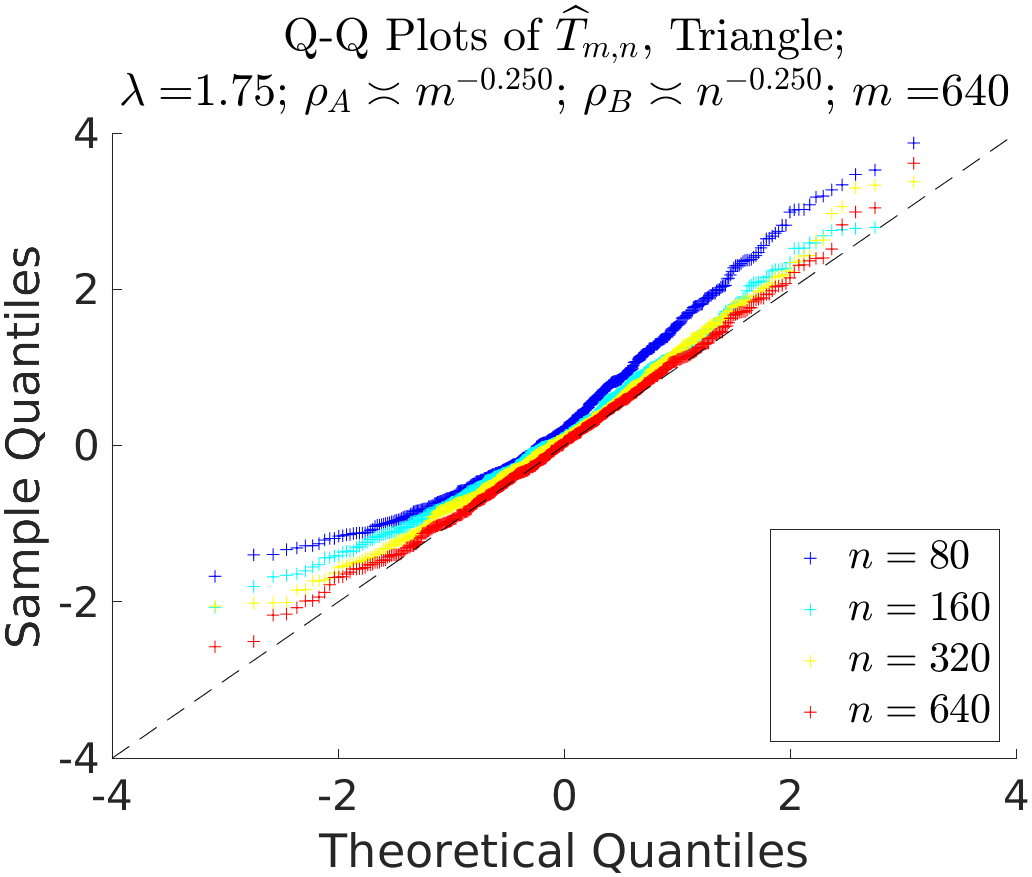}
        \includegraphics[width=0.29\linewidth]{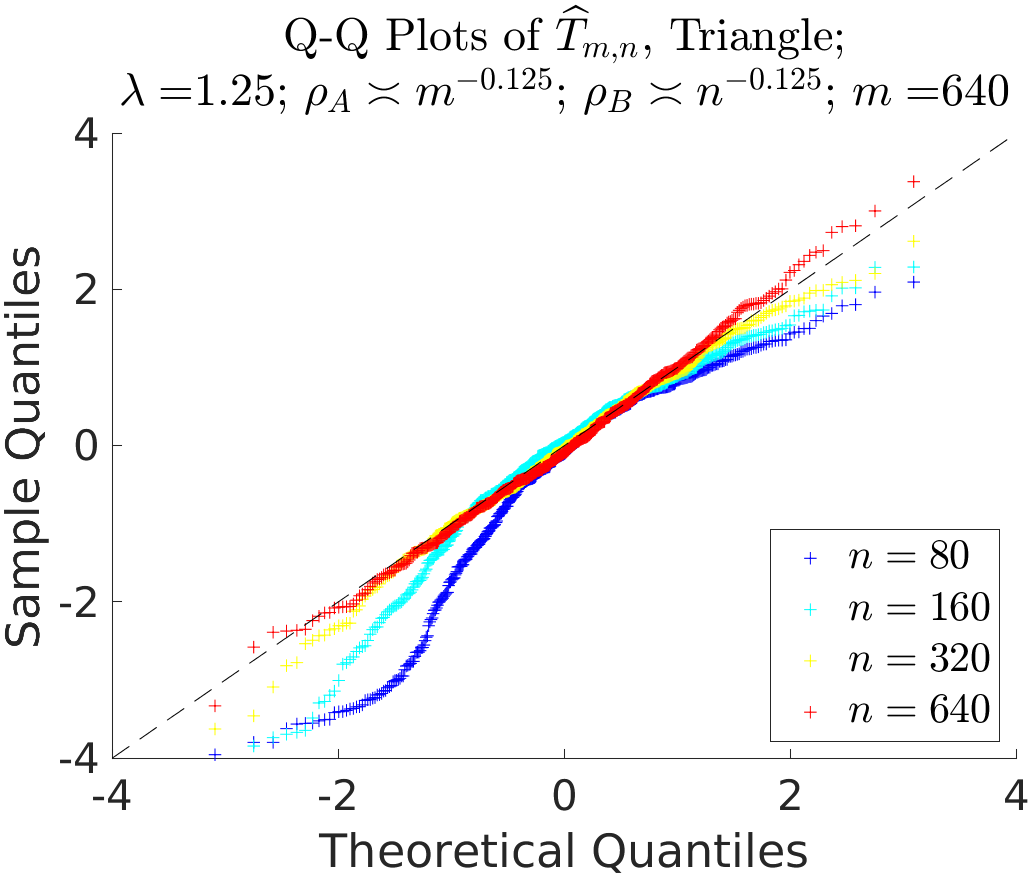}
        \includegraphics[width=0.29\linewidth]{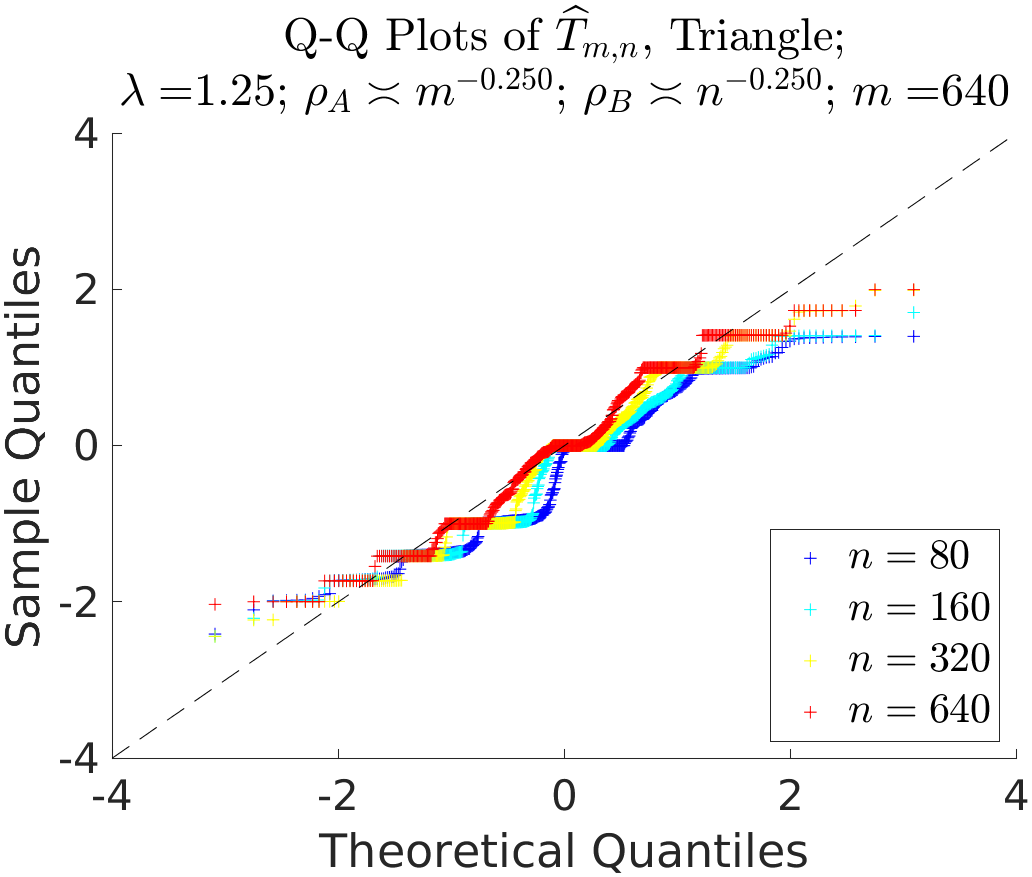}

    \end{adjustbox}
    \caption{Asymptotic normality of reduced network moments with adaptive variance estimation.  Motif is triangle.
    Plots 1 \& 2: model 1 vs model 2; plots 3 \& 4: model 1 vs model 1.
    }
    \label{new-fig::simu-5::degeneracy-triangle}
\end{figure}

Figure \ref{new-fig::simu-5::degeneracy-triangle} illustrates the result. 
We observe the anticipated asymptotic normality of the test statistic and the consistency of our adaptive variance estimator, regardless of degeneracy status.
Among the four scenarios tested, plot 4 adheres least to the assumptions of Theorem \ref{new-theorem::degeneracy::one-sample-universal-asymptotic-normality}. 
As anticipated, this leads to the slowest convergence of the studentized statistic towards $N(0,1)$.

}


\section{Data examples}

\subsection{Data example 1: Google+ ego-network data}
\label{subsec::data-example-1-google-plus-ego-networks}
In this example, we study the well-known SNAP-Gplus data set \citep{leskovec2012learning}.  
It consists of 132 ego-networks in severely varying sizes and densities.  The combined network consists of more than 100 thousand nodes and 1.36 million edges.  Existing works on this data set typically analyze the 132 ego-networks separately from each other \citep{leskovec2012learning, yang2013community}; while we are interested in exploring the structural similarity relationships between these ego-networks.  
We pre-processe the data by symmetrizing the ego-networks, eliminating 2 ego-networks with $\leq10$ nodes.  Our two-sample inference method then produces a similarity graph between the remaining 130 ego-nodes.  
Table \ref{tab::data::time-cost} row 1 reports the time costs of all methods.  It turns out that subsampling is the only benchmark that can finish running.  We use a negative-exponential transformation $\zeta(u):= \exp(-|u|)$ to convert Cornish-Fisher CI midpoints to network similarity measures. {\revisionfiltered Triangle and the V-shape are the motifs considered in this example.}

\begin{figure}[h!]
    \centering
    \makebox[\textwidth][c]{
    \includegraphics[width=0.32\textwidth]{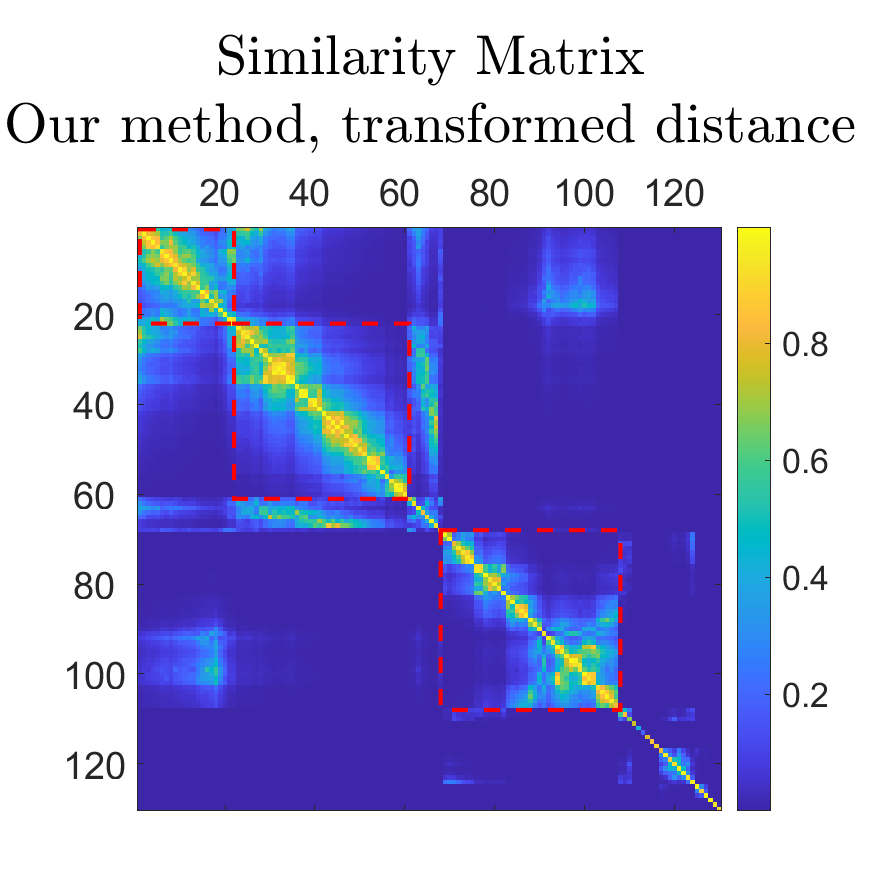}
    \includegraphics[width=0.32\textwidth]{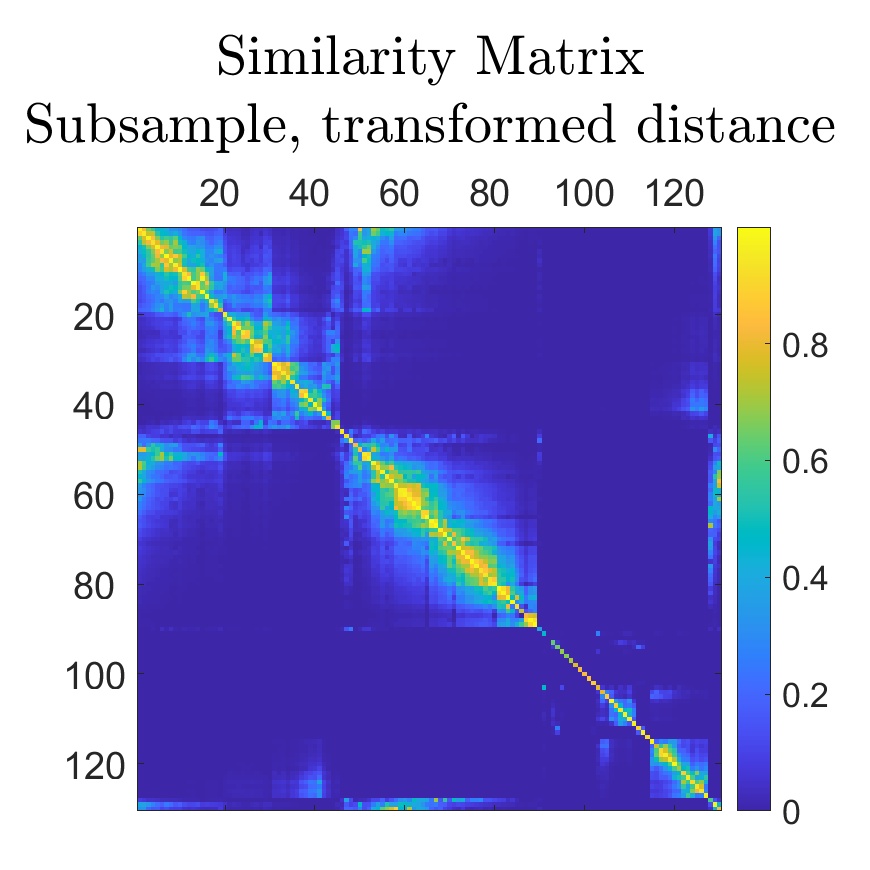}
    \includegraphics[width=0.37\textwidth]{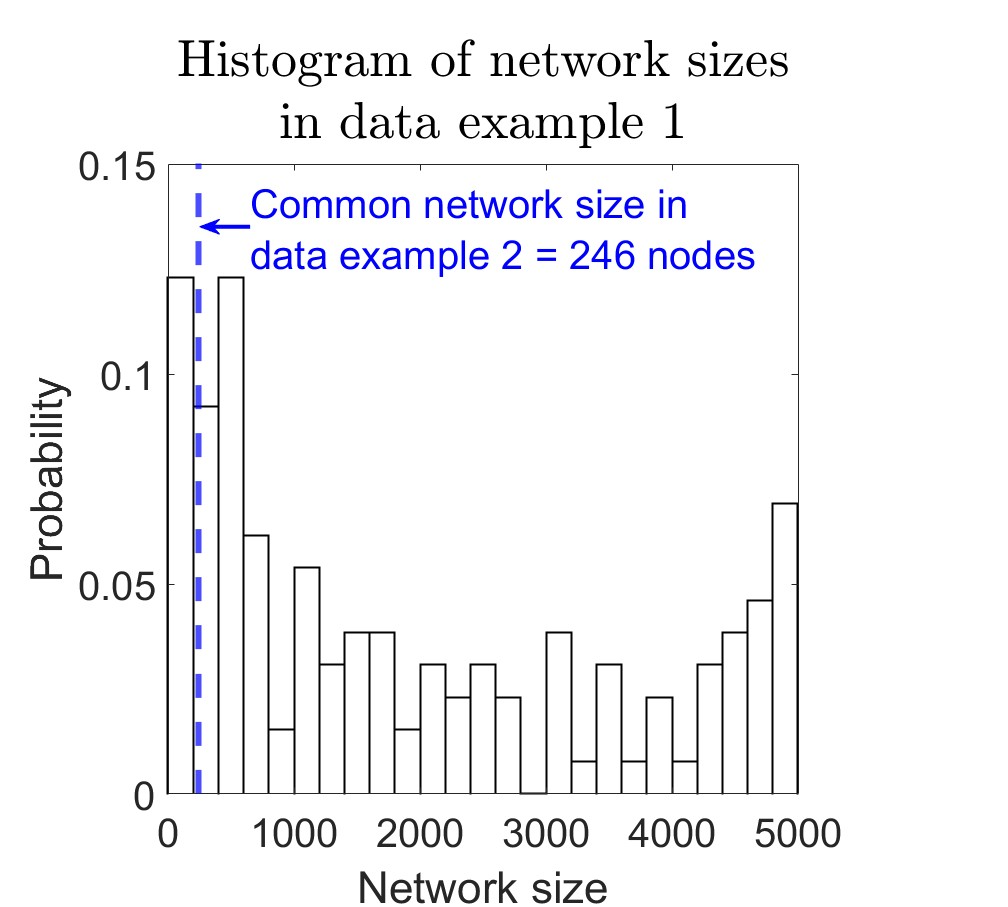}
    }

    \caption{Data example 1: Google+ ego-networks.  
    Similarity graphs by our method on the left, and subsampling in the middle.  In both plots, nodes were reordered using graphon slice similarity \citep{zhang2017estimating}.
    The right panel illustrates the distribution of ego-network sizes.
    }
    \label{fig::data-1}
\end{figure}

Figure \ref{fig::data-1} reports the result.  In the left panel, our method identifies 3 loosely clustered subgroups among ego-networks with further internal structures.  
To further improve presentation, we post-process the $130\times 130$ similarity graph by reordering nodes based on the estimated graphon slice similarity measure \citep{zhang2017estimating}.  We find this sorting method to be more effective when the similarity graph displays within-group heterogeneity.
Compared to our method, the transformed distance estimated by node subsampling seem to be systemtically lower than our method.  To intuitively understand why network subsampling may inflate type I error in finite-sample examples, consider two moderately large networks generated from the same model, but the model has much heterogeneity.  Consequently, there is a good chance that subsampling may sample different parts of the two networks and incorrectly reject the null hypothesis.

\begin{table}[h!]
    \centering
    \renewcommand{\arraystretch}{0.6}
    \caption{Time cost comparison table.  Unit is second.  Timeout is 12 hours = 43200 seconds.}
    \label{tab::data::time-cost}
    \begin{tabular}{c|ccccc}\hline
     &  Our method (hash) & Our method (test) & Subsample & Resample \\
     \hline
    Data example 1
        & $116.39$ & $18.81$ & $10884.62$ & (Timeout)\\
    Data example 2
        & $3.60$ & $64.36$ & $2488.21$& (Timeout)\\
        \hline
     &  NonparGT & NetLSD & NetComp \\
     \hline
    Data example 1  & (Numerical error) & (Timeout) & (Timeout)\\
    Data example 2  & $4327.09$ & (Numerical error) & $4304.51$\\
    \hline
    \end{tabular}
\end{table}

\subsection{Data example 2: Brain connectome data for schizophrenia research }
\label{subsec::data-example-2-schizophrenia}
This data set, collected by \cite{adhikari2019functional}, consists of functional brain connectivity networks among 103 patients with schizophrenia and 124 healthy people as normal controls.  
We follow the protocol by \cite{adhikari2019functional} to  pre-process the resting-state fMRI data.  All individuals' networks share a common set of 246 nodes that represent different regions of interests.  Each edge is a Fisher Z-transformed correlation between the blood-oxygen-level-dependent signals at the two terminal nodes.
We test all methods for pairwise comparison and use the technique similar to data example 1 to reorder nodes in the obtained similarity graph to facilitate result interpretation.

\begin{figure}[h!]
    \centering
    
    \makebox[\textwidth][c]{\includegraphics[width=0.3\textwidth]{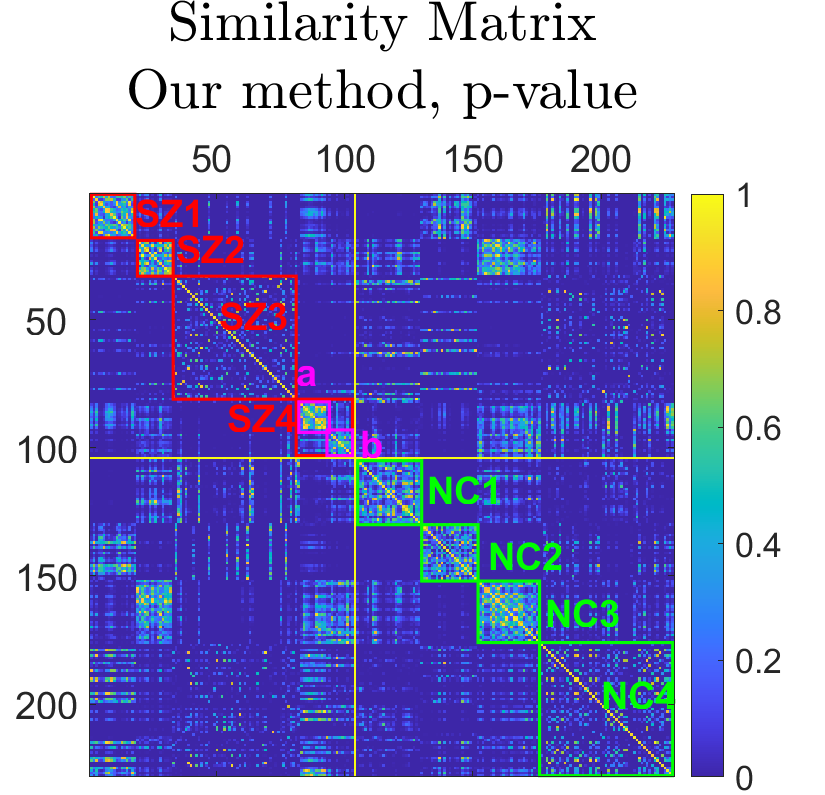}    \includegraphics[width=0.3\textwidth]{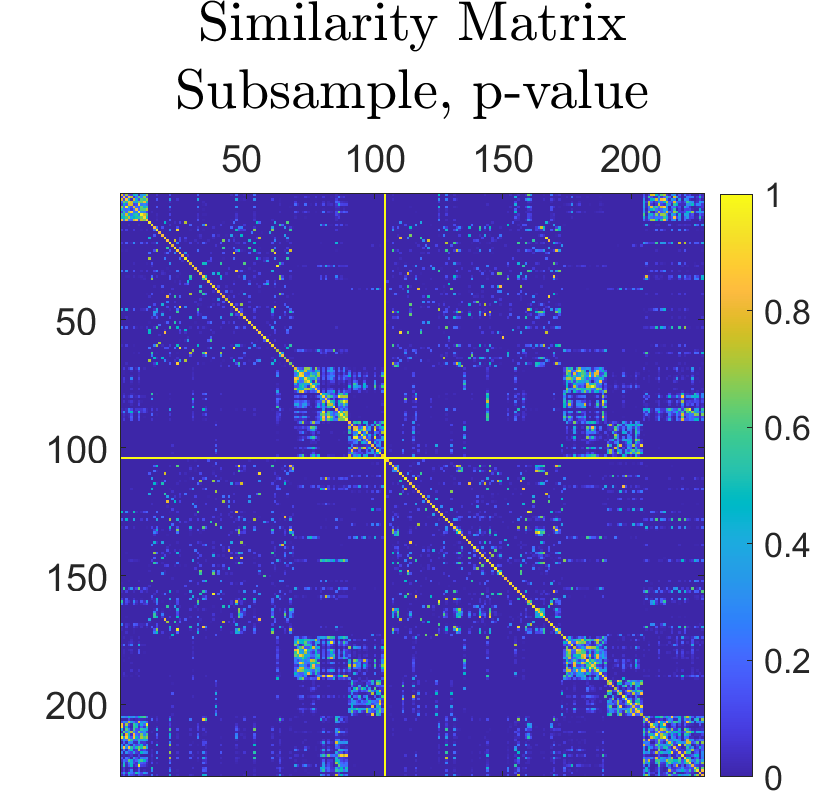}
    \includegraphics[width=0.3\textwidth]{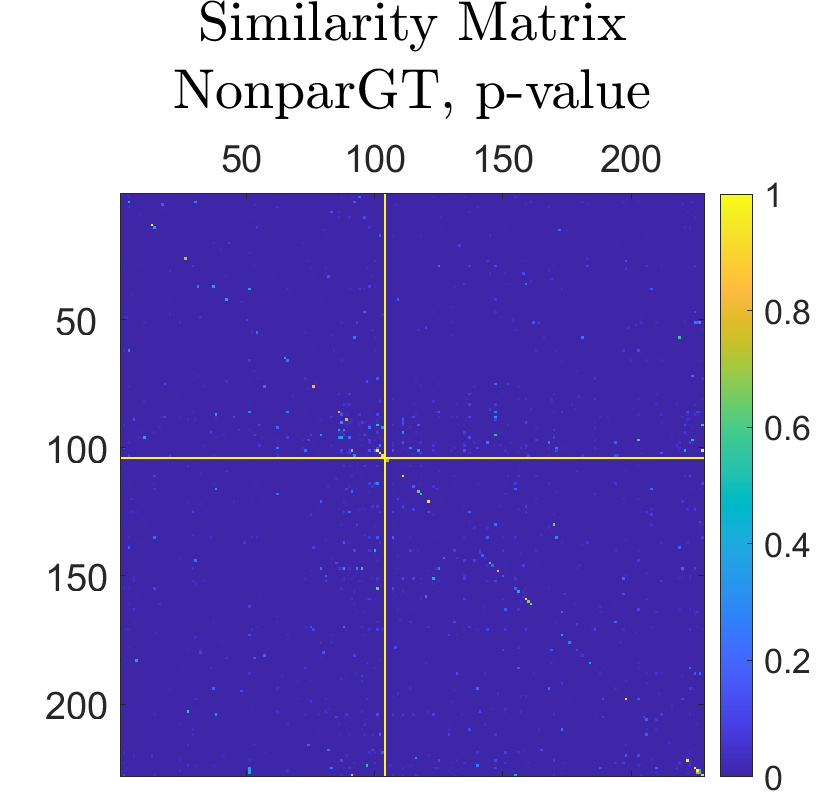}
    \includegraphics[width=0.3\textwidth]{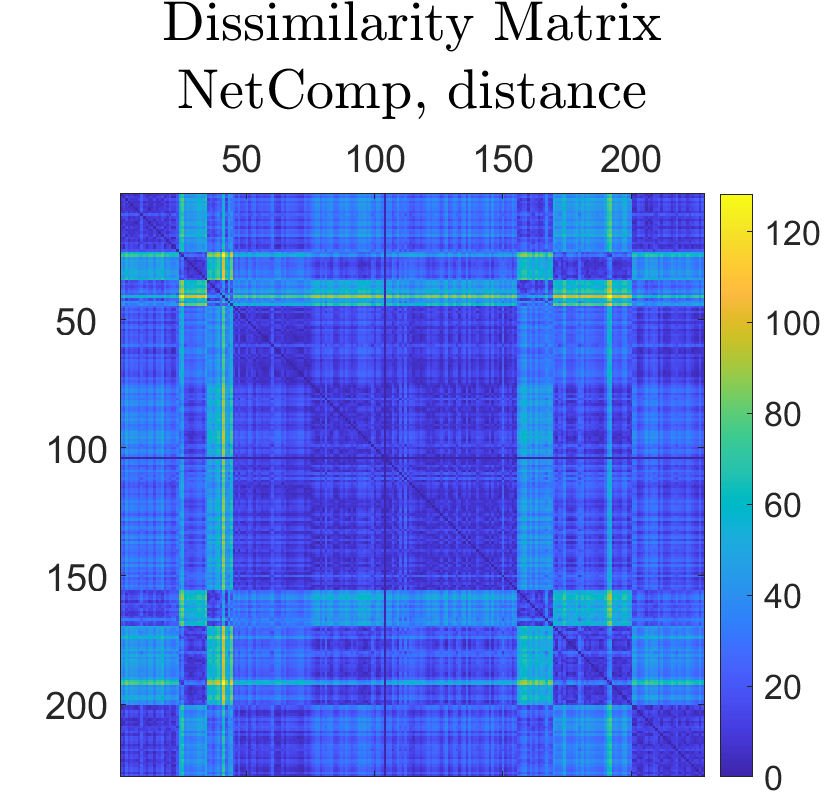}
    }\\
    \includegraphics[width = \textwidth]{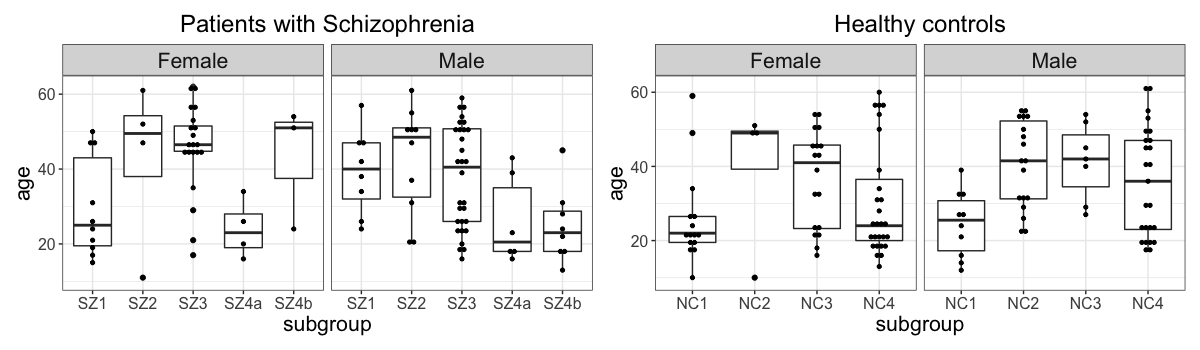}

    \caption{Data example 2: schizophrenia. 
    Row 1: similarity graph between brain images, constructed by our method using p-values.  Nodes 1--103 are patients and nodes 104--227 are healthy people.  We added artificial yellow lines to separate the two groups.  
    Row 2: age distributions in each subgroup identified by our method for female and male participants.
    }
    \label{fig::data-2}
\end{figure}

Figure \ref{fig::data-2} shows the result.
Our method identified several subgroups that are potentially sub-types of the disease worth further investigation.  
The result significant enriches over many network two-sample test methods in existing literature \citep{yuan2021practical,bravo2021principled,chen2020spectral}, which only produce a binary decision of whether the two groups of networks are structurally identical but could not discover within-group structures like our method provides.
Row 2 of Figure \ref{fig::data-2} suggests that among females, subgroups SZ2, SZ3 and NC2 mostly consist of mid-age to seniors, whereas NC1 is particularly young; among males, SZ4b, NC1 and NC3 exhibit different levels of concentration around their own particular age groups.
Moreover, the similarity graph produced by our method finds structural similarities between some patient-normal subgroup pairs, such as (SZ1, NC2) and (SZ2, NC3).  Therefore, it might be of interest for biomedical researchers to further compare these subgroups pairs and look for disease-linked differences.
Similar to data example 1, here, subsampling also identifies much less similarities between individual pairs.  
NonparGT identifies even less pairwise similarity, understandably since its null hypothesis requires the two network models to be completely identical.
NetComp does not produce an empirical p-value but an estimated distance between each network pair.  Its output a similarity graph with patterns different from ours and of independent interest.
Row 2 of Table \ref{tab::data::time-cost} records running time.  Our method again shows significant speed advantage by only computing the hashing once for each network, and all pair-wise comparisons can be done in $O(1)$ time.
To understand why our method spends much less time in hashing but more time in pairwise comparisons in data example 2 than that in data example 1, we refer to the network size histogram in Figure \ref{fig::data-1}. Most network sizes in data example 1 are in thousands, much larger than the common network size of 246 nodes in data example 2; while data examples 2 contains 75\% more networks than data example 1, and $1.75^2\approx 3$, which explains the inflation of pairwise comparison time.

\section{Future work}

Although this paper comprehensively addresses several crucial aspects of network two-sample testing, there remain many interesting directions for future research. 
To illustrate, we propose two potential extensions. 
The first is to extend our analysis to the joint distribution of various network moments. 
This topic, initially explored for one-sample inference under the Erdos-Renyi model by \cite{gao2017testinga, gao2017testingb}, warrants further investigation, especially on how to address the two-sample test problem under general models. 
Secondly, while our current work controls the FDR for the network query problem, it remains an intriguing yet challenging problem to devise a valid FDR control procedure for pairwise comparison among many networks.

\bigskip
\begin{center}
{\large\bf SUPPLEMENTARY MATERIAL}
\end{center}
The Supplementary Material contains the following contents:
1. definitions of the $\alpha$- and $\beta$-terms in Section \ref{subsec::characterizing the distribution::alpha} and their estimators
2. all proofs;
and
3. additional simulation results.

\clearpage

\if0\blind
{
  \title{\bf Supplementary Material to ``Higher-order accurate two-sample network inference and network hashing"}
  \author{Meijia Shao\vspace{-1ex}\\
    Department of Statistics, The Ohio State University\\\vspace{1ex}
    Dong Xia \\
    Department of Mathematics, Hong Kong University of Science and Technology\\\vspace{1ex}
    Yuan Zhang\\
    Department of Statistics, The Ohio State University\\\vspace{1ex}
    Qiong Wu\\
    Department of Biostatistics, Epidemiology and Informatics, University of Pennsylvania\\\vspace{1ex}
    Shuo Chen\\
    Division of Biostatistics and Bioinformatics, Department of Epidemiology and Public Health, School of Medicine, University of Maryland}
  \maketitle
} \fi

\if1\blind
{
  \bigskip
  \bigskip
  \bigskip
  \begin{center}
    {\LARGE\bf Supplementary Material to ``Higher-order accurate two-sample network inference and network hashing"}
\end{center}
  \medskip
} \fi

\bigskip

\spacingset{1.9} 

\section{Definitions and estimations of $\alpha$'s and $\beta$'s in Section \ref{subsec::characterizing the distribution::alpha} of Main Paper}
\label{calculation alpha}

The Edgeworth expansion terms ${\cal I}_0, Q_{m,n,\rho_A,\rho_B;1}$ and $Q_{m,n,\rho_A,\rho_B;2}$ depend on the expectation of quantities involving $\alpha_0,\cdots,\alpha_4$ and $\beta_0,\cdots,\beta_4$, whose definitions are enumerated as follows.  For completeness, we still write down the formulation of $\alpha_1$, which has been defined in Main Paper.
\begin{align}
    \alpha_0
    &
    :=
    2s(s+1)\rho_A^{-(s+2)}\mu_m\xi_{\rho_A;1}^2{\dx-2rs\rho_A^{-(s+1)}\xi_{A,\rho_A;1}}\label{eq:def_alpha0}
    \\
    \alpha_1&(X_i):=\rho_A^{-s}\cdot \left\{rg_{A;1}(X_i) -2s\mu_m\cdot\rho_A^{-1}\cdot g_{\rho_A;1}(X_i)\right\} \label{eq:def_alpha1}\\
    \alpha_2&(X_{i_1},X_{i_2})
    := 
    {\dx\frac{r(r-1)}{2}}\rho_A^{-s}g_{A;2}(X_{i_1},X_{i_2})
    -{\dx s}\rho_A^{-(s+1)}\mu_mg_{\rho_A;2}(X_{i_1},X_{i_2})\notag\\
    &
    +2{\dx \rho_A^{-s}\mu_m}s(s+1)\rho_A^{-2}g_{\rho_A;1}(X_{i_1})g_{\rho_A;1}(X_{i_2})
    -{\dx 2}rs\rho_A^{-(s+1)}g_{\rho_A;1}(X_{i_1})g_{A;1}(X_{i_2}) \label{eq:def_alpha2}
    \\
    \alpha_3&(X_i)
    :=
    {\dx -4}r^2s\rho_A^{{\dx-(2s+1)}}\xi_{A;1}^2g_{\rho_A;1}(X_i)
    +r^2\rho_A^{-2s}(g_{A;1}^2(X_i)-\xi_{A;1}^2)\notag\\
    &
    -16s^2(s+1)\rho_A^{-(2s+3)}\mu_m^2\xi_{\rho_A;1}^2g_{\rho_A;1}(X_i)
    +8rs^2\rho_A^{-(2s+2)}\mu_m g_{A;1}(X_i)\xi_{\rho_A;1}^2\notag\\
    &
    +4s^2\rho_A^{-(2s+2)}\mu_m^{\dx 2} (g_{\rho_A;1}^2(X_i)-\xi_{\rho_A;1}^2)
    \notag\\
    &
    -4rs\Big\{
    -(4s+2)\rho_A^{-(2s+2)}g_{\rho_A;1}(X_i)\mu_m\xi_{A,\rho_A;1}
    +r\rho_A^{-(2s+1)}g_{A;1}(X_i)\xi_{A,\rho_A;1}
    \notag\\
    &
    +\rho_A^{-(2s+1)}\mu_m\big( g_{A;1}(X_i)g_{\rho_A;1}(X_i) - \xi_{A,\rho_A;1} \big)
    \Big\} \label{eq:def_alpha3}
    \\
    \alpha_4&(X_{i_1}, X_{i_2})
    :=
    2r^2(r-1)\rho_A^{-2s}g_{A;1}(X_{i_1})g_{A;2}(X_{i_1},X_{i_2})
    +8s^2\rho_A^{-(2s+2)}\mu_m^2
    \notag\\
    &g_{\rho_A;1}(X_{i_1})g_{\rho_A;2}(X_{i_1},X_{i_2})
    -4r(r-1)s\rho_A^{-(2s+1)}\mu_m g_{\rho_A;1}(X_{i_1})g_{A;2}(X_{i_1},X_{i_2})
    \notag\\
    &-4rs\rho_A^{-(2s+1)}\mu_m g_{A;1}(X_{i_1})g_{\rho_A;2}(X_{i_1},X_{i_2}). \label{eq:def_alpha4}
\end{align}
Similarly, define
\begin{align}
    \beta_0
    &
    :=
    2s(s+1)\rho_B^{-(s+2)}\nu_n\xi_{\rho_B;1}^2-2rs\rho_B^{-(s+1)}\xi_{B,\rho_B;1}\label{eq:def_beta0}
    \\
    \beta_1&(Y_i):=\rho_B^{-s}\cdot \left\{rg_{B;1}(Y_i) -2s\nu_n\cdot\rho_B^{-1}\cdot g_{\rho_B;1}(Y_i)\right\} \label{eq:def_beta1}\\
    \beta_2&(Y_{i_1},Y_{i_2})
    := 
    {\dx\frac{r(r-1)}{2}}\rho_B^{-s}g_{B;2}(Y_{i_1},Y_{i_2})
    -{\dx s}\rho_B^{-(s+1)}\nu_n g_{\rho_B;2}(Y_{i_1},Y_{i_2})\notag\\
    &
    +2{\dx \rho_B^{-s}\nu_n}s(s+1)\rho_B^{-2}g_{\rho_B;1}(Y_{i_1})g_{\rho_B;1}(Y_{i_2})
    -{\dx 2}rs\rho_B^{-(s+1)}g_{\rho_B;1}(Y_{i_1})g_{B;1}(Y_{i_2}) \label{eq:def_beta2}
    \\
    \beta_3&(Y_i)
    :=
    {\dx -4}r^2s\rho_B^{{\dx-(2s+1)}}\xi_{B;1}^2g_{\rho_B;1}(Y_i)
    +r^2\rho_B^{-2s}(g_{B;1}^2(Y_i)-\xi_{B;1}^2)\notag\\
    &
    -16s^2(s+1)\rho_B^{-(2s+3)}\nu_n^2\xi_{\rho_B;1}^2g_{\rho_B;1}(Y_i)
    +8rs^2\rho_B^{-(2s+2)}\nu_n g_{B;1}(Y_i)\xi_{\rho_B;1}^2\notag\\
    &
    +4s^2\rho_B^{-(2s+2)}\nu_n^{\dx 2} (g_{\rho_B;1}^2(Y_i)-\xi_{\rho_B;1}^2)
    \notag\\
    &
    -4rs\Big\{
    -(4s+2)\rho_B^{-(2s+2)}g_{\rho_B;1}(Y_i)\nu_n\xi_{B,\rho_B;1}
    +r\rho_B^{-(2s+1)}g_{B;1}(Y_i)\xi_{B,\rho_B;1}
    \notag\\
    &
    +\rho_B^{-(2s+1)}\nu_n\big( g_{B;1}(Y_i)g_{\rho_B;1}(Y_i) - \xi_{B,\rho_B;1} \big)
    \Big\} \label{eq:def_beta3}
    \\
    \beta_4&(Y_{i_1}, Y_{i_2})
    :=
    2r^2(r-1)\rho_B^{-2s}g_{B;1}(Y_{i_1})g_{B;2}(Y_{i_1},Y_{i_2})
    +8s^2\rho_B^{-(2s+2)}\nu_n^2
    \notag\\
    &g_{\rho_B;1}(Y_{i_1})g_{\rho_B;2}(Y_{i_1},Y_{i_2})
    -4r(r-1)s\rho_B^{-(2s+1)}\nu_n g_{\rho_B;1}(Y_{i_1})g_{B;2}(Y_{i_1},Y_{i_2})
    \notag\\
    &-4rs\rho_B^{-(2s+1)}\nu_n g_{B;1}(Y_{i_1})g_{\rho_B;2}(Y_{i_1},Y_{i_2}). \label{eq:def_beta4}
\end{align}
All the above terms involve $g_{A;1}(\cdot), g_{A;2}(\cdot,\cdot), g_{\rho_A;1}(\cdot), g_{\rho_A;2}(\cdot,\cdot)$ and $g_{B;1}(\cdot), g_{B;2}(\cdot,\cdot), g_{\rho_B;1}(\cdot), g_{\rho_B;2}(\cdot,\cdot)$, which need to be estimated in practice. Without loss of generality, we only elaborate the estimators for all $A$-indexed quantities. 
\begin{align}
&\hat g_{A;1}(X_i):=\binom{m-1}{r-1}^{-1}\sum_{\{i_1<\cdots<i_{r-1}\}\subseteq[1:m]\backslash\{i\}}h(A_{i,i_1,\ldots,i_{r-1}})- \hat U_m \label{eq:hat_gA1}\\
&\hat g_{\rho_A;1}(X_i)  := (m-1)^{-1}\sum_{\substack{1\leq i'\leq m, i'\neq i}} A_{ii'} - \hat\rho_A \label{eq:hat_grhoA1}\\
&\hat g_{A;2}(X_{i_1},X_{i_2}) := \binom{m-2}{r-2}^{-1}\sum_{\substack{1\leq i_1'<\cdots<i_r'\leq m\\\{i_1,i_2\}\subseteq\{i_1',\ldots,i_r'\}}} h(A_{i_1',\ldots,i_r'})-\hat g_{A;1}(X_{i_1})-\hat g_{A;1}(X_{i_2})-\hat U_m \label{eq:hat_gA2}\\
&\hat g_{\rho_A;2}(X_{i_1},X_{i_2}) := A_{i_1i_2}-\hat g_{\rho_A;1}(X_{i_1})-\hat g_{\rho_A;1}(X_{i_2})-\hat \rho_A \label{eq:hat_grhoA2}
\end{align}
Then, $\xi_{A;1}^2$ and $\xi_{A,\rho_A;1}$ can be estimated, respectively, by 
\begin{align}
    \hat \xi_{A;1}^2:= \frac{1}{m}\sum_{i=1}^m \left\{ \hat g_{A;1}(X_i)\right\}^2 \quad {\rm and}\quad \hat \xi_{A,\rho_A;1}:= \frac{1}{m}\sum_{i=1}^m  \hat g_{A;1}(X_i) \hat g_{\rho_A;1}(X_i)   
    \label{eq:hat_xiA1}
\end{align}
Similarly, define $\hat \xi_{\rho_A;1}^2:=m^{-1}\sum_{i=1}^m \hat g^2_{\rho_A;1}(X_i)$. 
Plug the estimates $\hat U_m, \hat \rho_A, \hat \xi_{A;1}^2, \hat \xi_{\rho_A;1}^2, \hat \xi_{A,\rho_A;1}$, $\hat g_{A;1}(\cdot), \hat g_{A;2}(\cdot,\cdot), \hat g_{\rho_A;1}(\cdot), \hat g_{\rho_A;2}(\cdot,\cdot)$ into eq. (\ref{eq:def_alpha0}) through (\ref{eq:def_alpha4}).  We can now
define the empirical estimates $\hat\alpha_0, \hat \alpha_1(X_i), \hat\alpha_2(X_{i_1},X_{i_2}), \hat\alpha_3(X_i)$ and $\hat \alpha_4(X_{i_1},X_{i_2})$, e.g, 
$\hat\alpha_1(X_i) := r\hat\rho_A^{-s} \hat g_{A;1}(X_i) - 2s\hat\rho_A^{-(s+1)}\hat U_m\hat g_{\rho_A;1}(X_i)$. Moreover, we estimate $\sigma_{m,n}^2$ by 
$$
\hat S_{m,n}^2:=\frac{1}{m^2}\sum_{i=1}^m \hat \alpha_1^2(X_i)+\frac{1}{n^2}\sum_{j=1}^n \hat\beta_1^2(Y_j). 
$$

Equipped with above estimates $\hat S^2_{m,n}$, $\hat\alpha_0, \hat \alpha_1(X_i), \hat\alpha_2(X_{i_1},X_{i_2}), \hat\alpha_3(X_i)$ and $\hat \alpha_4(X_{i_1},X_{i_2})$, the empirical version of the Edgeworth expansion terms $\hat{ \cal I}_0, \hat Q_{m,n,\rho_A,\rho_B;1}$ and $\hat Q_{m,n,\rho_A,\rho_B;2}$ can be calculated accordingly. For instance, 
\begin{align}\label{eq:hat_calI0}
    \hat {\cal I}_0:=\hat S_{m,n}^{-1}(m^{-1}\hat \alpha_0-n^{-1}\hat \beta_0),
\end{align}
and we formulate $\hat Q_{m,n,\rho_A,\rho_B;1}$ and $\hat Q_{m,n,\rho_A,\rho_B;2}$ exactly similarly, but we omit the display of their lengthy expressions.  Note that the expectations $\ep[\alpha_1^3(X_1)], \ep[\alpha_1(X_1)\alpha_3(X_1)], \ep[\alpha_4(X_1,X_2)\alpha_1(X_2)]$ and $\ep[\alpha_1(X_1)\alpha_1(X_2)\alpha_2(X_1,X_2)]$ are all calculated by the empirical version as follows 
\begin{align}
    &\hat \ep[\alpha_1^3(X_1)]
    :=
    \frac{1}{m}\sum_{i=1}^m \hat \alpha_1^3(X_i) \label{eq:hat_ep_alpha1}
    \\
    &\hat \ep[\alpha_1(X_1)\alpha_3(X_1)]
    :=
    \frac1m \sum_{i=1}^m \hat \alpha_1(X_i) \alpha_3(X_i)
    \\
    &\hat \ep[\alpha_4(X_1,X_2)\alpha_1(X_2)]
    :=
    \frac1{m(m-1)}\sum_{1\leq \{i_1\neq i_2\}\leq m} \hat\alpha_4(X_{i_1},X_{i_2})\alpha_1(X_{i_2})
    \\
    &\hat \ep[\alpha_1(X_1)\alpha_1(X_2)\alpha_2(X_1,X_2)]
    := 
    {m\choose 2}^{-1}\sum_{1\leq i_1< i_2\leq m} \hat \alpha_1(X_{i_1})\hat \alpha_1(X_{i_2})\hat \alpha_2(X_{i_1},X_{i_2}). \label{eq:hat_ep_alpha1alpha2}
\end{align}
Additionally, $\xi^2_{\alpha;1}$ and $\xi^2_{\beta;1}$ can be estimated via 
\begin{align}\label{eq:hat_xi_alpha1}
    \hat\xi_{\alpha;1}^2=\frac{1}{m}\sum_{i=1}^m \hat\alpha_1(X_i)^2\quad {\rm and}\quad  \hat\xi_{\beta;1}^2=\frac{1}{n}\sum_{j=1}^n \hat\beta_1(Y_j)^2.
\end{align}

\subsection{Simulation 1: distribution approximation error}
Our first simulation aims to illustrate our method's higher-order accuracy in approximating the distribution of $\hat T_{m,n}+\delta_T$ where $C_\delta = 0.01$. 
In practice, we find the self-smoothing effect to be strong enough such that $F_{\hat T_{m,n}}(u)$ seems quite smooth.  For this reason, in all numerical work of this paper, we keep this small $C_\delta$ to avoid disrupting the distribution formula of $\hat T_{m,n}$ much for finite-sample examples.

We generate data from the two graphons $f_A(u,v) = 1.71\cdot (u +v)^2/2$ and $f_B(u,v) = 1.38\cdot \exp\{-(u +v)/3\}$,
and approximately the true $F_{\hat T_{m,n}}(u)$ by $n_{\rm MC}:=10^5$ Monte Carlo replications.
Accuracy is measured by the truncated Kolmogorov distance $\sup_{u\in[-2,2]}|\hat F_{\hat T_{m,n}+\delta_T}(u) - F_{\hat T_{m,n}+\delta_T}(u)|$. 
We compare to three benchmark methods: 
$N(0,1)$ approximation, 
node subsampling,
and node resampling.
Here, the node subsampling and resampling methods are in our own formulation, which extend \citet{bhattacharyya2015subsampling} and \citet{green2017bootstrapping}, respectively.  
The algorithm is given in Algorithm \ref{algorithm::node-resample}.
We set $m_{\rm sub}=m^{1/2},n_{\rm sub}=n^{1/2}$; and set $N_{\rm boot}=200$ for both subsampling and resampling.
\begin{algorithm}
    \caption{Node subsampling/resampling bootstrap for two-sample inference}
    \label{algorithm::node-resample}
    {\bf Input:} Networks $A,B$; bootstrap repetition $N_{\rm boot}$; if subsampling: subsample sizes $m_{\rm sub},n_{\rm sub}$\\
    {\bf Output:} Bootstrapped studentized empirical moment discrepancies $\{\hat T_{m,n}^{(b)}\}_{b=1,\ldots,N_{\rm boot}}$\\
    {\bf Steps:} For $b=1,\ldots,N_{\rm boot}$, do
    \begin{enumerate}
        \item Node subsample/resample $A,B$, obtain $A^{(b)}, B^{(b)}$. If resampling, randomly sample $m$ nodes ${\cal J}_A$ from $[1:m]$ with replacement; if subsampling, randomly sample $m_{\rm sub}$ nodes ${\cal J}_A$ from $[1:m]$ without replacement; In either case, set $A^{(b)}\leftarrow A_{{\cal J}_A}$; do the same for $B$
        \item Compute $\hat T_{m,n}^{(b)}$ using \eqref{def::S_m,n} and \eqref{def::T_m,n}, with $A^{(b)}, B^{(b)}$ as the input
    \end{enumerate}
\end{algorithm}
Taking subsampling as an example, a key distinction between our subsampling and \citet{bhattacharyya2015subsampling,lunde2019subsampling} is that we bootstrap the studentization $\hat T_{m,n}+\delta_T$ that we formulate in \eqref{def::S_m,n} and \eqref{def::T_m,n}, whereas they bootstrap the unstudentized $\hat D_{m,n}$ or its standardization.
As pointed out by \citet{wasserman2006all,hall2013bootstrap}, studentization is the key to achieve higher-order accuracy, which standardization does not achieve.
\cite{ghoshdastidar2017two} uses a conservative variance estimate, which leads to their test statistic that is not a studentized.  Their test method is based on the convergence of their test statistic, without characterizing the statistic's distribution, thus is not comparable.
To ease visualization, in this simulation, we set $m=n\in\{10,20,40,80,160\}$.
We did not test on larger networks because node resampling bootstrap costs more than 12 hours for $n=320$.
\begin{figure}[h!]
    \centering
    \includegraphics[width=0.3\textwidth]{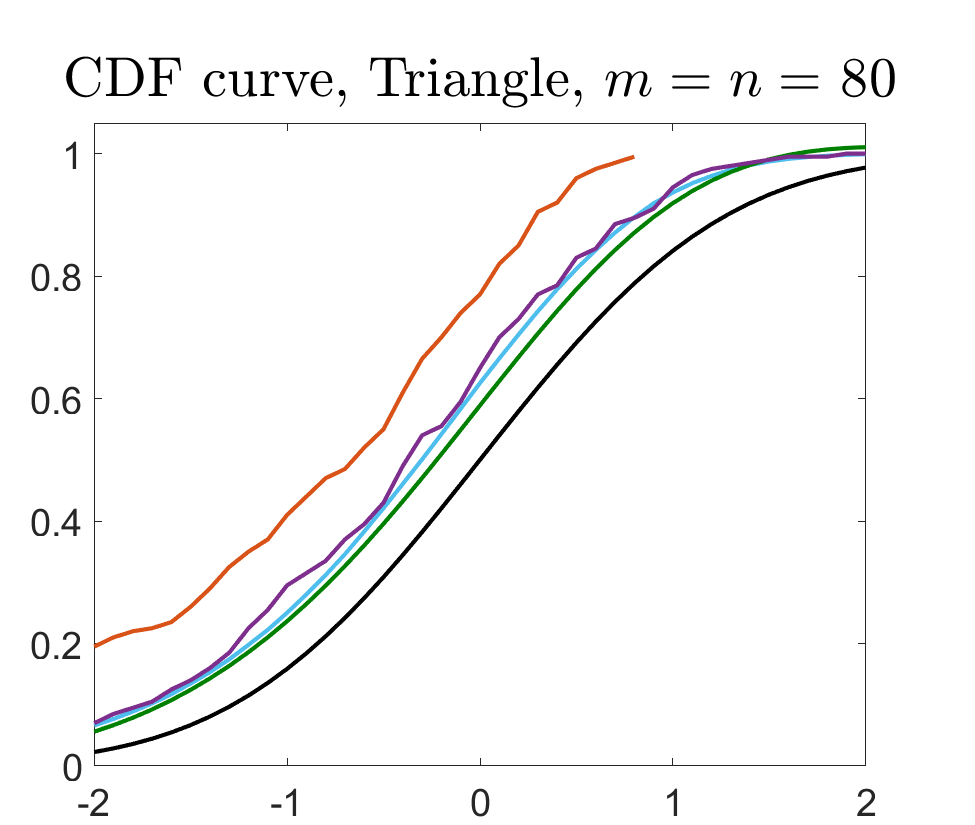}
    \includegraphics[width=0.3\textwidth]{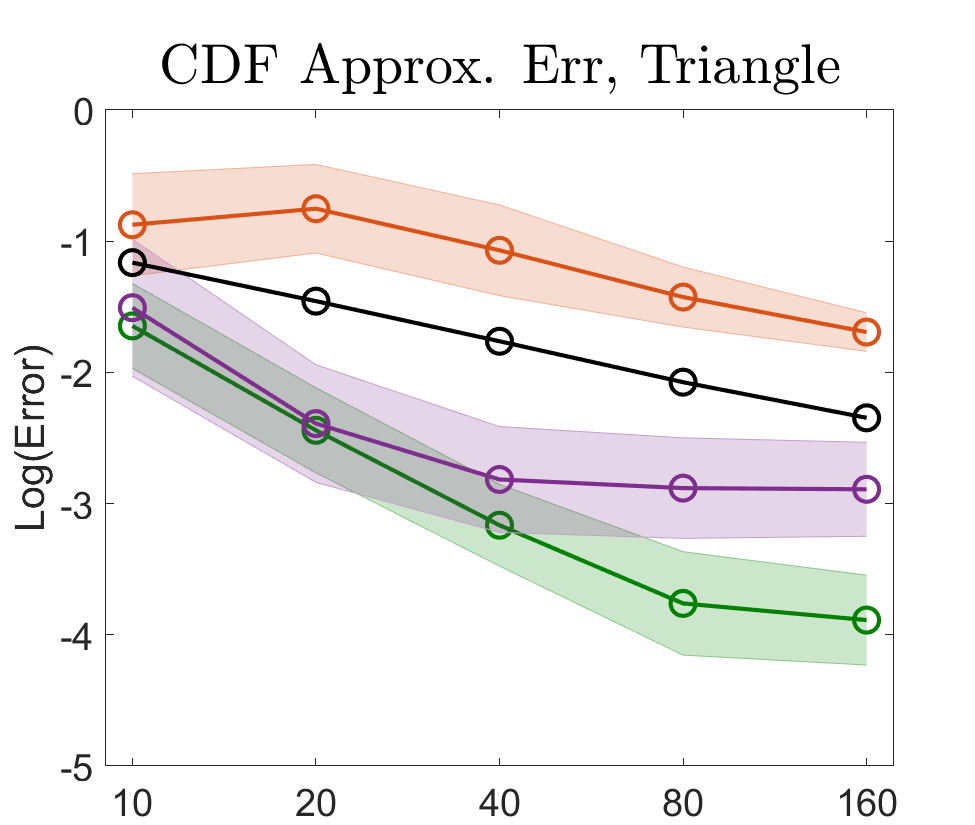}
    \includegraphics[width=0.3\textwidth]{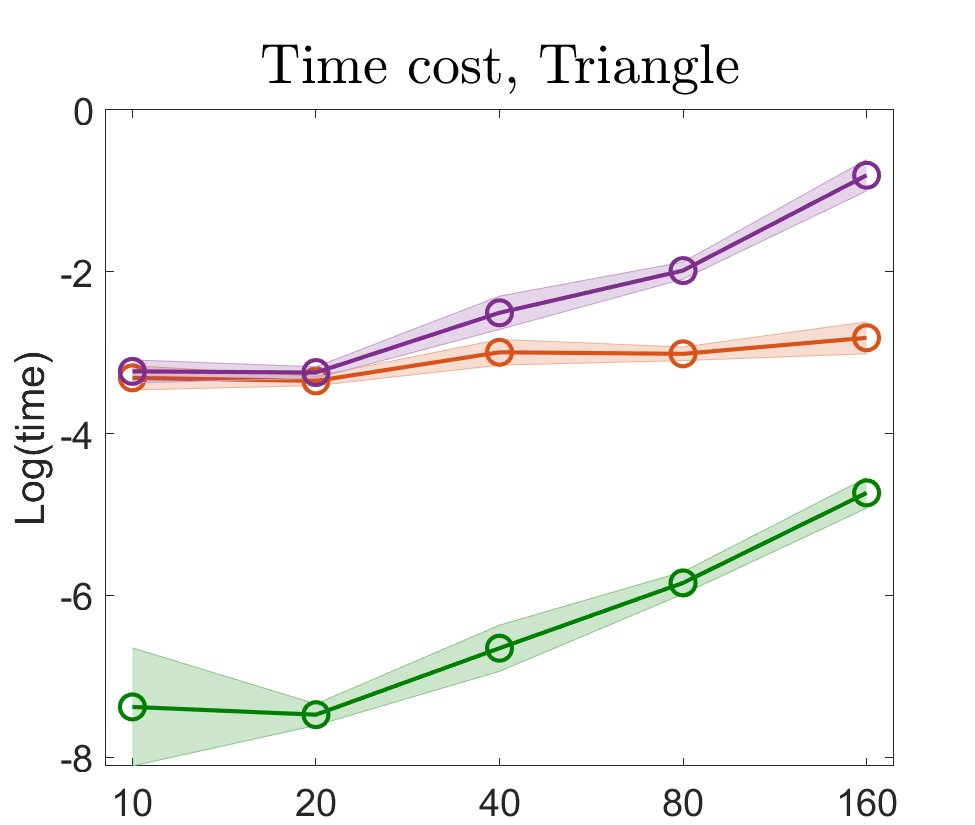}
    \\
    \includegraphics[width=0.3\textwidth]{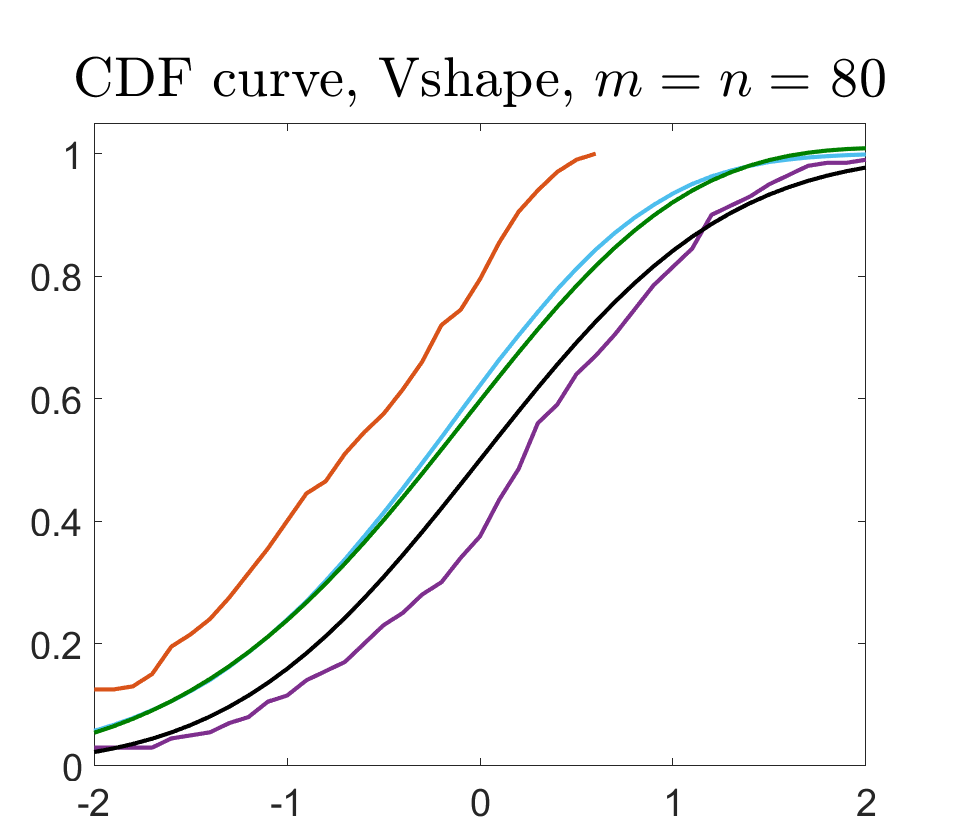}
    \includegraphics[width=0.3\textwidth]{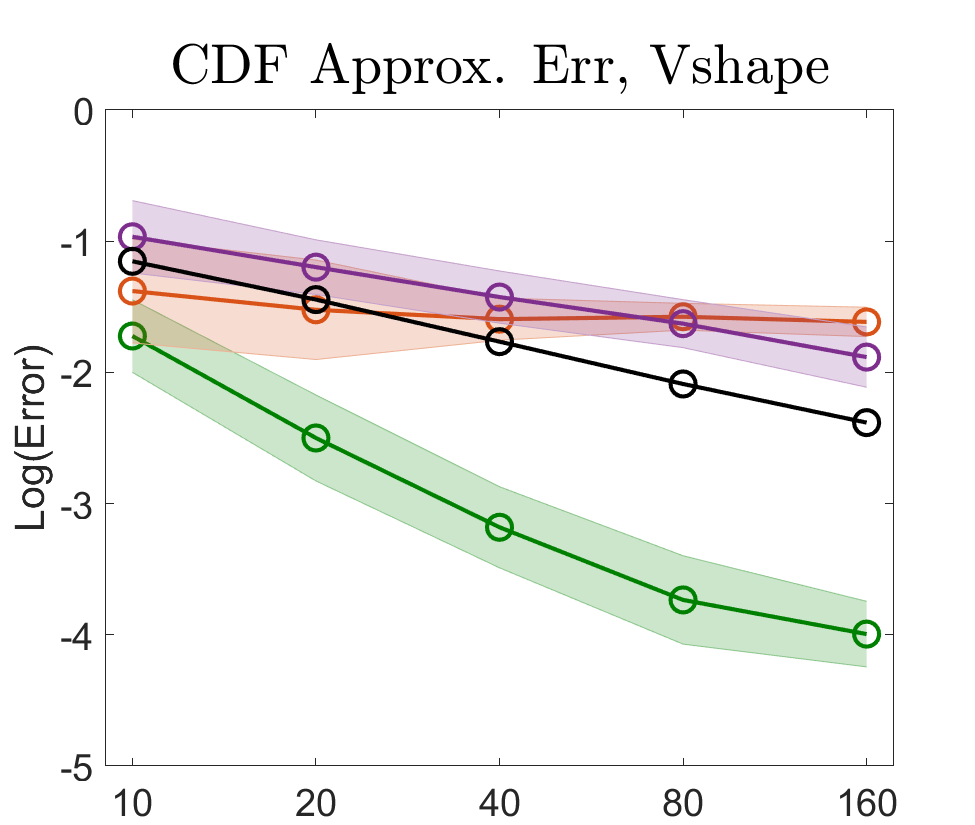}
    \includegraphics[width=0.3\textwidth]{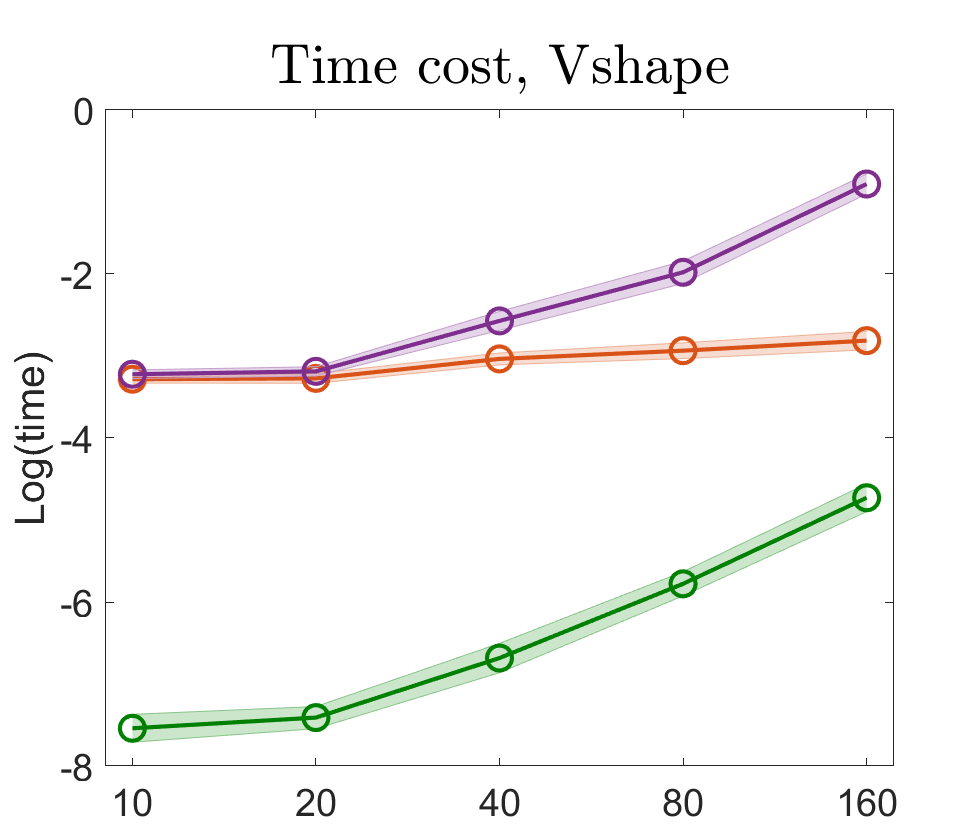}
    \caption{Distribution approximation accuracy comparison.  Left: true and estimated cumulative distribution functions; middle: comparison of logarithm Kolmogorov-distance; right: logarithm computation time, notice $N(0,1)$ costs no time to approximate $F_{\hat T_{m,n}}$.
    Cyan is the Monte Carlo evaluation of the true $F_{\hat T_{m,n}}$; green is our method;
    black is $N(0,1)$;
    and orange and purle are node subsampling and resampling, respectively, formulated by our Algorithm \ref{algorithm::node-resample}.
    }
    \label{fig::simulation-1}
\end{figure}

Figure \ref{fig::simulation-1} shows the result.  
The vertical error bar shows the standard deviation of our method over 100 repeated experiments.
The result echoes our method's higher-order accuracy that improves over $N(0,1)$ due to bias correction.
It also confirms our method's computational efficiency.

\subsection{Simulation 2: Coverage probability of Cornish-Fisher confidence interval}
\label{subsec::simulation 2::table for coverage probability}
As recognized by classical literature \cite{hall2013bootstrap, beran1987prepivoting}, one main merit of higher-order accurate CDF approximation is that it enables higher-order accurate control of the confidence interval's actual coverage probability.
In this experiment, we evaluate the discrepancy between the actual coverage probability and nominal confidence level as the performance measure.
We inherit the set up of Simulation 1, set $1-\alpha=90\%$, but experiment on more $(m,n)$ combinations $(m,n)\in\{10,20,40,80,160\}^2$.  We compare to the same benchmarks in Simulation 1.

\begin{figure}[h!]
    \centering
    \includegraphics[width=0.3\textwidth]{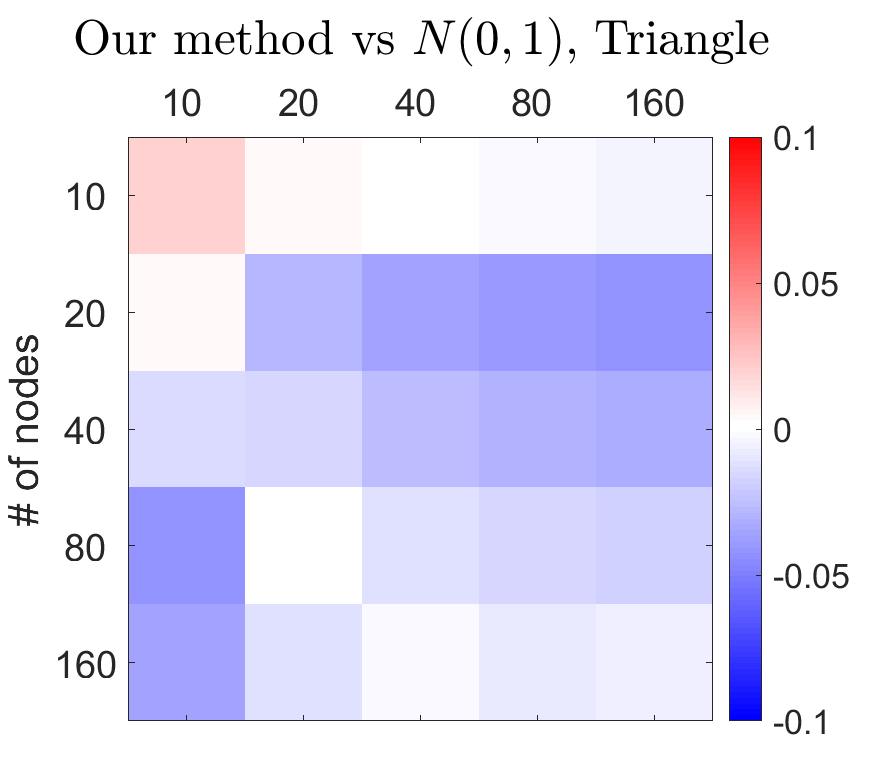}
    \includegraphics[width=0.3\textwidth]{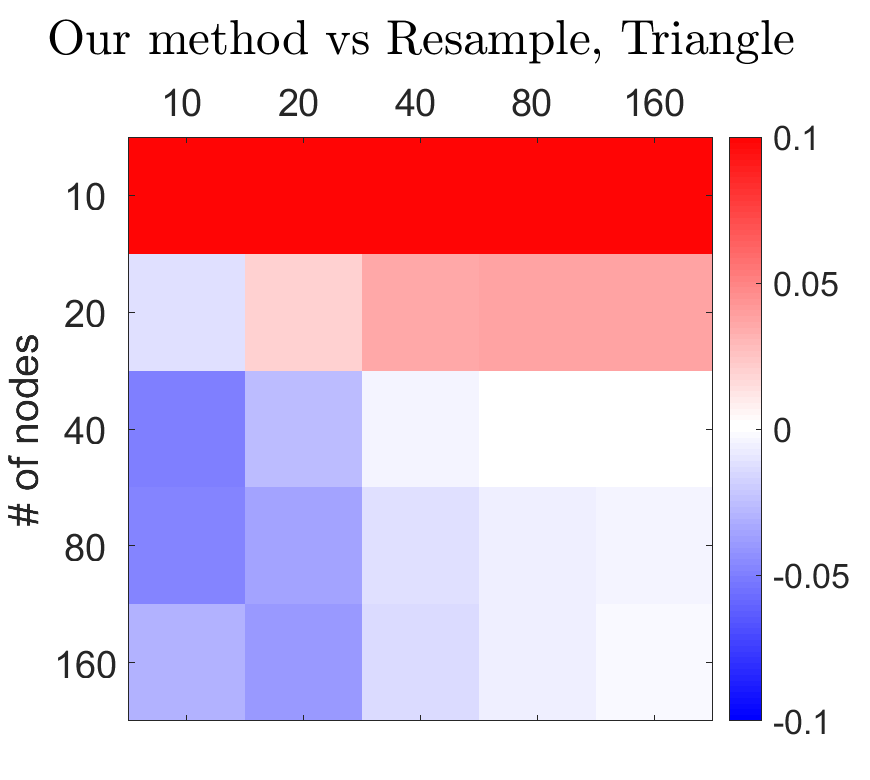}
    \includegraphics[width=0.3\textwidth]{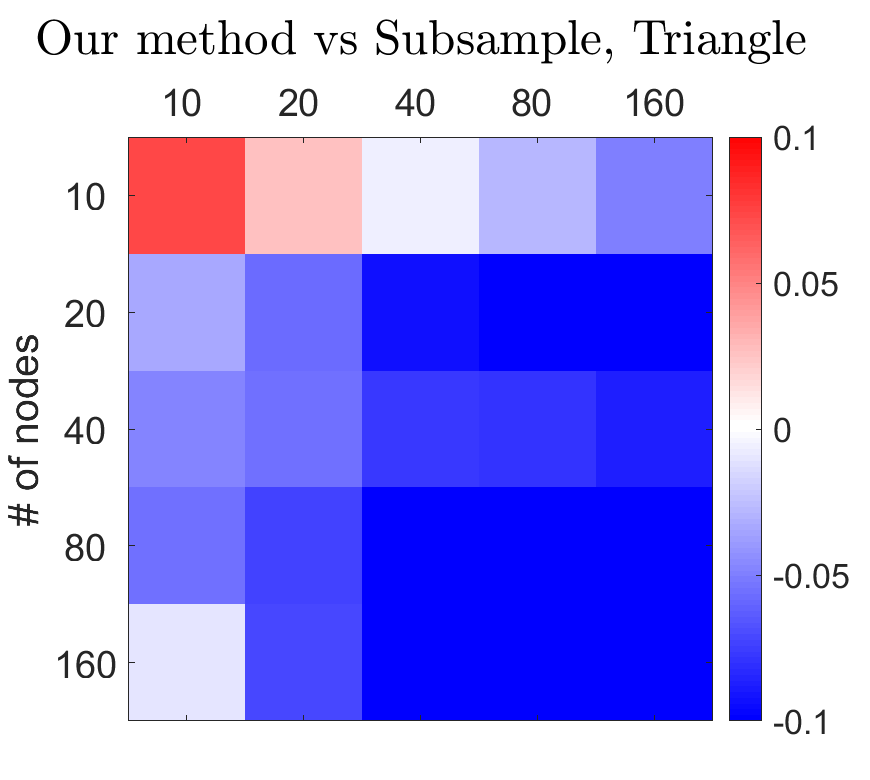}\\ \includegraphics[width=0.3\textwidth]{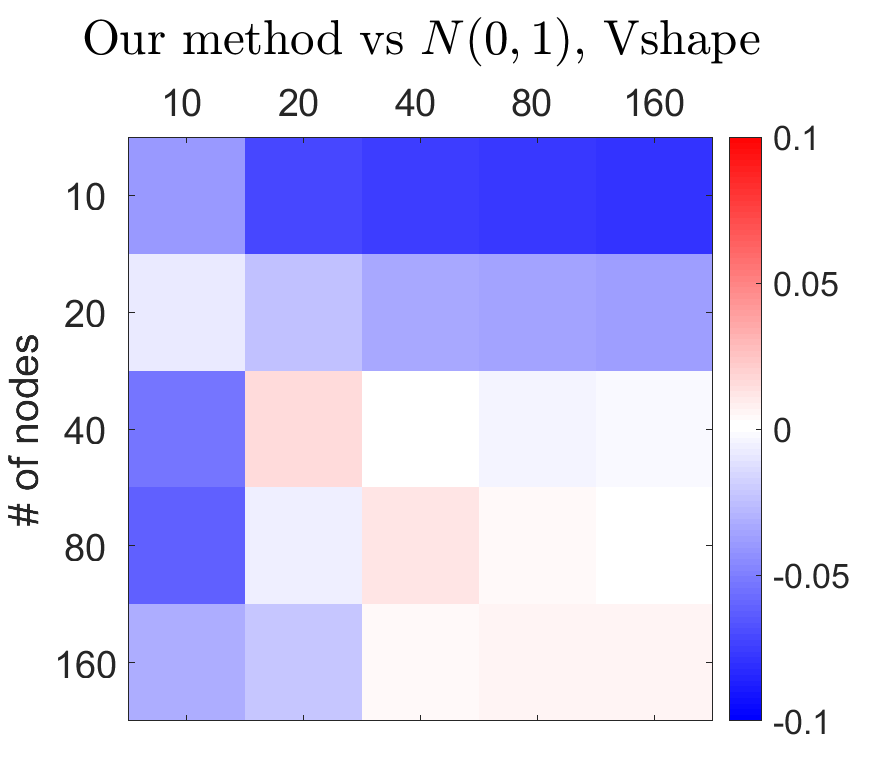} 
    \includegraphics[width=0.3\textwidth]{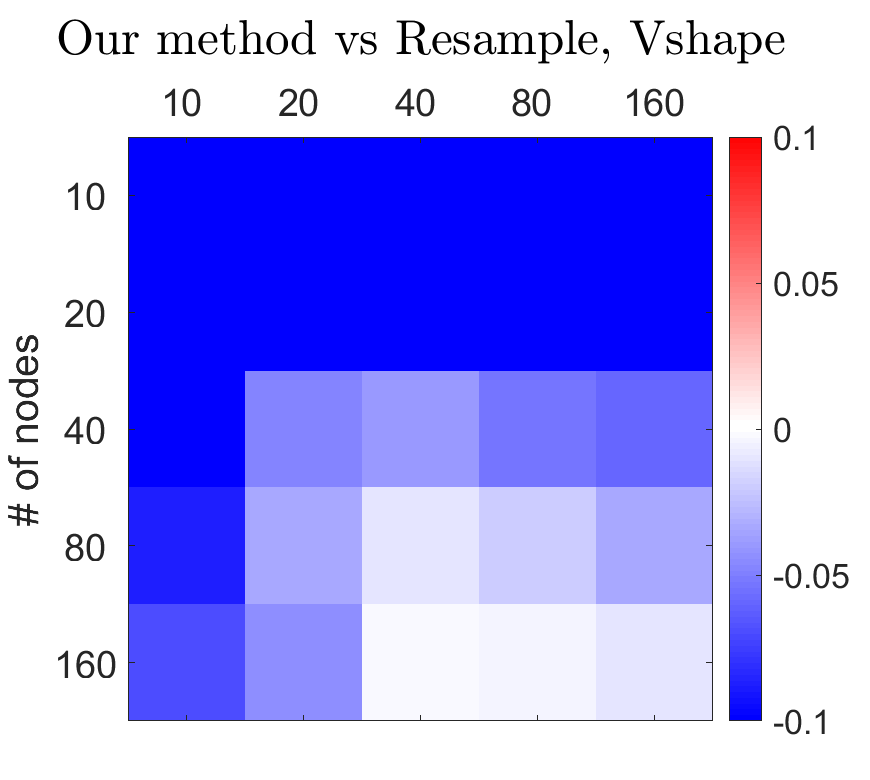} 
    \includegraphics[width=0.3\textwidth]{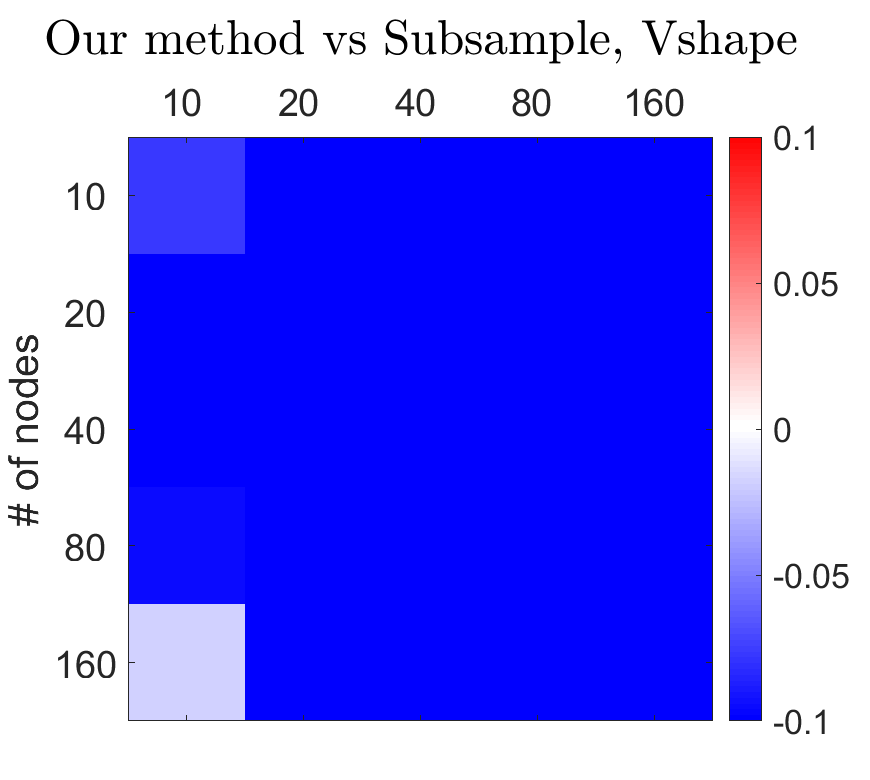}
    \caption{Comparison of CI coverage accuracy.  Bluer pixel indicates more advantage of our method over the compared method.
    Here we only show accuracy comparisons; evidence of marginal accuracy of our method in Section \ref{supple::table for coverage probability} in Supplementary Material.}
    \label{fig::simulation-2}
\end{figure}

Figure \ref{fig::simulation-2} shows that our method controls coverage probabilities more accurately around the nominal level $1-\alpha$ in most settings, except in the case where one network has only $m=10$ nodes.  
This is not surprising since the remainder terms ignored in the analytical approximation may be no longer ignorable when the network is very small.
We recall that both node bootstrap methods in this simulation are also devised based on our formulation of $\hat T_{m,n}$.
Node resampling shows good accuracy and robustness on some small network examples, but it is not scalable: it takes more than 12 hours to run node resampling experiments for $m\da n=320$ settings.
Due to page limit, we sink the numerical result tables to Section \ref{supple::table for coverage probability} in Supplementary Material.

\section{Additional simulation set-up information and results}

\subsection{Addition simulation results for the simulation in Section \ref{subsec::simulation 2::table for coverage probability} in Main Paper}
\label{supple::table for coverage probability}

Tables \ref{supp::tab::cover-prob-1} -- \ref{supp::tab::cover-prob-8} show the numerical results of empirical 90\% CI coverage probabilities of our method and $N(0,1)$ approximation.  The purpose of displaying these tables is to provide more details and confirm the marginal accuracy of our method, completing the message conveyed by Figure \ref{fig::simulation-2} in Main Paper.

\begin{table}[htbp!]
\centering
\caption{Empirical CI coverage probabilities, level~ $= 90\%$, Triangle, our method, Mean(SD)\\}
\label{supp::tab::cover-prob-1}
\begin{adjustbox}{center}
\begin{tabular}{c|ccccc}\hline
Network sizes & $n = 10$ & $n = 20$ & $n = 40$ & $n = 80$ & $n = 160$  \\\hline
$m = 10$  &$0.678(0.467)$&$0.671(0.470)$&$0.670(0.470)$&$0.672(0.470)$&$0.673(0.469)$\\\hline
$m = 20$  &$0.860(0.347)$&$0.858(0.349)$&$0.857(0.350)$&$0.857(0.350)$&$0.857(0.350)$\\\hline
$m = 40$  &$0.892(0.310)$&$0.892(0.311)$&$0.891(0.311)$&$0.892(0.311)$&$0.893(0.309)$\\\hline
$m = 80$  &$0.895(0.306)$&$0.901(0.298)$&$0.900(0.300)$&$0.899(0.302)$&$0.899(0.302)$\\\hline
$m = 160$  &$0.886(0.318)$&$0.904(0.295)$&$0.901(0.299)$&$0.901(0.299)$&$0.902(0.297)$\\\hline
\end{tabular}
\end{adjustbox}
\end{table}

\begin{table}[htbp!]
\centering
\caption{Empirical CI coverage probabilities, level~ $= 90\%$, Vshape, our method, Mean(SD)\\}
\label{supp::tab::cover-prob-2}
\begin{adjustbox}{center}
\begin{tabular}{c|ccccc}\hline
Network sizes & $n = 10$ & $n = 20$ & $n = 40$ & $n = 80$ & $n = 160$  \\\hline
$m = 10$  &$0.868(0.338)$&$0.872(0.334)$&$0.871(0.335)$&$0.870(0.336)$&$0.872(0.334)$\\\hline
$m = 20$  &$0.906(0.292)$&$0.912(0.283)$&$0.912(0.283)$&$0.913(0.282)$&$0.912(0.283)$\\\hline
$m = 40$  &$0.905(0.293)$&$0.917(0.276)$&$0.915(0.279)$&$0.916(0.277)$&$0.916(0.277)$\\\hline
$m = 80$  &$0.892(0.310)$&$0.915(0.279)$&$0.912(0.283)$&$0.910(0.286)$&$0.910(0.286)$\\\hline
$m = 160$  &$0.871(0.335)$&$0.913(0.283)$&$0.909(0.288)$&$0.907(0.290)$&$0.909(0.288)$\\\hline
\end{tabular}
\end{adjustbox}
\end{table}

\begin{table}[htbp!]
\centering
\caption{Empirical CI coverage probabilities, level~ $= 90\%$, Triangle, subsample, Mean(SD)\\}
\label{supp::tab::cover-prob-3}
\begin{adjustbox}{center}
\begin{tabular}{c|ccccc}\hline
Network sizes & $n = 10$ & $n = 20$ & $n = 40$ & $n = 80$ & $n = 160$  \\\hline
$m = 10$  &$0.751(0.432)$&$0.696(0.460)$&$0.664(0.472)$&$0.645(0.479)$&$0.624(0.484)$\\\hline
$m = 20$  &$0.826(0.379)$&$0.800(0.400)$&$0.763(0.425)$&$0.752(0.432)$&$0.745(0.436)$\\\hline
$m = 40$  &$0.845(0.362)$&$0.837(0.369)$&$0.815(0.389)$&$0.812(0.391)$&$0.806(0.396)$\\\hline
$m = 80$  &$0.840(0.366)$&$0.826(0.379)$&$0.795(0.403)$&$0.792(0.406)$&$0.790(0.407)$\\\hline
$m = 160$  &$0.877(0.328)$&$0.825(0.380)$&$0.773(0.419)$&$0.777(0.416)$&$0.781(0.413)$\\\hline
\end{tabular}
\end{adjustbox}
\end{table}

\begin{table}[htbp!]
\centering
\caption{Empirical CI coverage probabilities, level~ $= 90\%$, Vshape, subsample, Mean(SD)\\}
\label{supp::tab::cover-prob-4}
\begin{adjustbox}{center}
\begin{tabular}{c|ccccc}\hline
Network sizes & $n = 10$ & $n = 20$ & $n = 40$ & $n = 80$ & $n = 160$  \\\hline
$m = 10$  &$0.792(0.406)$&$0.734(0.442)$&$0.698(0.459)$&$0.688(0.463)$&$0.680(0.466)$\\\hline
$m = 20$  &$0.763(0.425)$&$0.707(0.455)$&$0.679(0.467)$&$0.674(0.469)$&$0.671(0.470)$\\\hline
$m = 40$  &$0.770(0.421)$&$0.701(0.458)$&$0.701(0.458)$&$0.712(0.453)$&$0.719(0.450)$\\\hline
$m = 80$  &$0.796(0.403)$&$0.702(0.457)$&$0.703(0.457)$&$0.716(0.451)$&$0.725(0.446)$\\\hline
$m = 160$  &$0.853(0.354)$&$0.723(0.448)$&$0.715(0.452)$&$0.734(0.442)$&$0.748(0.434)$\\\hline
\end{tabular}
\end{adjustbox}
\end{table}

\begin{table}[htbp!]
\centering
\caption{Empirical CI coverage probabilities, level~ $= 90\%$, Triangle, resample, Mean(SD)\\}
\label{supp::tab::cover-prob-5}
\begin{adjustbox}{center}
\begin{tabular}{c|ccccc}\hline
Network sizes & $n = 10$ & $n = 20$ & $n = 40$ & $n = 80$ & $n = 160$  \\\hline
$m = 10$  &$0.838(0.369)$&$0.848(0.359)$&$0.851(0.356)$&$0.853(0.355)$&$0.854(0.353)$\\\hline
$m = 20$  &$0.848(0.359)$&$0.878(0.328)$&$0.892(0.310)$&$0.894(0.308)$&$0.896(0.306)$\\\hline
$m = 40$  &$0.842(0.365)$&$0.866(0.340)$&$0.887(0.316)$&$0.892(0.310)$&$0.893(0.310)$\\\hline
$m = 80$  &$0.848(0.359)$&$0.862(0.345)$&$0.888(0.316)$&$0.893(0.310)$&$0.894(0.307)$\\\hline
$m = 160$  &$0.856(0.351)$&$0.856(0.351)$&$0.886(0.318)$&$0.893(0.309)$&$0.896(0.305)$\\\hline
\end{tabular}
\end{adjustbox}
\end{table}

\begin{table}[htbp!]
\centering
\caption{Empirical CI coverage probabilities, level~ $= 90\%$, Vshape, resample, Mean(SD)\\}
\label{supp::tab::cover-prob-6}
\begin{adjustbox}{center}
\begin{tabular}{c|ccccc}\hline
Network sizes & $n = 10$ & $n = 20$ & $n = 40$ & $n = 80$ & $n = 160$  \\\hline
$m = 10$  &$0.649(0.477)$&$0.694(0.461)$&$0.695(0.460)$&$0.689(0.463)$&$0.684(0.465)$\\\hline
$m = 20$  &$0.732(0.443)$&$0.784(0.411)$&$0.785(0.411)$&$0.771(0.420)$&$0.764(0.425)$\\\hline
$m = 40$  &$0.785(0.411)$&$0.835(0.371)$&$0.845(0.362)$&$0.831(0.375)$&$0.823(0.381)$\\\hline
$m = 80$  &$0.805(0.396)$&$0.852(0.355)$&$0.879(0.327)$&$0.869(0.337)$&$0.857(0.350)$\\\hline
$m = 160$  &$0.803(0.398)$&$0.844(0.363)$&$0.889(0.314)$&$0.890(0.313)$&$0.881(0.324)$\\\hline
\end{tabular}
\end{adjustbox}
\end{table}

\begin{table}[htbp!]
\centering
\caption{Empirical CI coverage probabilities, level~ $= 90\%$, Triangle, $N(0,1)$, Mean(SD)\\}
\label{supp::tab::cover-prob-7}
\begin{adjustbox}{center}
\begin{tabular}{c|ccccc}\hline
Network sizes & $n = 10$ & $n = 20$ & $n = 40$ & $n = 80$ & $n = 160$  \\\hline
$m = 10$  &$0.698(0.459)$&$0.676(0.468)$&$0.670(0.470)$&$0.670(0.470)$&$0.670(0.470)$\\\hline
$m = 20$  &$0.863(0.344)$&$0.830(0.376)$&$0.822(0.382)$&$0.818(0.386)$&$0.816(0.387)$\\\hline
$m = 40$  &$0.923(0.267)$&$0.877(0.329)$&$0.865(0.342)$&$0.862(0.345)$&$0.861(0.345)$\\\hline
$m = 80$  &$0.947(0.225)$&$0.901(0.299)$&$0.888(0.315)$&$0.883(0.322)$&$0.881(0.323)$\\\hline
$m = 160$  &$0.950(0.218)$&$0.916(0.277)$&$0.897(0.304)$&$0.892(0.310)$&$0.891(0.312)$\\\hline
\end{tabular}
\end{adjustbox}
\end{table}

\begin{table}[htbp!]
\centering
\caption{Empirical CI coverage probabilities, level~ $= 90\%$, Vshape, $N(0,1)$, Mean(SD)\\}
\label{supp::tab::cover-prob-8}
\begin{adjustbox}{center}
\begin{tabular}{c|ccccc}\hline
Network sizes & $n = 10$ & $n = 20$ & $n = 40$ & $n = 80$ & $n = 160$  \\\hline
$m = 10$  &$0.828(0.377)$&$0.800(0.400)$&$0.796(0.403)$&$0.793(0.405)$&$0.793(0.405)$\\\hline
$m = 20$  &$0.913(0.282)$&$0.864(0.343)$&$0.854(0.353)$&$0.851(0.356)$&$0.849(0.358)$\\\hline
$m = 40$  &$0.959(0.198)$&$0.901(0.299)$&$0.885(0.319)$&$0.881(0.324)$&$0.881(0.323)$\\\hline
$m = 80$  &$0.969(0.173)$&$0.920(0.272)$&$0.899(0.301)$&$0.894(0.308)$&$0.892(0.310)$\\\hline
$m = 160$  &$0.960(0.196)$&$0.935(0.247)$&$0.906(0.292)$&$0.899(0.301)$&$0.898(0.303)$\\\hline
\end{tabular}
\end{adjustbox}
\end{table}

\subsection{Detailed simulation set-up information and full-X-range plots for Simulation 3 in Section \ref{subsec::simulation3::10differentgraphon} in Main Paper}
\label{supple::different graphon}
In the simulation, we generated a large network database with 10 different graphon models, listed as follows.  All these graphons' $f(\cdot,\cdot)$ functions have been rescaled such that $\int_{[0,1]^2}f(u,v)\td u\td v=1$. Set $\rho_A=\rho_{B_1} =...=\rho_{B_K} = 0.4 $.
\begin{enumerate}
    \item {\tt SmoothGraphon-1}: $f(u,v) = u+v$;
    \item {\tt SmoothGraphon-2}: $f(u,v) =  (u+v)^2/2\cdot\tau$, where $\tau = 1.71$;
    \item {\tt SmoothGraphon-3}: $f(u,v) = e^{-(u+v)/2}\cdot\tau$, where $\tau = 1.61$;
    \item {\tt SmoothGraphon-4}: $f(u,v) = e^{-(u+v)/3}\cdot\tau$, where $\tau = 1.38$;
    \item {\tt SmoothGraphon-5}: $f(u,v) = \cos((u+v)/2)\cdot\tau$, where $\tau = 1.16$;
    $cos(1/(u^2+v^2))+0.15\}\cdot\tau$, where $\tau = 4.57$;
    \item {\tt BlockModel-1}: Stochastic block model with $K=2$ equal-sized communities and edge probabilities $B = (0.6,0.2;0.2,0.2)\cdot\tau$, where $\tau = 3.33$;
    \item {\tt BlockModel-2}: Stochastic block model with $K=2$ equal-sized communities and edge probabilities $B = (0.4,0.1;0.1,0.1)\cdot\tau$, where $\tau = 5.71$;
    \item {\tt BlockModel-3}: Stochastic block model with $K=2$ with $(3/4, 1/4)$ sized communities and edge probabilities $B = (0.6,0.2;0.2,0.2)\cdot\tau$, where $\tau = 2.35$;
    \item {\tt BlockModel-4}: Stochastic block model with $K=2$ with $(1/3, 2/3)$ sized communities and edge probabilities $B = (0.8,0.4;0.4,0.2)\cdot\tau$, where $\tau = 2.81$;
    \item {\tt BlockModel-5}: Stochastic block model with $(2/3, 1/3)$ sized communities and edge probabilities $B = (0.8,0.2;0.2,0.2)\cdot\tau$, where $\tau = 2.14$.
\end{enumerate}

The keyword network in the first experiment was generated from  {\tt SmoothGraphon-1}.
The two keyword networks in the second experiment were generated from the following graphon models:
\begin{itemize}
    \item Keyword 1: Same as {\tt BlockModel-1}.
    \item Keyword 2: {\tt SmoothGraphon-6}: $f(u,v):= \{(u^2+v^2)/3\cdot \cos(1/(u^2+v^2))+0.15\}\cdot\tau$, where $\tau = 4.57$.
\end{itemize}

\begin{figure}[h!]
    \centering
    \mbox{\includegraphics[width=0.3\textwidth]{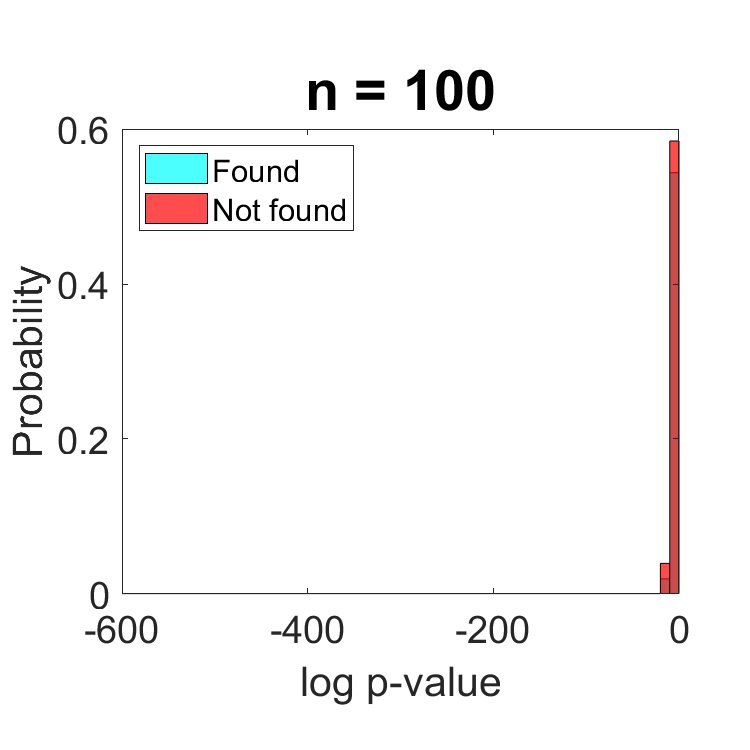}}  \mbox{\includegraphics[width=0.3\textwidth]{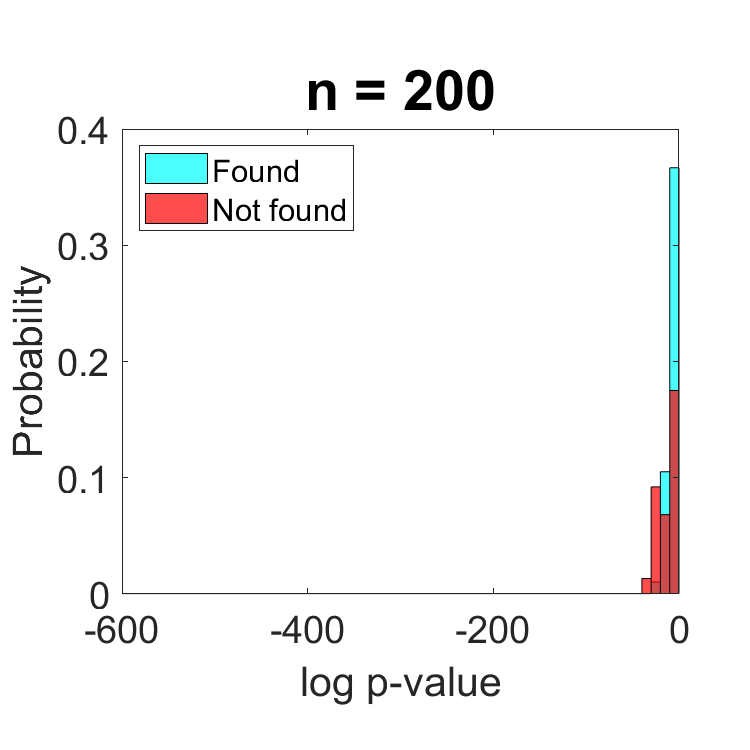}}  \mbox{\includegraphics[width=0.3\textwidth]{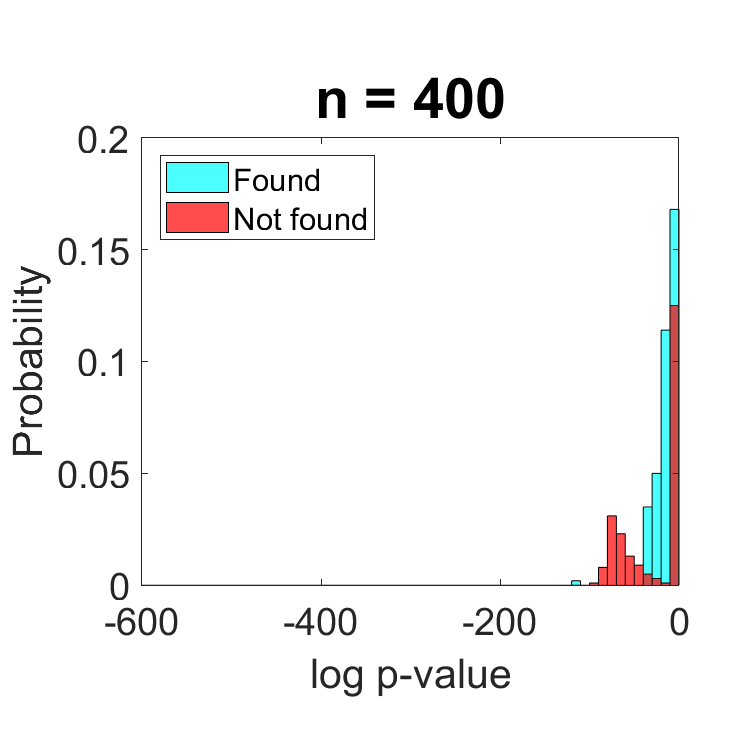}}\\
    \mbox{\includegraphics[width=0.3\textwidth]{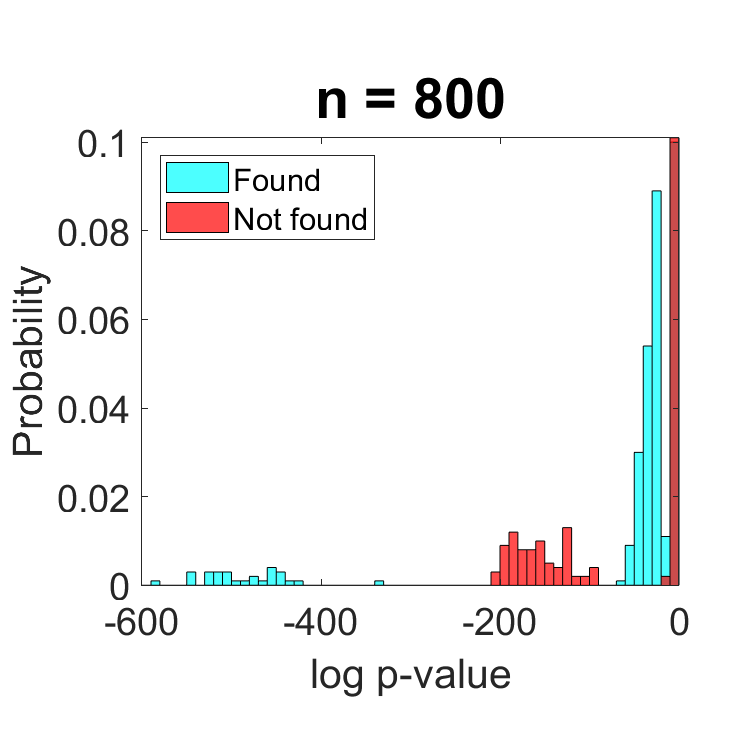}}  \mbox{\includegraphics[width=0.3\textwidth]{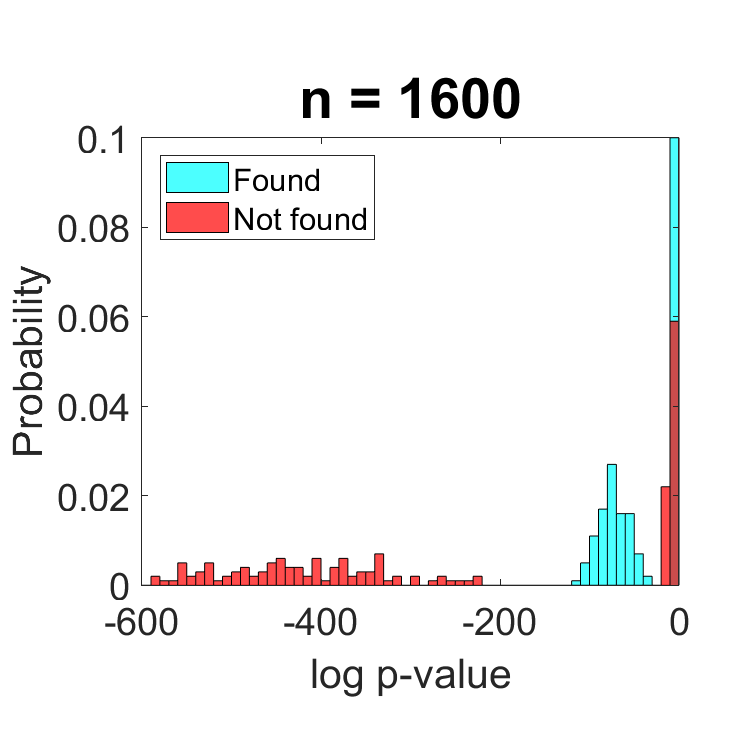}}\\
    \caption{Histograms of p-values from queries keyword 1 vs database (``found'', \textcolor{cyan}{cyan}) and keyword 2 vs database (``not found'', \textcolor{red}{red}), full-X-range plots for Row 2 of Figure \ref{fig::simulation-3} in Main Paper.}
    \label{supp::fig-simulation-3-full-range-plots}
\end{figure}

Figure \ref{supp::fig-simulation-3-full-range-plots} shows full-X-range plots of Row 2 in Figure \ref{fig::simulation-3} in Main Paper, with two more cases $n=200$, $n=800$, both omitted in Main Paper to meet page limit.  Particularly, the $n=1600$ plot in Figure \ref{supp::fig-simulation-3-full-range-plots} better matches the interpretation at the end of the second paragraph in Section 
\ref{subsec::simulation3::10differentgraphon}
in Main Paper, that there should be some cyan bars to the left of $\log(5\%)$ reference line, showing not-match.

\subsection{Additional simulation results for Section \ref{subsec::simulation-5::degeneracy}}

\begin{figure}[ht!]
    \centering
    \begin{adjustbox}{width=1.05\linewidth,center}
        \includegraphics[width=0.29\linewidth]{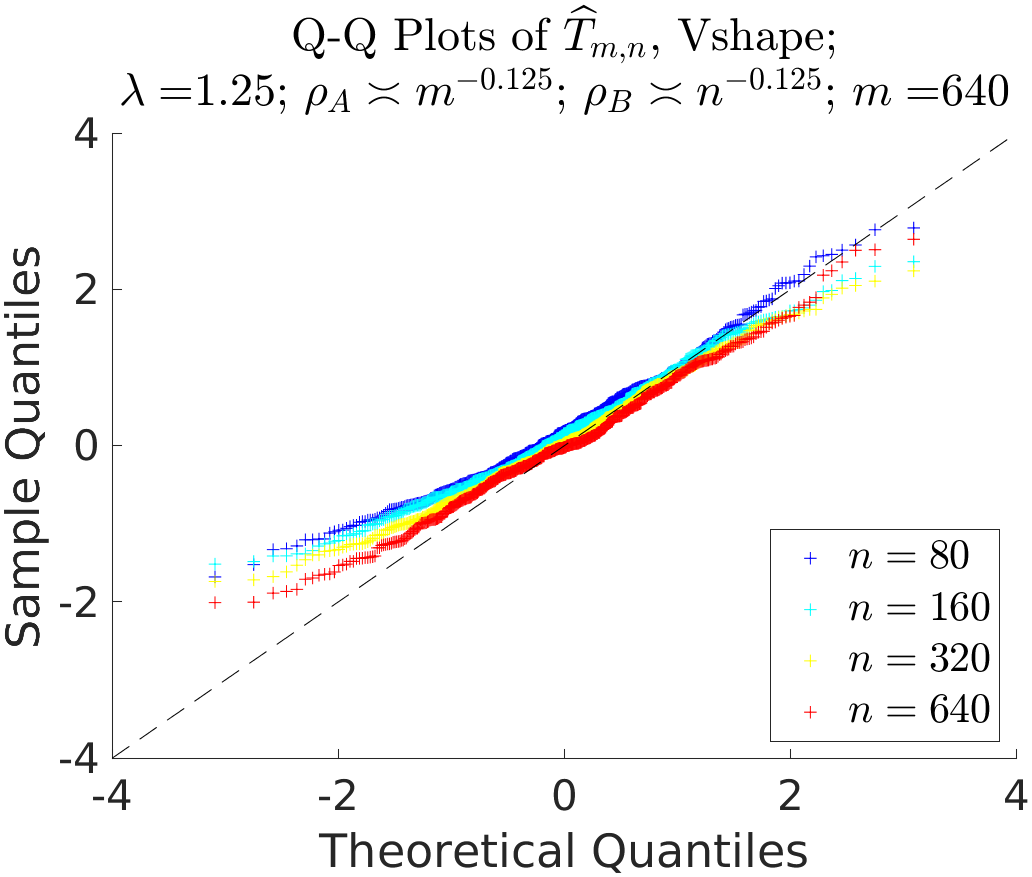}
        \includegraphics[width=0.29\linewidth]{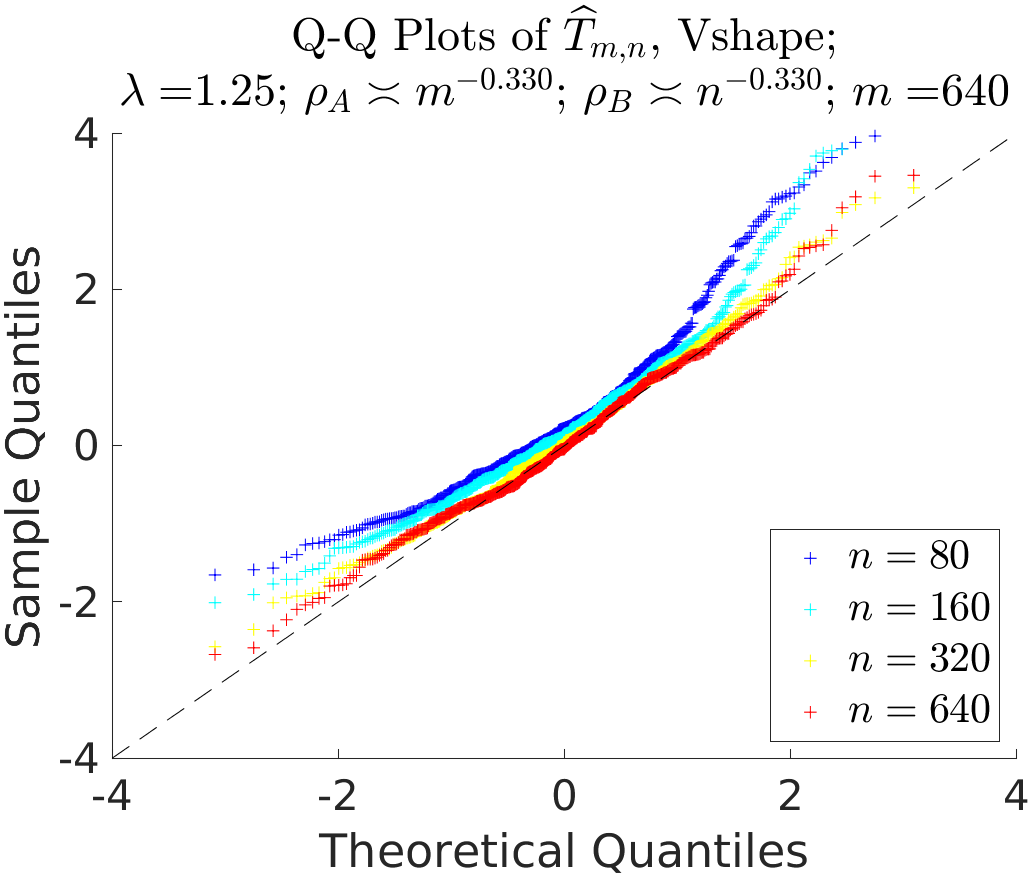}
        \includegraphics[width=0.29\linewidth]{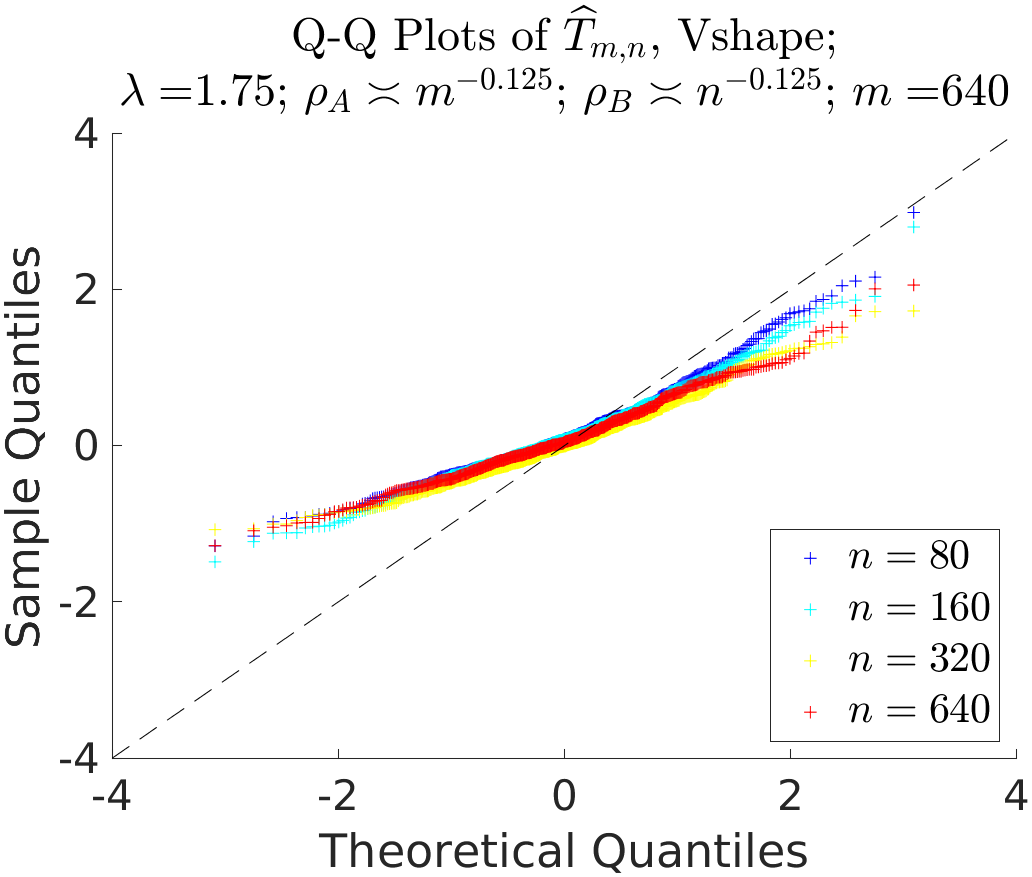}
        \includegraphics[width=0.29\linewidth]{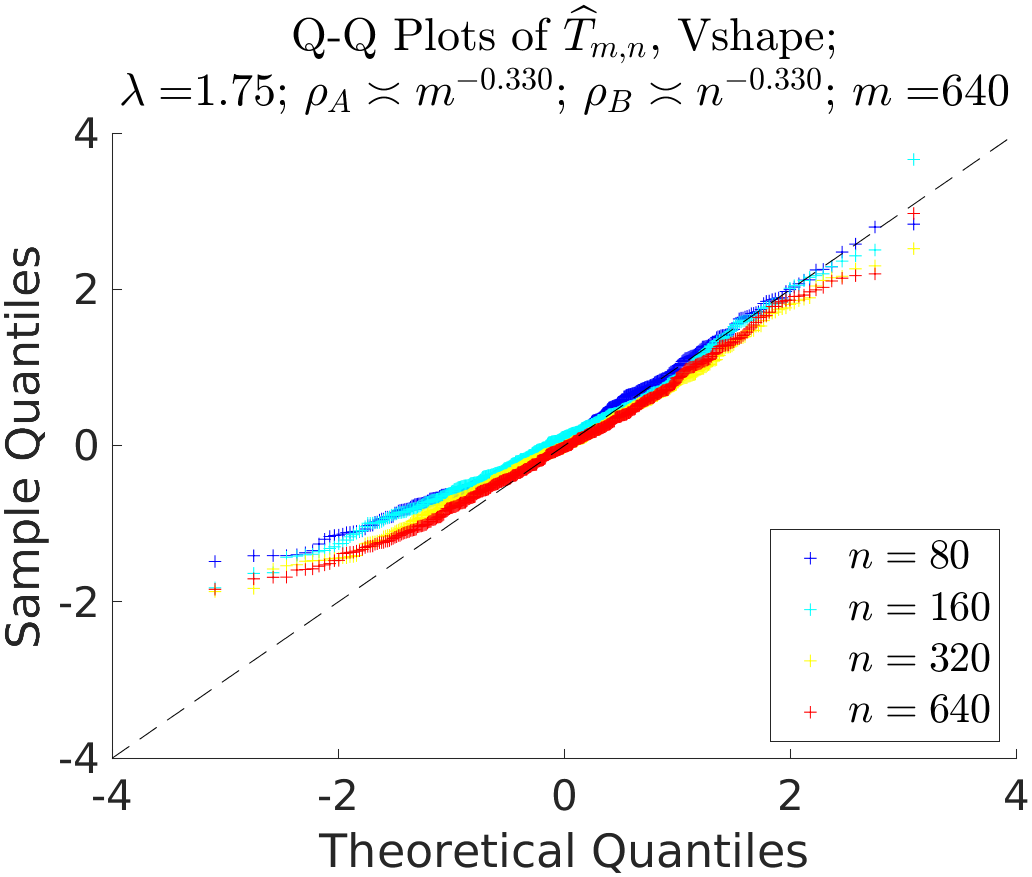}
    \end{adjustbox}
    \begin{adjustbox}{width=1.05\linewidth,center}
        \includegraphics[width=0.29\linewidth]{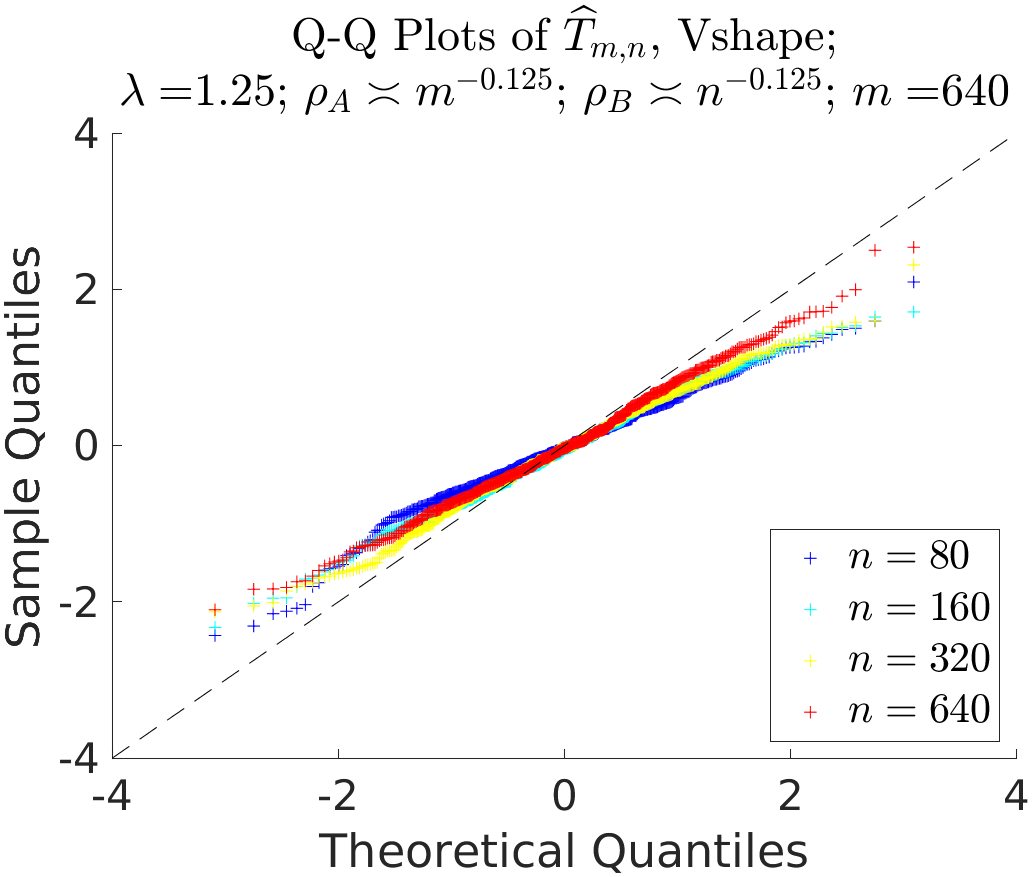}
        \includegraphics[width=0.29\linewidth]{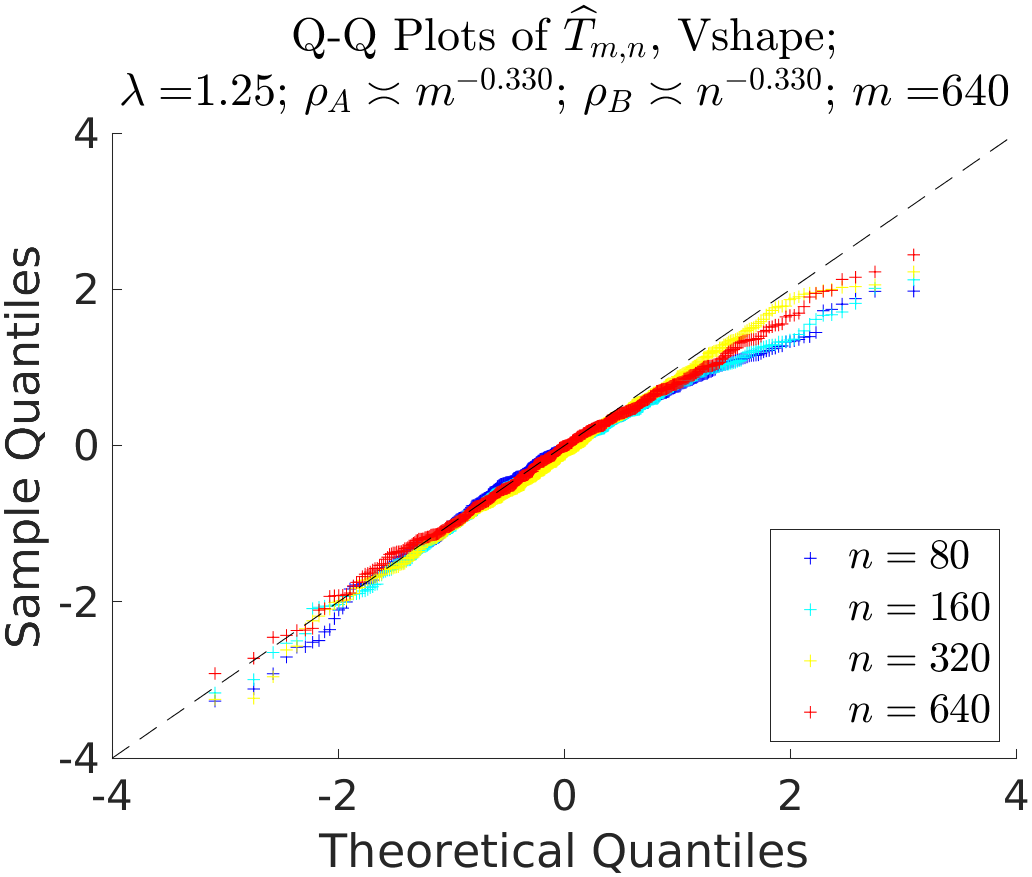}
        \includegraphics[width=0.29\linewidth]{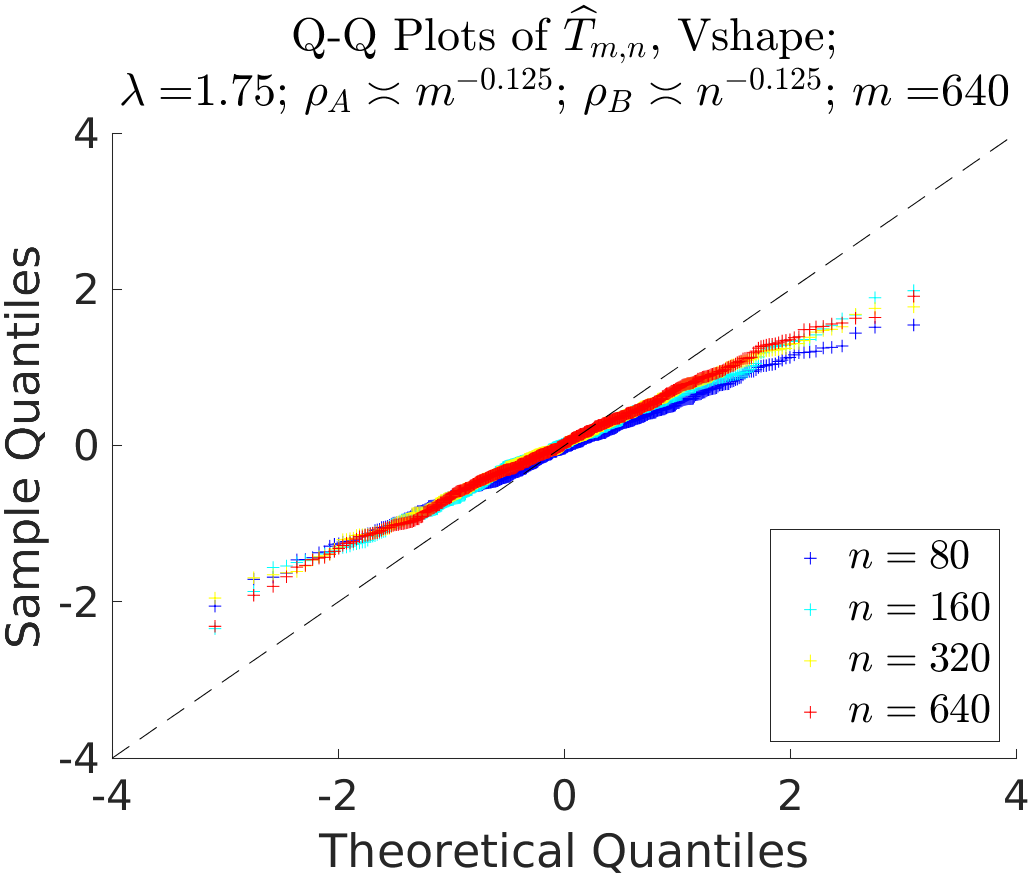}
        \includegraphics[width=0.29\linewidth]{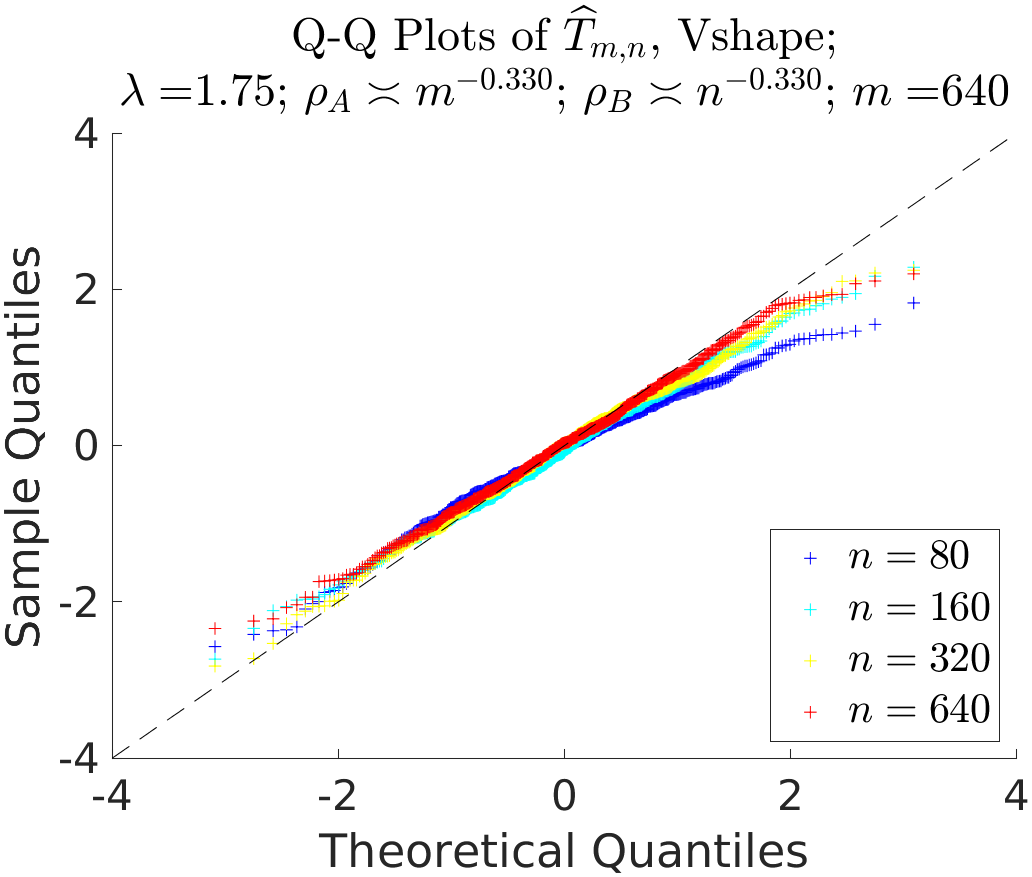}
    \end{adjustbox}
    \caption{Asymptotic normality of reduced network moments with adaptive variance estimation.  Motif is V-shape.
    Columns 1 \& 2: $\lambda=1.25$; columns 3 \& 4: $\lambda=1.75$.
    Columns 1 \& 3: $m\asymp m^{-1.25}$, $n\asymp n^{-1.25}$; columns 3 \& 4: $\rho_A\asymp m^{-1.75}$, $\rho_B\asymp n^{-1.75}$.
    Row 1: model 1 vs model 2; row 2: model 1 vs model 1.
    }
    \label{new-fig::simu-5::degeneracy-vshape}
\end{figure}

\spacingset{1}
{\footnotesize
\bibliographystyle{plainnat}
\bibliography{references}

\begin{thebibliography}{61}
\providecommand{\natexlab}[1]{#1}
\providecommand{\url}[1]{\texttt{#1}}
\expandafter\ifx\csname urlstyle\endcsname\relax
  \providecommand{\doi}[1]{doi: #1}\else
  \providecommand{\doi}{doi: \begingroup \urlstyle{rm}\Url}\fi

\bibitem[Abbe(2017)]{abbe2017community}
Emmanuel Abbe.
\newblock Community detection and stochastic block models: recent developments.
\newblock \emph{The Journal of Machine Learning Research}, 18\penalty0
  (1):\penalty0 6446--6531, 2017.

\bibitem[Adhikari et~al.(2019)Adhikari, Hong, Sampath, Chiappelli, Jahanshad,
  Thompson, Rowland, Calhoun, Du, and Chen]{adhikari2019functional}
Bhim~M Adhikari, L~Elliot Hong, Hemalatha Sampath, Joshua Chiappelli, Neda
  Jahanshad, Paul~M Thompson, Laura~M Rowland, Vince~D Calhoun, Xiaoming Du,
  and Shuo Chen.
\newblock Functional network connectivity impairments and core cognitive
  deficits in schizophrenia.
\newblock \emph{Human brain mapping}, 40\penalty0 (16):\penalty0 4593--4605,
  2019.

\bibitem[Agterberg et~al.(2020)Agterberg, Tang, and
  Priebe]{agterberg2020nonparametric}
Joshua Agterberg, Minh Tang, and Carey Priebe.
\newblock Nonparametric two-sample hypothesis testing for random graphs with
  negative and repeated eigenvalues.
\newblock \emph{arXiv preprint arXiv:2012.09828}, 2020.

\bibitem[Ahmed et~al.(2015)Ahmed, Neville, Rossi, and
  Duffield]{ahmed2015efficient}
Nesreen~K Ahmed, Jennifer Neville, Ryan~A Rossi, and Nick Duffield.
\newblock Efficient graphlet counting for large networks.
\newblock In \emph{2015 IEEE International Conference on Data Mining}, pages
  1--10. IEEE, 2015.

\bibitem[Arroyo et~al.(2021)Arroyo, Sussman, Priebe, and
  Lyzinski]{arroyo2021maximum}
Jes{\'u}s Arroyo, Daniel~L Sussman, Carey~E Priebe, and Vince Lyzinski.
\newblock Maximum likelihood estimation and graph matching in errorfully
  observed networks.
\newblock \emph{Journal of Computational and Graphical Statistics}, 30\penalty0
  (4):\penalty0 1111--1123, 2021.

\bibitem[Banerjee and Ma(2017)]{banerjee2017optimal}
Debapratim Banerjee and Zongming Ma.
\newblock Optimal hypothesis testing for stochastic block models with growing
  degrees.
\newblock \emph{arXiv preprint arXiv:1705.05305}, 2017.

\bibitem[Beran(1987)]{beran1987prepivoting}
Rudolf Beran.
\newblock Prepivoting to reduce level error of confidence sets.
\newblock \emph{Biometrika}, 74\penalty0 (3):\penalty0 457--468, 1987.

\bibitem[Beran(1988)]{beran1988prepivoting}
Rudolf Beran.
\newblock Prepivoting test statistics: a bootstrap view of asymptotic
  refinements.
\newblock \emph{Journal of the American Statistical Association}, 83\penalty0
  (403):\penalty0 687--697, 1988.

\bibitem[Bhattacharya et~al.(2022)Bhattacharya, Das, and
  Mukherjee]{bhattacharya2022motif}
Bhaswar~B Bhattacharya, Sayan Das, and Sumit Mukherjee.
\newblock Motif estimation via subgraph sampling: The fourth-moment phenomenon.
\newblock \emph{The Annals of Statistics}, 50\penalty0 (2):\penalty0 987--1011,
  2022.

\bibitem[Bhattacharyya and Bickel(2015)]{bhattacharyya2015subsampling}
Sharmodeep Bhattacharyya and Peter~J Bickel.
\newblock Subsampling bootstrap of count features of networks.
\newblock \emph{The Annals of Statistics}, 43\penalty0 (6):\penalty0
  2384--2411, 2015.

\bibitem[Bickel and Chen(2009)]{bickel2009nonparametric}
Peter~J Bickel and Aiyou Chen.
\newblock A nonparametric view of network models and newman--girvan
  modularities.
\newblock \emph{Proceedings of the National Academy of Sciences}, 106\penalty0
  (50):\penalty0 21068--21073, 2009.

\bibitem[Bickel et~al.(2011)Bickel, Chen, and Levina]{bickel2011method}
Peter~J Bickel, Aiyou Chen, and Elizaveta Levina.
\newblock The method of moments and degree distributions for network models.
\newblock \emph{The Annals of Statistics}, 39\penalty0 (5):\penalty0
  2280--2301, 2011.

\bibitem[Bravo-Hermsdorff et~al.(2021)Bravo-Hermsdorff, Gunderson, Maugis, and
  Priebe]{bravo2021principled}
Gecia Bravo-Hermsdorff, Lee~M Gunderson, Pierre-Andr{\'e} Maugis, and Carey~E
  Priebe.
\newblock A principled (and practical) test for network comparison.
\newblock \emph{arXiv preprint arXiv:2107.11403}, 2021.

\bibitem[Chen et~al.(2022)Chen, Luo, Chen, Shinohara, Shou, Initiative,
  et~al.]{chen2022privacy}
Andrew~A Chen, Chongliang Luo, Yong Chen, Russell~T Shinohara, Haochang Shou,
  Alzheimer’s Disease~Neuroimaging Initiative, et~al.
\newblock Privacy-preserving harmonization via distributed combat.
\newblock \emph{Neuroimage}, 248:\penalty0 118822, 2022.

\bibitem[Chen et~al.(2022+)Chen, Josephs, Lin, Zhou, and
  Kolaczyk]{chen2020spectral}
Li~Chen, Nathaniel Josephs, Lizhen Lin, Jie Zhou, and Eric~D Kolaczyk.
\newblock A spectral-based framework for hypothesis testing in populations of
  networks.
\newblock \emph{Statistica Sinica (In press)}, 2022+.

\bibitem[Chen et~al.(2022)Chen, Yao, Tijms, Wang, Wang, Song, Yang, Zhang,
  Zhao, Qu, et~al.]{chen2022four}
Pindong Chen, Hongxiang Yao, Betty~M Tijms, Pan Wang, Dawei Wang, Chengyuan
  Song, Hongwei Yang, Zengqiang Zhang, Kun Zhao, Yida Qu, et~al.
\newblock Four distinct subtypes of alzheimer's disease based on resting-state
  connectivity biomarkers.
\newblock \emph{Biological Psychiatry}, 2022.

\bibitem[Chen and Kato(2019)]{chen2019randomized}
Xiaohui Chen and Kengo Kato.
\newblock Randomized incomplete {$U$}-statistics in high dimensions.
\newblock \emph{The Annals of Statistics}, 47\penalty0 (6):\penalty0
  3127--3156, 2019.

\bibitem[DasGupta(2008)]{dasgupta2008asymptotic}
Anirban DasGupta.
\newblock \emph{Asymptotic theory of statistics and probability}, volume 180.
\newblock Springer, 2008.

\bibitem[Decelle et~al.(2011)Decelle, Krzakala, Moore, and
  Zdeborov{\'a}]{decelle2011inference}
Aurelien Decelle, Florent Krzakala, Cristopher Moore, and Lenka Zdeborov{\'a}.
\newblock Inference and phase transitions in the detection of modules in sparse
  networks.
\newblock \emph{Physical Review Letters}, 107\penalty0 (6):\penalty0 065701,
  2011.

\bibitem[Fan et~al.(2012)Fan, Han, and Gu]{fan2012estimating}
Jianqing Fan, Xu~Han, and Weijie Gu.
\newblock Estimating false discovery proportion under arbitrary covariance
  dependence.
\newblock \emph{Journal of the American Statistical Association}, 107\penalty0
  (499):\penalty0 1019--1035, 2012.

\bibitem[Friguet et~al.(2009)Friguet, Kloareg, and Causeur]{friguet2009factor}
Chlo{\'e} Friguet, Maela Kloareg, and David Causeur.
\newblock A factor model approach to multiple testing under dependence.
\newblock \emph{Journal of the American Statistical Association}, pages
  1406--1415, 2009.

\bibitem[Gao and Lafferty(2017{\natexlab{a}})]{gao2017testinga}
Chao Gao and John Lafferty.
\newblock Testing network structure using relations between small subgraph
  probabilities.
\newblock \emph{arXiv preprint arXiv:1704.06742}, 2017{\natexlab{a}}.

\bibitem[Gao and Lafferty(2017{\natexlab{b}})]{gao2017testingb}
Chao Gao and John Lafferty.
\newblock Testing for global network structure using small subgraph statistics.
\newblock \emph{arXiv preprint arXiv:1710.00862}, 2017{\natexlab{b}}.

\bibitem[Gao et~al.(2015)Gao, Lu, and Zhou]{gao2015rate}
Chao Gao, Yu~Lu, and Harrison~H Zhou.
\newblock Rate-optimal graphon estimation.
\newblock \emph{The Annals of Statistics}, 43\penalty0 (6):\penalty0
  2624--2652, 2015.

\bibitem[Gao et~al.(2018)Gao, Ma, Zhang, and Zhou]{gao2018community}
Chao Gao, Zongming Ma, Anderson~Y Zhang, and Harrison~H Zhou.
\newblock Community detection in degree-corrected block models.
\newblock \emph{Annals of Statistics}, 46\penalty0 (5):\penalty0 2153--2185,
  2018.

\bibitem[Genovese and Wasserman(2002)]{genovese2002operating}
Christopher Genovese and Larry Wasserman.
\newblock Operating characteristics and extensions of the false discovery rate
  procedure.
\newblock \emph{Journal of the Royal Statistical Society Series B: Statistical
  Methodology}, 64\penalty0 (3):\penalty0 499--517, 2002.

\bibitem[Ghoshdastidar and Von~Luxburg(2018)]{ghoshdastidar2018practical}
Debarghya Ghoshdastidar and Ulrike Von~Luxburg.
\newblock Practical methods for graph two-sample testing.
\newblock \emph{Advances in Neural Information Processing Systems}, 31, 2018.

\bibitem[Ghoshdastidar et~al.(2017)Ghoshdastidar, Gutzeit, Carpentier, and von
  Luxburg]{ghoshdastidar2017two}
Debarghya Ghoshdastidar, Maurilio Gutzeit, Alexandra Carpentier, and Ulrike von
  Luxburg.
\newblock Two-sample tests for large random graphs using network statistics.
\newblock In \emph{Conference on Learning Theory}, pages 954--977. PMLR, 2017.

\bibitem[Ghoshdastidar et~al.(2020)Ghoshdastidar, Gutzeit, Carpentier, and
  Von~Luxburg]{ghoshdastidar2020two}
Debarghya Ghoshdastidar, Maurilio Gutzeit, Alexandra Carpentier, and Ulrike
  Von~Luxburg.
\newblock Two-sample hypothesis testing for inhomogeneous random graphs.
\newblock \emph{The Annals of Statistics}, 48\penalty0 (4):\penalty0
  2208--2229, 2020.

\bibitem[Ginestet et~al.(2017)Ginestet, Li, Balachandran, Rosenberg, and
  Kolaczyk]{ginestet2017hypothesis}
Cedric~E Ginestet, Jun Li, Prakash Balachandran, Steven Rosenberg, and Eric~D
  Kolaczyk.
\newblock Hypothesis testing for network data in functional neuroimaging.
\newblock \emph{The Annals of Applied Statistics}, pages 725--750, 2017.

\bibitem[Green and Shalizi(2022)]{green2017bootstrapping}
Alden Green and Cosma~Rohilla Shalizi.
\newblock {Bootstrapping exchangeable random graphs}.
\newblock \emph{Electronic Journal of Statistics}, 16\penalty0 (1):\penalty0
  1058 -- 1095, 2022.
\newblock \doi{10.1214/21-EJS1896}.

\bibitem[Hall(2013)]{hall2013bootstrap}
Peter Hall.
\newblock \emph{The Bootstrap and Edgeworth Expansion}.
\newblock Springer Science \& Business Media, 2013.

\bibitem[Hall and Martin(1988)]{hall1988bootstrap}
Peter Hall and Michael~A Martin.
\newblock On bootstrap resampling and iteration.
\newblock \emph{Biometrika}, 75\penalty0 (4):\penalty0 661--671, 1988.

\bibitem[Hladky et~al.(2021)Hladky, Pelekis, and Sileikis]{hladky2021limit}
Jan Hladky, Christos Pelekis, and Matas Sileikis.
\newblock A limit theorem for small cliques in inhomogeneous random graphs.
\newblock \emph{Journal of Graph Theory}, 97\penalty0 (4):\penalty0 578--599,
  2021.

\bibitem[Hoeffding(1948)]{hoeffding1948class}
Wassily Hoeffding.
\newblock A class of statistics with asymptotically normal distribution.
\newblock \emph{The Annals of Mathematical Statistics}, 19\penalty0
  (3):\penalty0 293--325, 1948.

\bibitem[Hunter et~al.(2008)Hunter, Handcock, Butts, Goodreau, and
  Morris]{hunter2008ergm}
David~R Hunter, Mark~S Handcock, Carter~T Butts, Steven~M Goodreau, and Martina
  Morris.
\newblock ergm: A package to fit, simulate and diagnose exponential-family
  models for networks.
\newblock \emph{Journal of Statistical Software}, 24\penalty0 (3), 2008.

\bibitem[Jin et~al.(2018)Jin, Ke, and Luo]{jin2018network}
Jiashun Jin, Zheng Ke, and Shengming Luo.
\newblock Network global testing by counting graphlets.
\newblock In \emph{International Conference on Machine Learning}, pages
  2333--2341. PMLR, 2018.

\bibitem[Kolaczyk et~al.(2020)Kolaczyk, Lin, Rosenberg, Walters, and
  Xu]{kolaczyk2020averages}
Eric~D Kolaczyk, Lizhen Lin, Steven Rosenberg, Jackson Walters, and Jie Xu.
\newblock Averages of unlabeled networks: Geometric characterization and
  asymptotic behavior.
\newblock \emph{The Annals of Statistics}, 48\penalty0 (1):\penalty0 514--538,
  2020.

\bibitem[Leskovec and Mcauley(2012)]{leskovec2012learning}
Jure Leskovec and Julian Mcauley.
\newblock Learning to discover social circles in ego networks.
\newblock \emph{Advances in Neural Information Processing Systems}, 25, 2012.

\bibitem[Levin and Levina(2019)]{levin2019bootstrapping}
Keith Levin and Elizaveta Levina.
\newblock Bootstrapping networks with latent space structure.
\newblock \emph{arXiv preprint arXiv:1907.10821}, 2019.

\bibitem[Li and Li(2018)]{li2018two}
Yezheng Li and Hongzhe Li.
\newblock Two-sample test of community memberships of weighted stochastic block
  models.
\newblock \emph{arXiv preprint arXiv:1811.12593}, 2018.

\bibitem[Lunde and Sarkar(2019)]{lunde2019subsampling}
Robert Lunde and Purnamrita Sarkar.
\newblock Subsampling sparse graphons under minimal assumptions.
\newblock \emph{arXiv preprint arXiv:1907.12528}, 2019.

\bibitem[Lyzinski et~al.(2015)Lyzinski, Fishkind, Fiori, Vogelstein, Priebe,
  and Sapiro]{lyzinski2015graph}
Vince Lyzinski, Donniell~E Fishkind, Marcelo Fiori, Joshua~T Vogelstein,
  Carey~E Priebe, and Guillermo Sapiro.
\newblock Graph matching: Relax at your own risk.
\newblock \emph{IEEE Transactions on Pattern Analysis and Machine
  Intelligence}, 38\penalty0 (1):\penalty0 60--73, 2015.

\bibitem[Maesono(1997)]{maesono1997edgeworth}
Yoshihiko Maesono.
\newblock Edgeworth expansions of a studentized u-statistic and a jackknife
  estimator of variance.
\newblock \emph{Journal of Statistical Planning and Inference}, 61\penalty0
  (1):\penalty0 61--84, 1997.

\bibitem[Maugis et~al.(2020)Maugis, Olhede, Priebe, and
  Wolfe]{maugis2020testing}
P-AG Maugis, SC~Olhede, CE~Priebe, and PJ~Wolfe.
\newblock Testing for equivalence of network distribution using subgraph
  counts.
\newblock \emph{Journal of Computational and Graphical Statistics}, 29\penalty0
  (3):\penalty0 455--465, 2020.

\bibitem[Maugis(2020)]{maugis2020central}
PA~Maugis.
\newblock Central limit theorems for local network statistics.
\newblock \emph{arXiv preprint arXiv:2006.15738}, 2020.

\bibitem[Olhede and Wolfe(2014)]{olhede2014network}
Sofia~C Olhede and Patrick~J Wolfe.
\newblock Network histograms and universality of blockmodel approximation.
\newblock \emph{Proceedings of the National Academy of Sciences}, 111\penalty0
  (41):\penalty0 14722--14727, 2014.

\bibitem[Sabanayagam et~al.(2021)Sabanayagam, Vankadara, and
  Ghoshdastidar]{sabanayagam2021graphon}
Mahalakshmi Sabanayagam, Leena~Chennuru Vankadara, and Debarghya Ghoshdastidar.
\newblock Graphon based clustering and testing of networks: Algorithms and
  theory.
\newblock \emph{arXiv preprint arXiv:2110.02722}, 2021.

\bibitem[Serfling(2009)]{serfling2009approximation}
Robert~J Serfling.
\newblock \emph{Approximation theorems of mathematical statistics}.
\newblock John Wiley \& Sons, 2009.

\bibitem[Shao et~al.(2023)Shao, Xia, and Zhang]{shao2023u}
Meijia Shao, Dong Xia, and Yuan Zhang.
\newblock U-statistic reduction: Higher-order accurate risk control and
  statistical-computational trade-off, with application to network
  method-of-moments.
\newblock \emph{arXiv preprint arXiv:2306.03793}, 2023.

\bibitem[Tang et~al.(2017)Tang, Athreya, Sussman, Lyzinski, Park, and
  Priebe]{tang2017semiparametric}
Minh Tang, Avanti Athreya, Daniel~L Sussman, Vince Lyzinski, Youngser Park, and
  Carey~E Priebe.
\newblock A semiparametric two-sample hypothesis testing problem for random
  graphs.
\newblock \emph{Journal of Computational and Graphical Statistics}, 26\penalty0
  (2):\penalty0 344--354, 2017.

\bibitem[Tsitsulin et~al.(2018)Tsitsulin, Mottin, Karras, Bronstein, and
  M{\"u}ller]{tsitsulin2018netlsd}
Anton Tsitsulin, Davide Mottin, Panagiotis Karras, Alexander Bronstein, and
  Emmanuel M{\"u}ller.
\newblock Netlsd: hearing the shape of a graph.
\newblock In \emph{Proceedings of the 24th ACM SIGKDD International Conference
  on Knowledge Discovery \& Data Mining}, pages 2347--2356, 2018.

\bibitem[Wasserman(2006)]{wasserman2006all}
Larry Wasserman.
\newblock \emph{All of Nonparametric Statistics}.
\newblock Springer Science \& Business Media, 2006.

\bibitem[Wills and Meyer(2020)]{wills2020metrics}
Peter Wills and Fran{\c{c}}ois~G Meyer.
\newblock Metrics for graph comparison: a practitioner’s guide.
\newblock \emph{Plos one}, 15\penalty0 (2):\penalty0 e0228728, 2020.

\bibitem[Yang et~al.(2013)Yang, McAuley, and Leskovec]{yang2013community}
Jaewon Yang, Julian McAuley, and Jure Leskovec.
\newblock Community detection in networks with node attributes.
\newblock In \emph{2013 IEEE 13th international conference on data mining},
  pages 1151--1156. IEEE, 2013.

\bibitem[Yang et~al.(2014)Yang, Han, and Airoldi]{yang2014nonparametric}
Justin Yang, Christina Han, and Edoardo Airoldi.
\newblock Nonparametric estimation and testing of exchangeable graph models.
\newblock In \emph{Artificial Intelligence and Statistics}, pages 1060--1067.
  PMLR, 2014.

\bibitem[Young and Scheinerman(2007)]{young2007random}
Stephen~J Young and Edward~R Scheinerman.
\newblock Random dot product graph models for social networks.
\newblock In \emph{International Workshop on Algorithms and Models for the
  Web-Graph}, pages 138--149. Springer, 2007.

\bibitem[Yuan and Wen(2021)]{yuan2021practical}
Mingao Yuan and Qian Wen.
\newblock A practical two-sample test for weighted random graphs.
\newblock \emph{Journal of Applied Statistics}, pages 1--17, 2021.

\bibitem[Zhang and Xia(2022)]{zhang2020edgeworth}
Yuan Zhang and Dong Xia.
\newblock Edgeworth expansions for network moments.
\newblock \emph{The Annals of Statistics}, 50\penalty0 (2):\penalty0 726--753,
  2022.

\bibitem[Zhang et~al.(2017)Zhang, Levina, and Zhu]{zhang2017estimating}
Yuan Zhang, Elizaveta Levina, and Ji~Zhu.
\newblock Estimating network edge probabilities by neighbourhood smoothing.
\newblock \emph{Biometrika}, 104\penalty0 (4):\penalty0 771--783, 2017.

\bibitem[Zhao(2023)]{zhao2023graph}
Yufei Zhao.
\newblock \emph{Graph Theory and Additive Combinatorics: Exploring Structure
  and Randomness}.
\newblock Cambridge University Press, 2023.

\end{thebibliography}
}
\end{document}